\documentclass{aa}  
\usepackage{multirow}
\usepackage{fontawesome}
\usepackage{placeins}
\usepackage[colorlinks=true,linkcolor=blue,citecolor=blue,urlcolor=blue]{hyperref}
\usepackage{braket}
\usepackage{graphicx}
\usepackage{txfonts}
\usepackage{ulem}
\usepackage{lipsum}
\newcount\Comments  
\usepackage{color}
\definecolor{darkgreen}{rgb}{0,0.5,0}
\newcommand{\kibitz}[2]{\ifnum\Comments=1\textcolor{#1}{#2}\fi}

\usepackage{mathtools}
\usepackage[absolute]{textpos}
\DeclarePairedDelimiter\abs{\lvert}{\rvert}%
\DeclarePairedDelimiter\norm{\lVert}{\rVert}%
\makeatletter
\let\oldabs\abs
\def\abs{\@ifstar{\oldabs}{\oldabs*}}
\let\oldnorm\norm
\def\norm{\@ifstar{\oldnorm}{\oldnorm*}}
\makeatother
\def\MS{\text{M}_{\odot}}

\def\gs{\textsc{GravSphere}}
\def\gsa{\textsc{PyGravSphere}}
\begin{document} 
\newcommand\ww{7.3cm}

\title{GravSphere2: A higher order Jeans method for mass modeling spherical stellar systems}
\titlerunning{GravSphere 2: An extended higher-order Jeans modeling approach}

\author{Andr\'es Ba\~nares-Hern\'andez
\inst{1,2}\thanks{\href{mailto:a.banareshernandez@gmail.com}{\texttt{a.banareshernandez@gmail.com}}}
\and
Justin I. Read
\inst{3}
\and Mariana P. Júlio \inst{4,5}
}
\institute{Instituto de Astrof\'\i sica de Canarias, La Laguna, Tenerife, E-38200, Spain
\and
Departamento de Astrof\'\i sica, Universidad de La Laguna \and  Department of Physics, University of Surrey, Guildford, GU2 7XH, United Kingdom \and Leibniz-Institut für Astrophysik Potsdam (AIP), An der Sternwarte 16, D-14482 Potsdam, Germany \and Institut für Physik und Astronomie, Universität Potsdam, Karl-Liebknecht-Straße 24/25, D-14476 Potsdam, Germany}

\authorrunning{Ba\~nares-Hern\'andez et al.}

\date{\today}

  \abstract
   {}
{Mass-modeling methods are used to infer the gravitational field of stellar systems, from globular clusters to giant elliptical galaxies. While many methods already exist, most require assumptions on the form of the underlying distribution function or binning the data, leading to some loss of information. Furthermore, when only line-of-sight (LOS) data are available, many methods suffer from the well-known mass-anisotropy degeneracy. To overcome these limitations, we developed a new and publicly available mass modeling method, \textsc{GravSphere2}. It combines individual stellar velocities from LOS and proper motion (PM) measurements to solve the Jeans equations up to fourth order, without any data binning. Using flexible functional forms for the velocity anisotropy profiles at second and fourth order, we show how including additional constraints from a new observable, fourth-order PMs, allows us to obtain a full solution along the three dimensions and breaking the mass-anisotropy degeneracy at all orders. We tested our method on mock data for dwarf galaxies, showing how \textsc{GravSphere2} improves on previous methods.}
{\textsc{GravSphere2} introduces four key improvements over previous Jeans mass modeling methods in the literature: (i) we included fourth-order velocity moment equations in both the LOS and PM directions, for the first time, using them to break model degeneracies; (ii) we used a fully general treatment of both the second and fourth-order velocity anisotropies; (iii) we introduced a ``bin-free'' approach where we fit individual tracer velocities and positions using flexible and self-consistent probability density functions that include kurtosis; and (iv) we improved the likelihood sampling by using the nested sampler \textsc{dynesty}.}
{\gs2 was able to recover the mass density, stellar velocity anisotropy, and the logarithmic slope of the mass density profile within its quoted 95\% confidence intervals across almost all mocks over a wide radial range ($0.1 \lesssim r/R_{1/2} \lesssim  10$, where $R_{1/2}$ is the projected half-light radius). As the number of tracers is lowered (even down to just ten tracers) it gracefully degrades, with larger uncertainties but no induced bias. We find that \gs2 outperforms simple mass estimators, suggesting that it is worth using even when only a few LOS velocities are available. Using 1,000 tracers without PMs, \gs2 recovers the logarithmic density slope at $R_{1/2}$ with $12\%$ (25\%) statistical errors for 
cuspy (cored) mock data, enabling us to make a distinction between the two. When including PMs, this result can be improved to $8\%$ (12\%). With only 100 tracers and no PMs, we were still able to recover slopes with $\sim 30\%$ (20\%) errors. \gs2 will become a valuable new tool to hunt for massive black holes and invisible dark matter in spherical stellar systems, from globular clusters and dwarf galaxies to giant ellipticals and galaxy clusters.}
{}
\keywords{
  galaxies: kinematics and dynamics -- galaxy: globular clusters: general -- galaxies: dwarf --
  galaxies: Local Group}
 
   \maketitle
\section{Introduction}
\label{sec:intro}

Many stellar systems are approximately spherical, including globular clusters (GCs) and nearby dwarf spheroidal galaxies (dSphs). Both GCs and dSphs contain seemingly invisible mass components that can be inferred from the anomalous motions of their stars and gas (see e.g., \citealt{1978PhDT.......195B, 1980ApJ...238..471R, 1987ARA&A..25..425T, 2002MNRAS.333..697L, van2010new, battaglia2013internal}, for more recent examples in GCs and dSphs;  see e.g., \citealt{Evans:2021bsh, 2024MNRAS.529..331D,  2024ApJ...975..268S, 2025A&A...693A.104B} and \citealt{2019MNRAS.484.1401R,2020ApJ...904...45H,2022NatAs...6..659B, 2023A&A...676A..63B,2024MNRAS.535.1015D,2025A&A...699A.347A, 2025A&A...700A..77P, 2025arXiv250820711V}). For GCs, the invisible mass most likely constitutes stellar remnants or an intermediate-mass black hole (IMBH), although some GCs may also contain dark matter. By contrast, dSphs appear to be dominated at all radii by dark matter. With increasingly available spectroscopic and astrometric data from recent surveys, GCs promise an unprecedented natural laboratory to study stellar remnants and hunt for IMBHs \citep{2017IJMPD..2630021M, 2020A&ARv..28....4N, 2020ARA&A..58..257G, 2023arXiv231112118A, 2023MNRAS.522.5740V, 2024Natur.631..285H,2025A&A...693A.104B}, while dSphs promise the same for dark matter \citep{2017arXiv170701348T, 2018MNRAS.481..860R,2019MNRAS.484.1401R,2022NatAs...6..659B,2024MNRAS.535.1015D,2024ApJ...970....1V, 2025arXiv250820711V,2025PhRvL.134o1001Z}.

The above proliferation in data has led to an associated proliferation in mass modeling methodologies. All methods seek to solve the steady-state Collisionless Boltzmann Equation (CBE), also known as the Vlasov Equation (e.g., \citealt{Binney2008}), expressed as

\begin{equation}
\label{eq:vlasov}
    \frac{\text{d} f}{\text{d} t} = \frac{\partial f}{\partial t} + \nabla_{\mathbf{x}} f \cdot 
    \mathbf{v} - \nabla_{\mathbf{v}} f \cdot \nabla_{\mathbf{{x}}} \Phi = 0,
\end{equation}
where $\Phi$ is the gravitational potential resulting from  
all mass components in the system. Then, satisfying the Poisson equation with mass density, $\rho$, we have
\begin{equation}
\label{eq:poisson}
    \nabla^2_{\mathbf{{x}}} \Phi = 4 \pi G \rho,
\end{equation}
and $f$ is the phase-space distribution function.

By solving for the gravitational potential and subtracting the expected potential from all visible components we can infer the presence of invisible masses for future study, such as black holes and dark matter (e.g., \citealt{van2010new, 2009ApJ...704.1274W}), which is usually done in the steady state approximation, assuming $\frac{\partial f}{\partial t} = 0$.

There are number of methods for solving the CBE. Jeans methods solve moments of the CBE \citep{1922MNRAS..82..122J}, having the advantages that they are fast\footnote{This is particularly timely when assessing increasingly large samples with thousands (in some cases approaching millions, \citealt{2025ApJ...983...95H}) of tracers.} and minimize potential biases that could creep in due to assumptions about the form of  the distribution function, $f$ (e.g., \citealt{2008MNRAS.390...71C, 2013MNRAS.436.2598W, 2013MNRAS.429.3079M, 2015arXiv150405533C, 2017MNRAS.471.4541R}). The main disadvantages, however, are that the data typically need to be binned, while $f$ is not determined. This leads to some loss of information and makes it hard to incorporate stars of uncertain membership, binaries, and/or rigorously model selection effects \citep[e.g.][]{2021MNRAS.501..978R}. Furthermore, since $f$ is not determined, a physical $f$ is not guaranteed. This can result in models that formally require a negative $f$ in some regions of phase space \citep{1992ApJ...391..531D, 2013MNRAS.432.3361R} which  should then be discarded \citep[e.g.,][]{2006ApJ...642..752A}.

Distribution function methods assume some form for $f$ and then fit this to the data (e.g., \citealt{2015MNRAS.454..576G, 2019MNRAS.482.1525V,2021MNRAS.501..978R,2023MNRAS.522.5320D}). In such cases, $f$ can also be implicitly expressed as a superposition of orbits with Schwarzschild methods (e.g., \citealt{1979ApJ...232..236S, 2008ApJ...682..841C, 2008MNRAS.385..647V, 2013MNRAS.433.3173B}) or with  made-to-measure methods, in which $N$-body simulations are evolved to fit some data constraints (e.g., \citealt{1996MNRAS.282..223S, 2007MNRAS.376...71D, 2009MNRAS.395.1079D, 2013MNRAS.430.1928H}). In addition, \cite{2025MNRAS.538.1442L} proposed a novel nonparametric distribution function method. Distribution function methods have the advantages that no data binning is required and $f$ is determined by the data. However, if the functional form for $f$ (or its implicit specification via other methods) does not encompass the true solution, then bias can creep in. 

Finally, numerous machine-learning methods have been proposed that learn from training data to either accelerate distribution function fitting or incorporate cosmological priors into the mass modeling pipeline (e.g., \citealt{2023MNRAS.519.4384E, 2025arXiv250303812N, 2025arXiv250500763L, 2025arXiv250307717S}).

Given the proliferation of mass modeling methods, we may wonder why yet another method is needed. However, there are two key problems that remain to be addressed, even for spherical stellar systems. Firstly, while the amount and quality of data continue to improve, there will always be interesting systems with very few stellar tracers (e.g., \citealt{2021MNRAS.505.5686C, 2024ApJ...961...92S, 2025arXiv250504475P}). Typically, simple mass estimators are used for low tracer numbers, but these have been shown to be potentially significantly biased \citep{2019MNRAS.484..245L}. More sophisticated, unbiased, mass modeling methods for low tracer numbers are only just being developed (e.g., \citealt{2025arXiv250303812N}). 

Secondly, there is a well-known degeneracy between the cumulative mass (which we are typically most interested in) and the velocity anisotropy profile, which is a measure of how circular or radial the orbit distribution is (e.g., \citealt{1987MNRAS.224...13D, 1987ApJ...313..121M, 1990AJ.....99.1548M, Wilkinson:2001ut, 2003MNRAS.343..401L, 2009MNRAS.395...76D, 2017MNRAS.471.4541R,2021MNRAS.501..978R}). For Jeans modeling, this occurs because the Jeans equations do not constitute a closed system, meaning that for a given mass model and tracer profile, a unique solution is not guaranteed (e.g., \citealt{Binney2008}). This typically arises when only projected line-of-sight (LOS) velocities are available.

Two key methods have been proposed to break the mass-anisotropy degeneracy. Firstly, we can incorporate higher moments of the velocity distribution as data constraints. At infinite order, these fully specify the phase-space distribution function and ensure its positivity  \citep{1990AJ.....99.1548M, 1992ApJ...391..531D, 2013MNRAS.432.3361R}. After the usually-used second moment, the next nontrivial moment for a spherical, nonrotating, pressure-supported system is the fourth moment. This encodes information about deviations from Gaussianity, including the "tailedness" of the distribution and the kurtosis. It can be incorporated directly into the Jeans equations, thereby introducing a new fourth-order anisotropy parameter \citep{1990AJ.....99.1548M, 2003MNRAS.343..401L, 2005MNRAS.363..918L, 2013MNRAS.432.3361R, 2024arXiv240412671W}. Alternatively, we can perform luminosity-weighted integrals over the fourth moment to derive the virial shape parameters (VSPs) that marginalize out the higher-order anisotropy \citep{1990AJ.....99.1548M, 2014MNRAS.441.1584R, 2017MNRAS.471.4541R}. Secondly, we can include proper motions (PMs) to gain additional phase-space information that is independent of the LOS velocities (e.g., \citealt{van2010new, 2013MNRAS.436.2598W}). Until recently, such PMs have only been available for GCs (e.g., \citealt{2003ApJ...595..187M, anderson2010new, 2022MNRAS.514..806V,2023MNRAS.522.5320D,2024Natur.631..285H,2025A&A...693A.104B}). However, data are becoming increasingly available  for nearby dwarf galaxies, with recent mass models incorporating PMs recently reported for the Draco and Sculptor dSphs \citep{2024ApJ...970....1V, 2025arXiv250820711V}.

Although Jeans mass modeling methods typically use data binning \citep[e.g.][]{2017MNRAS.471.4541R}, this is not actually a requirement. Several authors have explored assuming a local Gaussian distribution function constrained by the second moment to create a bin-free Jeans method (e.g., \citealt{2013MNRAS.436.2598W, 2013MNRAS.429.3079M}). With the widespread availability of individually resolved stellar velocities, this approach is preferable and has all the advantages of having $f$ that distribution function methods enjoy.\footnote{This is true, up to a given order of moments, provided that the Jeans method has those moments self-consistently implemented in the velocity distribution function.}

In this work, we present the \gs2 mass modeling method. This builds on \gs \: \citep{2017MNRAS.471.4541R, 2021MNRAS.505.5686C} with the specific goals of: (i) reducing bias; (ii) maximizing the information content in the data; (iii) avoiding the need for data binning; and (iv) having the capacity able to successfully model, without bias, very low numbers of tracers. To achieve these goals, we self-consistently solved the second- and fourth-order Jeans equations, following the framework in \cite{2013MNRAS.432.3361R}, extending it to include a new observable: fourth-order PMs. We show that this approach breaks the degeneracies in the higher order Jeans equations by solving along their three projected components. Unlike previous approaches, which have made more restrictive assumptions \citep[e.g.,][]{1990AJ.....99.1548M, 2003MNRAS.343..401L, 2005MNRAS.363..918L, 2013MNRAS.432.3361R, 2024arXiv240412671W}, we also included fully independent and radially varying anisotropies at both orders. We modeled the individual stellar velocities, rather than binning the data, by introducing a local non-Gaussian velocity probability distribution function (PDF) constrained by the second and fourth moments. We also modeled the positions of stars without binning, when the relevant data were available. We tested these improvements using mock data for dSphs and put together the Python package, \textsc{GravSphere2}, which is now publicly available\footnote{Available at: \href{https://github.com/dadams42/GravSphere2}{\texttt{https://github.com/dadams42/GravSphere2}}}.

This paper is organized as follows. In Sect.~\ref{sec:2_all}, we introduce the modeling methodology behind \gs2, including our method for solving the fourth-order Jeans equations, which we applied following \cite{2013MNRAS.432.3361R} and extended to include PMs (Sect.~\ref{sec:3d_gs2}). We also describe its implementation using non-Gaussian local velocity PDFs for individual tracers, following \citealt{2020MNRAS.499.5806S} (Sect.~\ref{sec:PDF}), and the functional forms we assume for our mass model and the second and fourth-order stellar velocity anisotropy profiles (Sect.~\ref{sec:massani}). In Sects.~\ref{sec:gcdata} and \ref{sec:sddata}, we introduce the mock data  we used to test \gs2. In Sect.~\ref{sec:llhood}, we describe how we integrate the approach introduced in Sect.~\ref{sec:3d_gs2} to fit these data, specifying the form of the likelihood function and our assumed priors.  Sect.~\ref{sec:results} shows the results of our tests of \gs2 on the mock data. We compared these results with previous methods that were applied to these same data in the literature, including previous versions of \gs. We also explore the limit of low tracer numbers (i.e., corresponding to the faintest or most distant systems) and we compare the performance of \gs2 with simple mass estimators that are widely used in the literature. In Sect.~\ref{sec:disconc}, we briefly discuss the significance of this work and add our concluding remarks, highlighting upcoming works that make use of the new \gs2 method.

\section{Modeling stellar kinematics with GravSphere2}
\label{sec:2_all}

\textsc{GravSphere} is a numerical Jeans solver that was originally introduced in  \cite{2017MNRAS.471.4541R}. It has been further extended and extensively tested with mock data \citep{2017MNRAS.471.4541R, 2018MNRAS.481..860R, 2019MNRAS.485.2010G, 2020MNRAS.498..144G, 2021MNRAS.501..978R, 2021MNRAS.505.5686C, 2022MNRAS.510.2186G, 2024MNRAS.535.1015D, 2025arXiv250303812N, 2025arXiv250418617T}, including simulations deliberately chosen to explore expected bias in dynamical models due to breaking of key assumptions, such as spherical symmetry (triaxial mocks) and dynamical pseudo-equilibrium (tidally stripped mocks). A public version of this \gs1 code, called \textsc{pyGravSphere}, was introduced in \citet{2020MNRAS.498..144G}.\footnote{Available at: \href{https://github.com/AnnaGenina/pyGravSphere}{\texttt{https://github.com/AnnaGenina/pyGravSphere}}.} Its most recent version was introduced and made publicly available in \cite{2021MNRAS.505.5686C}\footnote{Available at: \href{https://github.com/justinread/gravsphere}{\texttt{https://github.com/justinread/gravsphere}}.}. It improved the velocity binning with the \textsc{binulator} feature,  improved the mass model and improved the way that the VSPs contributed to the likelihood. Henceforth, we refer to this version as \gs1.5. \cite{Julio:2023oyg} also introduced and made publicly available a version of \gs1.5 modified to test scalar field dark matter models in dwarf galaxies.\footnote{Available at:  \href{https://github.com/marianajulio/alternative_models_for_gravsphere}{\texttt{https://github.com/marianajulio/alternative \\ \_models\_for\_gravsphere}}.} A version of \gs1.5  with an explicit treatment of rotation has also been presented in \cite{2025arXiv251006905J}. Lastly, self-consistent inclusion of millisecond pulsar accelerations as independent kinematic tracers to constrain mass-anisotropy models in star clusters and improved photometric modeling with an  $\alpha \beta \gamma$ profile were also introduced in \cite{2025A&A...693A.104B}, who used it to probe the presence of an IMBH and stellar remnant populations in Omega Centauri.

\textsc{GravSphere} has been a popular tool which has been used in the study of the dark matter distribution of nearby dwarf galaxies (e.g., \citealt{2018MNRAS.481..860R, 2019MNRAS.484.1401R, 2019MNRAS.485.2010G, 2020JCAP...09..004A,2021MNRAS.501.1188A, 2021MNRAS.505.5686C, 2021arXiv211209374Z, Julio:2023oyg, 2024MNRAS.535.1015D, 2025PhRvL.134o1001Z, 2025arXiv250504475P, 2025arXiv251006905J}), galaxy clusters \citep{2023A&A...677A..24L}, and to model GCs \citep{2025A&A...693A.104B}. In this section, we introduce the improved methodology of \textsc{GravSphere2} for the mass modeling of stellar kinematics. We briefly summarize the key elements of previous incarnations of \gs, focusing here on the novel aspects with respect to these earlier versions \citep{2017MNRAS.471.4541R, 2018MNRAS.481..860R, 2020MNRAS.498..144G, 2021MNRAS.501..978R, 2021MNRAS.505.5686C}.  Crucially, these improvements include the removal of binning in favor of realistic self-consistent velocity distribution functions (Sect.~\ref{sec:PDF}), the incorporation of the fourth-order Jeans equations to treat higher moments, with a more general treatment than previous approaches (Sect.~\ref{sec:3d_gs2}), and the inclusion of higher order PMs. The details of mass-anisotropy models are addressed in Sect.~\ref{sec:massani}.

\subsection{An extended Jeans analysis up to fourth order}
\label{sec:3d_gs2}

\gs2 numerically solves the Jeans equations for a spherical, nonrotating system of tracers, with radial density profile, $\nu_\star(r)$, in a dynamical pseudo-equilibrium \citep[e.g.,][]{Binney2008}, expressed as

\begin{equation}
\label{eq:sph_jeans}
\frac{1}{\nu_{\star}} \frac{\partial \nu_{\star} \sigma^2_r}{\partial r} 
+ \frac{2 \beta \sigma^2_r}{r} = -\frac{G M}{r^2},
\end{equation}
where 
\begin{equation}
\label{eq:beta}
\beta \equiv 1 - \frac{\sigma^2_t}{2 \sigma^2_r}
\end{equation}
is the velocity anisotropy\footnote{Unlike in, e.g., \cite{2025A&A...693A.104B}, $\sigma_t$ corresponds to the total 2D tangential velocity dispersion, rather than the 1D averaged one (thus the factor of two). This is purely notational, but we adopted this change because we found it more convenient to work with this quantity for the subsequent parts of this section.} and we assume Newtonian weak-field gravity with an enclosed mass, $M(r)$, within the 3D radial coordinate, $r$.

As discussed in \cite{2025A&A...693A.104B}, the assumption of nonrotation can be relaxed to some extent by solving for the second velocity moment directly. This is what we aim to do in this study. This is done by subtracting only global averages to the velocity distribution, leaving potential rotation signatures present. These are then reabsorbed when solving the Jeans equation, leading to a quadrature term that is implicitly added to the pure velocity dispersion (see, e.g., the discussion in \citealp{2016ApJ...821...44F}).

The solution for the radial velocity dispersion  \citep{2005MNRAS.363..705M,1994MNRAS.270..271V} is then given by\begin{equation}
\label{eq:tot_disp}
\sigma^2_r (r) = 
    \frac{1}{\nu_{\star}(r) g(r)} \int_r^{\infty} \frac{G M (r') \nu_{\star}(r')}{r'^2} g(r') dr',
\end{equation}
where
\begin{equation}
    \label{eq:g}
    g(r) \equiv \exp \Bigg(2 \int \frac{\beta(r)}{r} dr \Bigg).
\end{equation}
Projecting along the observable LOS and tangential and radial PM components, respectively, we obtain
\begin{equation}
    \label{eq:los_disp}
    \sigma^2_{\rm LOS} (R)
    = \frac{2}{\Sigma_{\star}(R)} \int_R^{\infty} 
    \bigg( 1 -  \frac{R^2}{r^2} \beta(r) \bigg)
    \frac{\nu_{\star}(r) \sigma^2_r(r) r}{\sqrt{r^2 - R^2}} dr,
\end{equation} 

\begin{equation}
    \label{eq:disp_t}
        \sigma^2_{\rm PM, \: t} (R)
    = \frac{2}{\Sigma_{\star}(R)} \int_R^{\infty} 
    \bigg( 1 - \beta(r) \bigg)
    \frac{\nu_{\star}(r) \sigma^2_r(r) r}{\sqrt{r^2 - R^2}} dr,
\end{equation}
and
\begin{equation}
\label{eq:disp_r}
        \sigma^2_{\rm PM, \: R} (R)
    = \frac{2}{\Sigma_{\star}(R)} \int_R^{\infty} 
    \bigg( 1 - \beta(r) +  \frac{R^2}{r^2} \beta(r) \bigg)
    \frac{\nu_{\star}(r) \sigma^2_r(r) r}{\sqrt{r^2 - R^2}} dr,
\end{equation}
where $\Sigma_{\star}(R)$ is the projected stellar tracer surface density.

By considering these three separate components, provided that one has enough data, it is possible to alleviate the well-known problem of the degeneracy between the cumulative mass, $M(r)$, and stellar velocity anisotropy, $\beta(r)$ \citep{1990AJ.....99.1548M,Wilkinson:2001ut,2003MNRAS.343..401L,2009MNRAS.395...76D,2017MNRAS.471.4541R}. Further, we can resort to higher moments of the velocity distribution to place additional constraints that have also been found to alleviate this problem. One popular choice is to use VSPs \citep{1990AJ.....99.1548M, 2014MNRAS.441.1584R,2017MNRAS.471.4541R}, which are two numbers obtained by integrating the luminosity-weighted, fourth-order LOS velocity moment to infinity, thereby marginalizing out any higher order velocity anisotropy. This approach is used in \gs1 and \gs1.5.

The key advantage of using VSPs over explicit modeling of the higher order moment equations is that no higher order analog of $\beta(r)$ is required. However, a key disadvantage is that the VSPs require integrals to infinity over the surface brightness profile and fourth velocity moment, which may both be poorly constrained at large radii. Furthermore, their large radius behavior could be adversely affected by disequilibrium due to tides. Both effects can lead to bias \citep[e.g.,][]{2017MNRAS.471.4541R,2018MNRAS.481..860R,2024MNRAS.535.1015D,2025arXiv250418617T}. For this reason, in \gs2 we take a different approach that (as we  show in Sect.~\ref{sec:results}) is less biased. Instead of using VSPs, we explicitly solve the fourth-order Jeans equations, introducing a flexible functional form for the associated higher order anisotropy that we can then marginalize over. This is more similar to the approach introduced in \cite{2013MNRAS.432.3361R}. Explicitly, we solve the fourth-order moment Jeans equations \citep{1990AJ.....99.1548M}, expressed as 
\begin{equation}
\label{eq:sph_41jeans}
\frac{1}{\nu_{\star}} \frac{\partial \nu_{\star} \braket{v^4_r}}{\partial r} 
- \frac{3}{r} \braket{v_r^2 v_t^2} + \frac{2}{r} \braket{v_r^4}
 = - 3 \sigma_r^2 \frac{G M}{r^2},
\end{equation}

\begin{equation}
\label{eq:sph_42jeans}
\frac{1}{\nu_{\star}} \frac{\partial \nu_{\star} \braket{v^2_r v^2_t}}{\partial r} 
- \frac{1}{r} \braket{v_t^4} + \frac{4}{r} \braket{v_r^2 v^2_t}
 = - \sigma_t^2 \frac{G M}{r^2},
\end{equation}
where the bra-kets ``$\braket{}$'' are used to denote the different velocity moments of the distribution.

As shown in \cite{2013MNRAS.432.3361R}, this system of equations can be further simplified, without loss of generality by first introducing the fourth-order anisotropy analog,
\begin{equation}
\label{eq:4ani}
\beta' \equiv 1 - \frac{3}{2} \frac{\braket{v^2_r v^2_t}}{\braket{v^4_r}},
\end{equation}
which allows the first fourth-order Jeans expression in Eq.~\eqref{eq:sph_41jeans} to be recast in the more familiar form of

\begin{equation}
\label{eq:sph_412jeans}
\frac{1}{\nu_{\star}} \frac{\partial \nu_{\star} \braket{v^4_r}}{\partial r} 
+ \frac{2 \beta'}{r} \braket{v_r^4}
 = - 3 \sigma_r^2 \frac{G M}{r^2},
\end{equation}
which, by comparison to the (second-order) Jeans Eq.~\eqref{eq:sph_jeans}, can be readily solved for the fourth-order radial velocity moment, yielding
\begin{equation}
\label{eq:tot4_disp}
\braket{v^4_r}(r) = 
    \frac{1}{\nu_{\star}(r) g'(r)} \int_r^{\infty} 3 \sigma^2_r(r') \frac{G M (r') \nu_{\star}(r')}{r'^2} g'(r') dr',
\end{equation}
where
\begin{equation}
    \label{eq:gprime}
    g'(r) \equiv \exp \Bigg(2 \int \frac{\beta'(r)}{r} dr \Bigg).
\end{equation}
Therefore, by first solving Eq.~\eqref{eq:sph_jeans} for $\sigma^2_r$ and then using this to obtain the solution $\braket{v^4_r}$ for Eq.~$\eqref{eq:sph_412jeans},$ we can self-consistently solve for the second- and fourth-order Jeans equations. This is the first key change we made in \gs2. 

As for the second-order counterparts, we now want to obtain the projected quantities that are directly related to observations. 
To express these purely in terms of $\braket{v^4_r}$, following \cite{2013MNRAS.432.3361R},  first, it follows from Eq.~\eqref{eq:4ani} that
\begin{equation}
    \label{eq:4anip}
     \frac{\partial \nu_{\star} \braket{v^2_r v^2_t}}{\partial r} = \frac{2}{3} \bigg[ 
     (1 -   \beta') \frac{\partial \nu_{\star} \braket{v^4_r}}{\partial r} - \frac{\partial \beta'}{\partial r} \nu_{\star} \braket{v^4_r}
     \bigg].
\end{equation}
Next, we made a substitution using the second fourth-order Jeans Eq.~\eqref{eq:sph_41jeans} for the first term on the right-hand side,

\begin{equation}
\label{eq:jsub}
\frac{\partial \nu_{\star} \braket{v^2_r v^2_t}}{\partial r} =
 - \frac{2}{3} (1 - \beta') \bigg[ 
 \frac{2 \beta'}{r}  \nu_{\star}  \braket{v_r^4}
  + 3  \nu_{\star}  \sigma_r^2 \frac{G M}{r^2}
 \bigg]
 - \frac{2}{3}\frac{\partial \beta'}{\partial r} \nu_{\star} \braket{v^4_r},
\end{equation}
which can then be substituted into the second fourth-order Jeans Eq.~\eqref{eq:sph_42jeans} to yield an expression for the tangential fourth-order moment in terms of the radial one as\footnote{It should be understood that $\beta, \beta', M, \braket{v_r^4}, \sigma_r$ are all, in general, functions of $r$ (we suppressed these implicit dependencies for notational convenience).}
\begin{equation}
\label{eq:4tan}
\braket{v^4_t}(r) = \frac{4}{3} \bigg[ (1 - \beta')(2 - \beta') - \frac{r}{2} \frac{\partial \beta'}{\partial r}
\bigg]\braket{v^4_r}
+ 2 (\beta' - \beta) \sigma_r^2 \frac{GM}{r}.
\end{equation}
By considering the geometry of the problem, expanding to fourth order, and simplifying in terms of radial moments for the projected LOS component, we find
\begin{align}
    \label{eq:4los}
    \braket{v^4_{\rm LOS}}(R) = \frac{2}{\Sigma_{\star}(R)} \int_R^{\infty} \frac{F_{\rm LOS} (r, R) \nu_{\star}(r) r}{\sqrt{r^2 - R^2}} dr,
\end{align}
where
\begin{align}
    \label{eq:flos}
    F_{\rm LOS} (r, R) \equiv \bigg( 1 -  2 \beta' \frac{R^2}{r^2}  + \frac{1}{2} \beta' (1 + \beta') \frac{R^4}{r^4}
     - \frac{1}{4} \frac{\partial \beta'}{ \partial r} \frac{R^4}{r^3}\bigg) \braket{v^4_r}
      \nonumber\\
      + \frac{3}{4} (\beta' - \beta) \sigma_r^2 GM \frac{R^4}{r^5}.
\end{align}
As noted in \cite{2013MNRAS.432.3361R}, this approach constitutes a significant theoretical improvement over second-order Jeans methods by ensuring physical consistency of a given mass model with higher order velocity moments. This is not trivially warranted, as Jeans models can in some cases lead to unphysical solutions implying negative distribution functions or be inconsistent with the velocity distribution that is implicitly assumed to map velocity dispersions \citep{1992ApJ...391..531D, 2013MNRAS.432.3361R}. We note that this is explicitly addressed in  Sect.~\ref{sec:PDF}. It also has the promise of serving as a means to further constrain mass models by considering additional higher order moments \citep{2013MNRAS.432.3361R}.

One potential disadvantage, however, which was also noted in \cite{2013MNRAS.432.3361R}, is that the constraining power of this approach can be compromised due to the fact that a fully general treatment leads to the introduction of the fourth-order anisotropy analog, $\beta',$ meaning that one may not  be able to unambiguously constrain solutions to the fourth-order Jeans equation for a given mass model and (second-order) anisotropy, $\beta,$ leading to a problem that is similar to the original mass-anisotropy degeneracy. One approach to this problem is to impose a prescription that fixes $\beta'.$ This can be done, for example, by assuming distribution function models which predict these \citep{2002MNRAS.333..697L,2005MNRAS.363..918L}. More generally, as explored in \cite{1992ApJ...391..531D, 2013MNRAS.432.3361R}, assuming a particular form for the distribution function amounts to imposing a prescription to solve all the moments of the Jeans equations (beyond the fourth order included). One problem with this approach, however, is that it might restrict the space of solutions unnecessarily and bias the models, unless one is able to be sufficiently judicious and general in specifying a form for the distribution function.

To tackle this potential problem and to make full use of the available kinematic information, similarly to how we can use PM velocity dispersions to constrain the second-order system, we extend this methodology by also considering the fourth-order velocity moments for the tangential and radial PM components, constraining also the fourth-order system. We show that this information can be then used to further constrain the system while allowing for a general treatment of $\beta'.$ As in Eq.~\eqref{eq:4los}, this leads to an expression for the tangential and radial fourth-order moments,  which can be obtained (respectively) as

\begin{align}
    \label{eq:4pmt}
    \braket{v^4_{\rm PM, t}}(R) = \frac{2}{\Sigma_{\star}(R)} \int_R^{\infty} 
    \frac{F_{\rm PM,t} (r, R) \nu_{\star}(r) r}{\sqrt{r^2 - R^2}} dr,
\end{align}
and

\begin{align}
    \label{eq:4pmr}
    \braket{v^4_{\rm PM, R}} (R)= \frac{2}{\Sigma_{\star}(R)} \int_R^{\infty} \frac{F_{\rm PM,R} (r, R)\nu_{\star}(r) r}{\sqrt{r^2 - R^2}} dr,
\end{align}
where
\begin{align}
    \label{eq:fpmt}
    F_{\rm PM,t} (r, R) \equiv \frac{1}{2}\bigg[ (1 - \beta')(2 - \beta') - \frac{r}{2} \frac{\partial \beta'}{\partial r} \bigg] \braket{v^4_r}
      \nonumber\\
      + \frac{3}{4} (\beta' - \beta) \sigma_r^2 \frac{GM}{r},
\end{align}
and
\begin{align}
    \label{eq:fpmr}
    F_{\rm PM, R} (r, R) \equiv \bigg(1 - 2 \frac{R^2}{r^2} + \frac{R^4}{r^4} \bigg) F_{\rm PM,t} (r, R)
   \nonumber\\
    + \bigg( 2 (1 - \beta') \frac{R^2}{r^2} - (1 - 2 \beta') \frac{R^4}{r^4} \bigg) \braket{v_r^4}.
\end{align}
In this way, we can exploit the increasing abundance of PM data found in nearby galactic systems and star clusters to constrain the fourth-order anisotropy, leading to a full three-dimensional and self-consistent modeling of both the second and fourth-order moments along the three observable LOS and PM components. To our knowledge, this is a new point that has not been published in the literature to date.\footnote{We note that \cite{2017ApJ...841...90E} independently introduced fourth-order PMs to quantify departures from dynamical equilibrium. However, to the authors' knowledge, our work is the first in which these are explicitly solved for from the Jeans equations and used to break the mass-anisotropy degeneracy and it is therefore a novelty in this respect.}

Lastly, we emphasize the following point: while PMs constitute an important and novel addition to the higher order treatment of kinematics, this does not mean that our approach will only be useful when these are available. In particular, the fact that quantities such as VSPs exist (i.e., fourth-order anisotropy-independent constraints on the higher moments) demonstrates that the Jeans equations should automatically encode this information, yielding comparable constraints if we marginalize over some general functional forms for $\beta$ and $\beta'$. In fact, given that virial constraints amount to positivity constraints on velocity moments and that the virial theorem can itself be derived from the Jeans equations \citep{1992ApJ...391..531D}, any quantities derived from these constraints should already be encoded in the Jeans equations. This means that the Jeans equations, solved at fourth order, should have at least as much information as using the Jeans equations at second order with VSPs. This appears to be an underappreciated fact: without additional assumptions, quantities such as VSPs should not provide any more information than velocity moments themselves from the Jeans equations (at a given order). Thus, any anisotropy-independent constraints derived from these are not inherently more constraining, but rather reflect a property of the Jeans equations themselves, which can be recovered after marginalization over the anisotropy. Therefore, we do not need to resort to these quantities to break the mass-anisotropy degeneracy and it may be possible to obtain better performance by working directly with the fourth-order Jeans equations instead. 

In fact, VSPs can be susceptible to significant biases, since they rely on assumptions about the behavior of the light profile and fourth velocity moment at large radii where data constraints are poor. By contrast, in the approach we take here, we managed to avoid these potential biases; however, we had to instead assume some functional form for $\beta'$ to marginalize over. In Sect.~\ref{sec:results}, we   use mock data for  the LOS velocities without PMs to demonstrate that the approach we take in \gs2 is indeed less biased than \gs1 or \gs1.5.

\subsection{Velocity distribution function}
\label{sec:PDF}

Given the methodology discussed in Sect.~\ref{sec:3d_gs2} to derive  the velocity moments from a given mass model, we  need a prescription to constrain these moments from observable velocities.
A common approach is to bin the data into discrete portions of a certain spatial width. While this has been shown to be sufficient to well-recover mass models from dwarf galaxy mock data \citep{2017MNRAS.471.4541R, 2018MNRAS.481..860R, 2019MNRAS.485.2010G, 2020MNRAS.498..144G, 2021MNRAS.501..978R, 2021MNRAS.505.5686C, 2024MNRAS.535.1015D, 2025arXiv250303812N, 2025arXiv250418617T}, it can lead to a loss of information and potential biases. It introduces an unavoidable trade-off between spatial and velocity resolution, which may be particularly acute when aiming to extract information of higher order moments, and it makes it challenging to properly marginalize over uncertain membership for the tracers, survey selection effects, binary stars, and so on.

To solve all of the above problems, in \gs2, we modeled each stellar velocity individually, without binning. In the context of Jeans analyses, comparable approaches have been introduced in the past, such as \textsc{DiscreteJAM} \citep{2008MNRAS.390...71C,2013MNRAS.436.2598W,2015arXiv150405533C}, 
\textsc{MAMPOSSt} \citep{2013MNRAS.429.3079M}, and, more recently, in \cite{2024arXiv240412671W}\footnote{The approach from \cite{2024arXiv240412671W}, which was recently and independently developed, is the one most similar to the one we use in \gs2. Some of the main differences between these include the improvements which we introduce, such as a fully general treatment of anisotropies (we marginalized $\beta$ and $\beta'$ as two independent and radially dependent, i.e.,  nonconstant, quantities) and the use of PMs to constrain both the second and fourth-order Jeans equations. We also implement uncertainties differently in our PDF modeling, which may explain why these authors find greater susceptibility to error-driven biases in their results. 
}. In our case, rather than assuming Gaussianity for the velocity PDF (projected or otherwise), we used the PDF models introduced in \cite{2020MNRAS.499.5806S}. These are based on analytical convolutions of a Gaussian with a uniform distribution kernel for models of negative excess kurtosis and with a Laplace kernel for positive excess kurtosis models. Taken together, they allow for meaningful deviations from Gaussianity with kurtoses $\kappa \equiv \braket{v^4} / \sigma^4$ in the range $1.8 < \kappa < 6$, with $\kappa = 3$ corresponding to a Gaussian (which defines excess kurtosis). For a symmetric distribution of mean zero for the scaled velocity variable, $w \equiv v/\sigma$, and measurement error, $s\equiv \delta v /\sigma$ (assumed to be symmetric and Gaussian), this yields the error-convolved PDF of the form

\begin{equation}
\label{eq:unik}
f_{\rm vel}(w) = \frac{b}{2a}\bigg[ \Phi \bigg(\frac{bw+a}{t} \bigg) - \
\Phi \bigg(\frac{bw-a}{t} \bigg)\bigg],
\end{equation}
for the uniform kernel ($1.8 < \kappa < 3$), where 
\begin{equation}
\label{eq:defphi}
\Phi (x) \equiv \frac{1}{\sqrt{2 \pi}} \int_{-\infty}^{x} dy \: e^{-y^2/2},
\end{equation}
\begin{equation}
t^2 = 1 + b^2 s^2, 
\end{equation}
and the parameters $a$ and $b$ are fully specified by the variance and kurtosis (before convolving with errors), respectively, as
\begin{equation}
\label{eq:univk}
\braket{w^2} = \frac{1}{b^2}\bigg(1 + \frac{a^2}{3} \bigg), \:
\kappa = 3 - \frac{2a^4}{15}\bigg(1 + \frac{a^2}{3} \bigg)^{-2}.
\end{equation}
Setting $\braket{w^2} = 1,$ so that $\braket{v^2} = \sigma^2,$ implies 
\begin{equation}
\label{eq:univk2}
b^2 = 1 + \frac{a^2}{3} .
\end{equation}
For the Laplacian kernel ($3 < \kappa < 6$), the error-convolved PDF takes the form
\begin{align}
\label{eq:lapk}
f_{\rm vel}(w) = \frac{b}{4a} \bigg[\exp\bigg(\frac{t^2 - 2abw}{2a^2}\bigg) \: {\rm erfc}\bigg(\frac{t^2 - abw}{\sqrt{2} ta}\bigg) 
  \nonumber\\ 
  + \exp\bigg(\frac{t^2 + 2abw}{2a^2}\bigg) \: {\rm erfc}\bigg(\frac{t^2 + abw}{\sqrt{2} ta}\bigg) \bigg],
\end{align}
with variance and kurtosis
\begin{equation}
\label{eq:lapvk}
\braket{w^2} = \frac{1}{b^2}\Big(2{a^2} + 1 \Big), \:
\kappa = 3 + 12a^4 \Big(2a^2 + 1 \Big)^{-2},
\end{equation}
so that, again, setting unit variance yields
\begin{equation}
\label{eq:univk3}
b^2 = 2a^2 + 1.
\end{equation}
This allows for a fully self-consistent treatment between the assumed PDF and the stellar kinematic model.\footnote{Our numerical implementation of these functions is based on the code introduced in \cite{2020MNRAS.499.5806S}, which is publicly available at:
\href{https://github.com/jls713/gh_alternative}{\texttt{https://github.com/jls713/gh\_alternative}}.
}

\subsection{Mass-anisotropy modeling}
\label{sec:massani}

As in \cite{2025A&A...693A.104B}, for the photometric tracer density profile, we use the generalized $\alpha \beta \gamma$ profile (e.g., \citealt{1990ApJ...356..359H, 1996MNRAS.278..488Z}). This is given by a double power-law model,
\begin{equation}
\label{eq:plum}
    \nu_{\star}(r) = \frac{\rho_{p}}{(r/r_{p})^{\gamma} (1 + (r/r_{p})^{\alpha})^{(\beta - \gamma)/\alpha}},
\end{equation}
where we introduced the three exponent variables $\alpha, \beta, \gamma,$ and the scale radius and density $r_p, \rho_p,$ respectively.  

Following recent versions of \textsc{GravSphere} (e.g., \citealt{2021MNRAS.505.5686C}),  we used the \textsc{coreNFWtides} profile \citep{2018MNRAS.481..860R} to model dark matter profiles of tidally stripped galactic halos expected from hydrodynamical simulations simulating Local Group environments. Previous versions of \gs \: have used nonparametric modeling with sequential power laws \cite{2017MNRAS.471.4541R, 2020MNRAS.498..144G}. As compared to these earlier choices, the  \textsc{coreNFWtides} profile has number of advantages: it does not suffer from biases due to monotonicity priors inherent to the previous method \citep{2025arXiv250418617T}; rather, it is designed to model mass profiles realistic galactic systems, and it makes recovering cosmologically valuable parameters such as the halo mass and concentration  straightforward. The model has an enclosed mass profile given by

\begin{align}
\label{eq:cnfwt}
    M_{\rm cNFWt}(r) = \begin{cases}
    \hspace{0.1cm} M_{\rm cNFW}(r) \hfill r \leq r_{t},\\
    \hspace{0.1cm} M_{\rm cNFW}(r) \: + \\ 
    4 \pi \rho_{\rm cNFW}(r_t) \frac{r_t^3}{3 - \delta} 
    \Big[ \Big(\frac{r}{r_t}\Big)^{3-\delta} - 1 \Big]\hspace{0.65cm} r > r_{t},
\end{cases}    
\end{align}
where 
\begin{align}
\label{eq:cnfw}
    M_{\rm cNFW}(r) = M_{\rm NFW}(r) f^n
\end{align}
is the \textsc{coreNFW} profile \citep{Read:2015sta} and we have the original Navarro-Frenk-White (NFW) profile \citep{1996ApJ...462..563N}, expressed as
\begin{align}
\label{eq:nfw}
    M_{\rm NFW}(r) = M_{200} g_c \bigg[ \ln \bigg( 1 + \frac{r}{r_s}\bigg) - 
    \frac{r}{r_s} \bigg (1 + \frac{r}{r_s}\bigg)^{-1} \bigg],
\end{align}
with a function that modulates the degree of core formation, 
\begin{align}
\label{eq:fn}
    f^n = \bigg[ \tanh \bigg( \frac{r}{r_c}\bigg)\bigg]^n,
\end{align}
and the densities given by
\begin{align}
\label{eq:rcnfw}
    \rho_{\rm cNFW}(r) = f^n \rho_{\rm NFW} + \frac{n f^{n-1} (1 - f^2)}{4 \pi r^2 r_c} M_{\rm NFW},
\end{align}
\begin{align}
\label{eq:rnfw}
    \rho_{\rm NFW}(r) = \rho_s \bigg( \frac{r}{r_s} \bigg) \bigg(1+ \frac{r}{r_s}\bigg)^{-2},
\end{align}
with the scale density and length given (respectively) as
\begin{align}
\label{eq:scales}
    \rho_s = \rho_{\rm crit} \Delta c^3_{200} g_c/3, \: r_s = r_{200}/c_{200},
\end{align}
with: 
\begin{align}
\label{eq:gc}
    g_c = \frac{1}{\ln(1 + c_{200})  - \frac{c_{200}}{1 + c_{200}}},
\end{align}
and 
\begin{align}
\label{eq:r200}
r_{200} = \bigg(\frac{3 M_{200}}{4 \pi \Delta \rho_{\rm crit}} \bigg)^{1/3},
\end{align}
with $c_{200}$ being the dimensionless concentration parameter and $r_{200}, \: M_{200}$ as the virial radius and mass, respectively, defined at a region where the average density exceeds the present-day cosmological critical density $\rho_{\rm crit} = 135.05 \: {\rm M}_{\odot}\:{ \rm kpc}^{-3}$ by the critical overdensity factor $\Delta = 200$.\footnote{The precise value of the adopted cosmological critical density will vary depending on the adopted cosmology. Our value corresponds to the `canonical value' with Hubble parameter $H_0 \sim 70$~km~s$^{-1}$/Mpc, so that  $\rho_{\rm crit} = 3H_0^2/8\pi G$. It should  be noted that various authors use different values of $\Delta$ to define the virial radius.} The core formation parameter $n$ yields full core formation when $n = 1,$ retrieves an NFW cusp when $n = 0,$ and can model potentially cuspier distributions for negative values. The core radius, $r_c$, determines the extent of the inner core region. The tidal radius, $r_t $, and slope exponent, $\delta$, regulate the amount of stripping in the outer region.

Following previous versions of \textsc{GravSphere}, we use the anisotropy profile assuming the following the generalized form from \cite{2007A&A...471..419B}, 
\begin{equation}
    \label{eq:beta_ansatz}
    \beta(r)  = \beta_{0} 
    + (\beta_{\infty} - \beta_0) \frac{1}{1 + \Big(\frac{r_\beta}{r}\Big)^{\eta}},
\end{equation}
where $\beta_0$ is the anisotropy at the center, $\beta_{\infty}$ its limit as it asymptotically approaches infinity, $r_\beta$ is the transition scale between these two regimes, and $\eta$ is the exponent which modulates the steepness of the transition. 
We also assumed the same functional form for the fourth-order anisotropy, $\beta',$ allowing for a fully general treatment of this quantity, unlike previous works that have made more restrictive assumptions about its form \citep[e.g.,][]{1990AJ.....99.1548M, 2003MNRAS.343..401L, 2005MNRAS.363..918L, 2013MNRAS.432.3361R, 2024arXiv240412671W}.

\section{Data, likelihood, and priors}
\label{sec:data}

\subsection{Gaia Challenge mock data}
\label{sec:gcdata}

To test the overall performance of \textsc{GravSphere2,} we started with a subset of the \textsc{Gaia Challenge} catalog of mock galaxies\footnote{See the \textsc{Gaia Challenge} website  for full access to the data and further information: \href{https://astrowiki.surrey.ac.uk/}{\texttt{https://astrowiki.surrey.ac.uk/}}.}, constructed from spherical dynamical equilibrium models. The results are shown in Sect.~\ref{sec:gc}. They have the advantage of having been extensively used for prior testing in the literature, allowing for a straightforward comparison with previous studies, including previous versions of \textsc{GravSphere} (e.g., \citealt{2019MNRAS.485.2010G, 2021MNRAS.501..978R, 2021MNRAS.505.5686C}). As in \cite{2021MNRAS.505.5686C}, we focus on a subset of two mock galaxies originally from \cite{2011ApJ...742...20W} which have been the ones for which several previous methods (including \gs) have been affected by significant bias (e.g., see \citealt{2021MNRAS.501..978R}). These correspond to a dark matter profile  with the $\alpha \beta \gamma$ functional form (same as Eq.~\ref{eq:plum}) embedded in a Plummer-like\footnote{In practice, these profiles closely resemble a Plummer sphere but show minor deviations in the form of an $\alpha \beta \gamma$ profile (Eq.~\ref{eq:plum}), which ensures positivity of the phase-space distribution function.} stellar tracer distribution, with a varying Osipkov-Merritt anisotropy profile \citep{1979PAZh....5...77O,1985MNRAS.214P..25M}. These two mock galaxies consist of a cored dark matter model (PlumCoreOM) and a cuspy one (PlumCuspOM). 
We consider cases with 100, 1,000, and 10,000 tracers, with and without including PMs. In all cases we simultaneously fit the binned surface density profile using 10,000 tracers, assuming also velocities with 2~km/s errors.

\subsection{Simulated dwarf galaxy data}
\label{sec:sddata}

The second case we considered is the recent suite of simulated dwarf galaxies from \cite{2025arXiv250418617T}. These are initialized with cuspy NFW profiles with massive Plummer-like stellar components and are evolved in a Milky Way-like potential to simulate the properties of nearby present-day dSphs. This allows us to probe more realistic galactic environments and aspects such as departures of equilibrium and departures from spherical symmetry due to tidal effects, or induced rotation, which may be present in simulations. 

For the purposes of this study, we  focus on the Fornax analog of this catalog which was extensively studied, has a large number of tracer particles and was found to produce significant biases when applying \gsa\   (\gs1).\footnote{In particular, we are referring to variant with a light Milky Way potential, as opposed to the heavy one, chosen due to the properties we outlined and the fact that it was also more extensively studied by these authors in this context.} We studied the case with all the 34,158 bound tracers,\footnote{As per the subhalo finder, and applying also a 3$\sigma$-velocity clipping to the complete sample. For datasets where we use PMs, we require tracers to pass a 3.5$\sigma$-clipping test along each direction, reducing the sample to 34,081 tracers.} using only the LOS velocities used in \cite{2025arXiv250418617T}. We also considered a reduced samples of 1,000 and 100 tracers. Additionally, we considered a sample of 100 and 1,000 tracers where PMs are included. We also considered LOS samples with 10, 25, and 50 tracers for our comparisons with mass estimators. For the tracer profile fit, we used 10,000 tracer positions in all cases, except for the fit with all tracers from the simulation. Velocities were perturbed with 2~km/s errors in all cases.

\subsection{Likelihood and priors}
\label{sec:llhood}

The full log-likelihood we adopt for the fitting process is given by
\begin{align}
    \label{eq:full_ll}
    \ln \mathcal{L}_{\rm tot} (\boldsymbol{\theta}) &=
    \ln \mathcal{L}_{\rm LOS} +
       \ln \mathcal{L}_{\rm PM, t} 
       + \ln \mathcal{L}_{\rm PM, R}
       + \ln \mathcal{L}_{\Sigma_{\star}}, 
\end{align}
where $\boldsymbol{\theta}$ includes all the model parameters, which were simultaneously fitted. This includes the anisotropy parameters (that are distinct for the second and fourth-order anisotropy profiles), the mass model parameters, and the photometric tracer profile parameters. The first three terms correspond to the likelihood due to individual velocity PDFs (Eqs.~\eqref{eq:unik} and \eqref{eq:lapk}) for the LOS and tangential and radial PM components for the data we considered, whilst the last term corresponds to the term from the photometric tracer profile. \gs2 incorporates two options in this respect: one for a binned photometric light profile with Gaussian errors and a completely `bin-free' option where individual stellar tracer positions are used instead. We note that \gs2 does not bin any data. For the case of the binned profile it is assumed that this is provided as a data constraint from observations. Unlike previous versions of \gs, it is not necessary to derive a binned profile from individual stars as we can simply use the "bin-free" approach instead. In the latter case, similarly to the case of kinematic tracers, we can write the total likelihood as the product of those of individual photometric tracers, whose (positional) PDFs take the form

\begin{align}    \label{eq:prob_ind}
    {f}_{\rm pos}(R) =  \frac{R \Sigma_{\star}( R)}
    {\int_{0}^{\infty}  R' \Sigma_{\star}( R') \: dR'}.
\end{align}

In all cases, we verified that our fits were able to reproduce the photometric tracer profiles well and recover the true solution. We show a representative example fit in Appendix~\ref{fig:priorless} (Fig.~\ref{fig:priorless_phot}), showing excellent recovery which is far in excess of the range allowed by our priors.
The advantages of the $\alpha \beta \gamma$ functional form will be more evident when reproducing more complex distributions which cannot be reduced to Plummer profiles (or a summation of these), this includes cuspier distributions or those with shallow slopes at higher radii. One example of this is the GC Omega Centauri, where this functional form was introduced for \gs \: \citep{2025A&A...693A.104B}.

As in \cite{2025A&A...693A.104B}, we worked with the symmetrized anisotropy (for both the second- and fourth-order anisotropy profiles), expressed as
\begin{equation}
\label{eq:beta_t}
\widetilde{\beta} \equiv \frac{\beta}{2 - \beta} ,
\end{equation}
which has the property that is bounded as $-1\leq \widetilde{\beta} \leq 1,$ to give an unbiased sampling of the full parameter space.

For the \textsc{Gaia Challenge} data (Sect.~\ref{sec:gcdata}), the priors we adopt for the mass and anisotropy profile are similar (albeit broader for some quantities) to those Section 4.1.2 of \cite{2021MNRAS.505.5686C}, whilst for the photometric component we allow a flat prior with limits set with 50\% scatter around the true solution, which we fit as a binned surface density profile using 10,000 tracers. A summary of all priors adopted is shown in Table~\ref{tab:priors}.

For the Fornax-like simulated dwarf galaxy (Sect.~\ref{sec:sddata}), our mass and anisotropy priors were also the same as for the Gaia Challenge mocks, except for the fact the we used unbounded values for the limiting anisotropies ($-1 < \widetilde{\beta}_{0, \infty} < 1$) and somewhat less broad priors for the concentration parameter ($1 < c_{200} < 50$). For the stellar tracer model, we did not use binned photometry and instead use the full "bin-free" method, using Eq.~\eqref{eq:prob_ind}. We allowed for a scatter of 1 dex from the 2D half-light radius ($R_{1/2}$) of $\sim 770$~pc for the length scale in the photometric light profile (which we sample logarithmically as a flat prior)\footnote{This can be obtained directly from the photometric tracers (either from the profile or their median position), but is only meant to set a relevant scale and its precise value is not needed as the results are insensitive to it.} and $0.1< \alpha <4, \: 3.1< \beta <7, \: 0< \gamma <2.9$. These were chosen to allow for convergence (needed to set a proper normalization) while allowing a wide range of profiles of varying steepness. We recall here that $\gamma$ and $\beta$ are the negative inner and outer logarithmic slopes of the light profile (respectively), where a Plummer profile corresponds to $\alpha, \: \beta, \: \gamma = 2,5,0).$ Lastly, we sampled the mass of the stellar component with a flat prior with 25\% scatter around $4.3 \times 10^7 \: \MS,$ as in \cite{2025arXiv250418617T}, assuming a constant mass-to-light ratio and that the stellar mass distribution has the same radial density profile as the tracer stars. This is usually a good approximation provided that the tracer profile is complete and that there is no meaningful mass segregation or observational bias in the photometric selection. The latter is typically a good approximation for nearby dwarf galaxies, which have relaxation times longer than a Hubble time due to the presence of dark matter (e.g., \citealt{2022MNRAS.510.3531B}). If this is not the case (as in star clusters), we needs to include the presence of additional components of varying segregation (e.g., see \citealt{2023MNRAS.522.5320D, 2025A&A...693A.104B}) in the mass model. The full implementation of our likelihood modeling with the priors and data we used; the \gs2 code itself has been made publicly available, as noted in the introduction (Sect.~\ref{sec:intro}). 

We also note that our priors are generally very broad and have little effect on our results. Some, such as imposing isotropy in the center and barring large negative anisotropies, which we do for the Gaia Challenge sample following \cite{2021MNRAS.505.5686C}, are physically motivated constraints. We show explicitly in Appendix~\ref{app:priorless} that the space of mass and tracer density models spanned by our priors is much broader than those favored by the data, demonstrating that our priors are sufficiently broad so as to have little effect on our results. For completeness, we also included marginalized parameter constraints (corner plots) of each of the 19 parameters for all our models in a supplementary repository.\footnote{\href{https://github.com/dadams42/GS2-results}{\texttt{https://github.com/dadams42/GS2-results}}}

All fits were performed using the nested sampling package \textsc{dynesty} \citep{2020MNRAS.493.3132S, 2022zndo...6609296K}. This is based on nested sampling techniques \citep{2004AIPC..735..395S,10.1214/06-BA127} and the `dynamic nested sampling technique' \citep{cite-key} (which we use here). \textsc{dynesty} has been optimized for the inference of posterior distributions (as opposed to evidence estimation), using the bounding method from \cite{2009MNRAS.398.1601F}. The sampling method used is based on \cite{10.1214/aos/1056562461,2015MNRAS.450L..61H,2015MNRAS.453.4384H}. For further information, see:  \href{https://dynesty.readthedocs.io/en/latest/index.html}{\texttt{https://dynesty.readthedocs.io/en/latest/index.h\\tml}}. An example with the specific implementation used for \gs2 can be found in the publicly available code release. This sampling method has been shown to outperform Markov chain Monte Carlo Methods (MCMC), particularly when dealing with complex, high-dimensional, and multimodal posterior spaces (e.g., \citealt{2020MNRAS.493.3132S}). This is consistent with our own experiments comparing with previous versions of \gs\ that used \textsc{emcee} instead \citep{Foreman_Mackey_2013}. Following \cite{2025A&A...693A.104B}, we thus opted to use this improved sampling technique for \gs2. 

Run times will vary depending on the number of tracer velocities used (including PMs), whether fitting the photometry star-by-star, binned, or fixing it, and also on the sampling resolution used. For our cases, we found that the default value of 500 live points sufficed. As a rule of thumb, a few times the number of dimensions squared usually suffices, but this will depend on how multimodal the posterior is, as well as how much volume the modes occupy (smaller volumes require higher sampling resolution). In all cases, models could be run within approximately $1 - 3$ days on a normal laptop. The dominant source of computational cost in all versions of \gs \: is the evaluation of numerical integrals for solving the Jeans equations. For a given likelihood evaluation, \gs2 takes roughly twice as long as \gs1, since one solves the Jeans equations twice, once for second-order moments (as in previous versions), and another time for fourth-order ones. This still makes \gs2 a highly competitive method in terms of run-time (cf. \citealt{2021MNRAS.501..978R}). We note that the run-times reported for \gs1 in \cite{2021MNRAS.501..978R} appear substantially less than what one might expect given the above. This is at least partly due to the fact that we use a more sophisticated and robust sampling method (\textsc{dynesty} rather than \textsc{emcee}). As noted earlier, runs can also be accelerated by reducing the number of live points, and we opted for a conservative choice in this respect. Likelihood evaluation times will also scale roughly as $\sim N_{\rm int}^2$, where $N_{\rm int}$ is the number of points used for each numerical integral when solving the Jeans equations.

\begin{table}[t]
\caption{Priors implemented in \textsc{dynesty} for our analysis.} 
    \centering
    \setlength{\tabcolsep}{0.7em}
\renewcommand{\arraystretch}{1.5}
  \setlength{\arrayrulewidth}{.30mm}
    \begin{tabular}{c|c c c}
     \hline\hline
        \textbf{Parameter} & \textbf{Units} & \textbf{Prior} & \textbf{Range}  \\ \hline 
        $\widetilde{\beta}_0, \: \widetilde{\beta}_{0}'$ & none & flat &  
        \begin{tabular}{@{}c@{}}$[-0.01, \: 0.01]$ (G)
       \\  $[-1, \: 1]$ (F)
       \end{tabular}
        \\
        $\widetilde{\beta}_{\infty}, \: \widetilde{\beta}_{\infty}'$ & none & flat & 
        \begin{tabular}{@{}c@{}}$[-0.1, \: 1]$ (G)
       \\  $[-1, \; 1]$ (F)
       \end{tabular}
        \\
        $ r_{\beta}, \: r_{\beta}'$ & $\log_{10}$ kpc & flat &  $[-2, \: 1]$
        \\
        $\eta, \: \eta'$ & none & flat &  $[1, \: 3]$
        \\
        $M_{200}$ & $\log_{10} \MS$ & flat &  $[7.5,  \: 11.5]$
        \\
        $c_{200}$ & None & flat &  
        \begin{tabular}{@{}c@{}}$[1, \: 100]$ (G)
       \\  $[1, \: 50]$ (F)
       \end{tabular}
       \\
       $r_c$ & $\log_{10}$ kpc & flat &  $[-2, \: 1]$
        \\
        $n$ & none & flat &  $[0, \: 1]$
        \\
       $r_t$ & $\log_{10}$ kpc & flat &  $[0, \: \log_{10}(20)]$
        \\
        $\delta$ & none & flat &  $[3.01, \: 5]$
        \\
        $M_{\star}$ & $ 10^7 \: \MS$ & flat &  $4.3 \pm 25 \%$ (F)
        \\
         $\rho_p$ & kpc$^{-3}$ & flat &  ${\rm Truth} \pm 50\%$ (G)
        \\
        $r_p$ & 
        \begin{tabular}{@{}c@{}} kpc (G)
       \\ $\log_{10}$ kpc (F)
       \end{tabular}
        & flat &  
        \begin{tabular}{@{}c@{}}${\rm Truth} \pm 50\%$ (G)
       \\  $R_{1/2} \pm 1$ dex (F)
       \end{tabular}
        \\
        $\alpha$ & 
        None
        & flat &  
        \begin{tabular}{@{}c@{}}${\rm Truth} \pm 50\%$ (G)
       \\ $[0.1, \: 4]$ (F)
       \end{tabular}
        \\
        $\beta$ & 
        None
        & flat &  
        \begin{tabular}{@{}c@{}}${\rm Truth} \pm 50\%$ (G)
       \\ $[3.1, \: 7]$ (F)
       \end{tabular}
        \\
        $\gamma$ & 
        None
        & flat &  
        \begin{tabular}{@{}c@{}}${\rm Truth} \pm 50\%$ (G)
       \\ $[0, \: 2.9]$ (F)
       \end{tabular}
        \\
        \hline\hline
    \end{tabular}
\tablefoot{The parenthesized "G" denotes priors specific to the Gaia Challenge mocks, while "F" corresponds to the simulated Fornax dwarf analog. For both cases, there are 19 fitting parameters in total (for Fornax, we marginalize over the additional stellar mass $M_{\star}$, but do not use a density scale for the photometry, since we fit stars individually, rather than from a profile). Primed quantities (rows $1-4$) denote the parameters of the fourth-order counterpart of the anisotropy. For the Gaia Challenge mocks, the exponents $\alpha, \beta, \gamma$ are also ensured not to exceed the broad convergence limits used for the Fornax case.   The full implementation of these priors, with initialization files and data, are included in the \textsc{GravSphere2} public release repository on \href{https://github.com/dadams42/GravSphere2}{\texttt{GitHub}}. }
\label{tab:priors}
\end{table}

\section{Results}
\label{sec:results}


\subsection{Gaia Challenge mocks}
\label{sec:gc}

\begin{figure*}[ht]
\centering
\begin{minipage}[c]{0.75\textwidth}
\includegraphics[width=\ww]{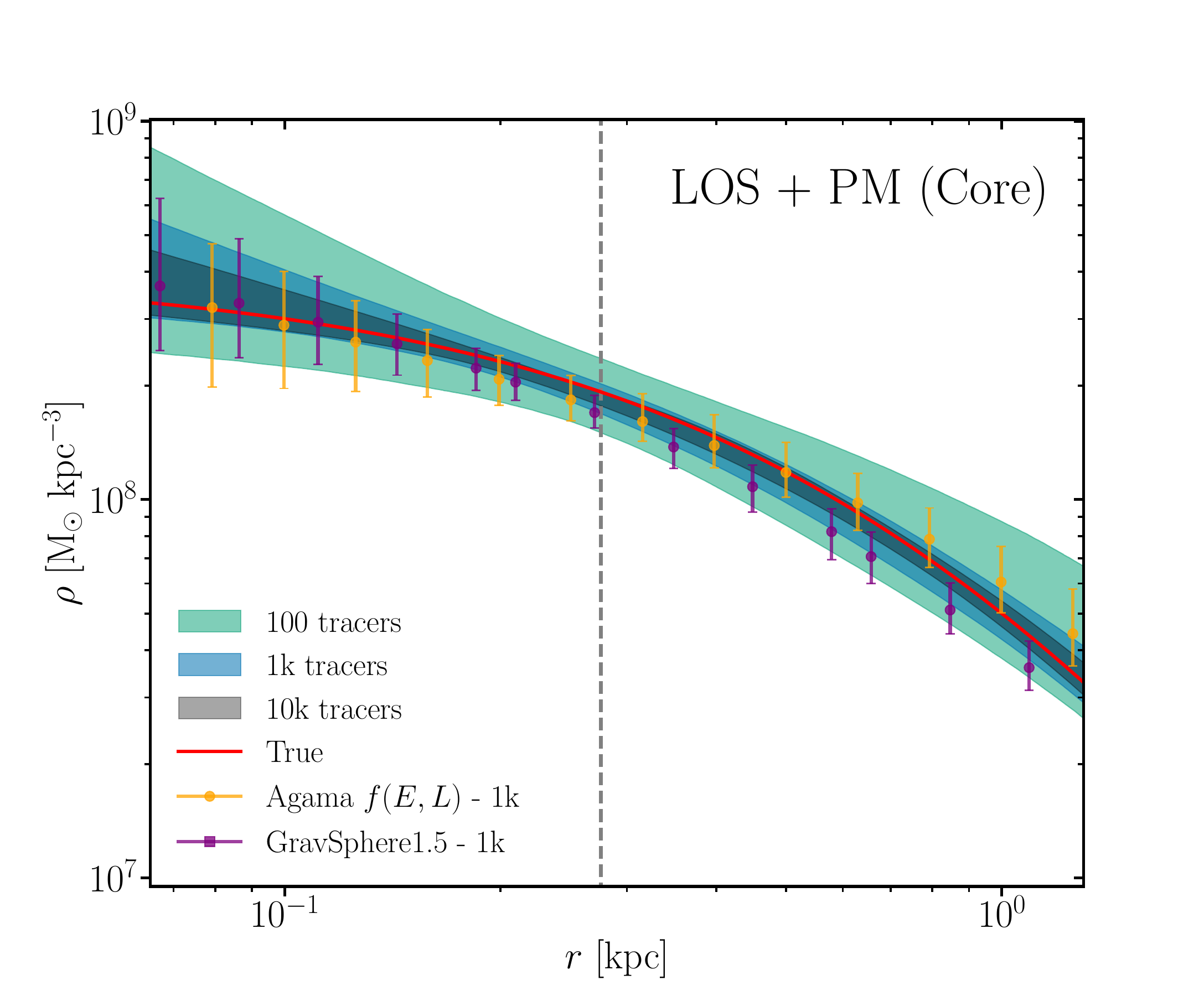}
\hspace{-0.7cm}
\includegraphics[width=\ww]{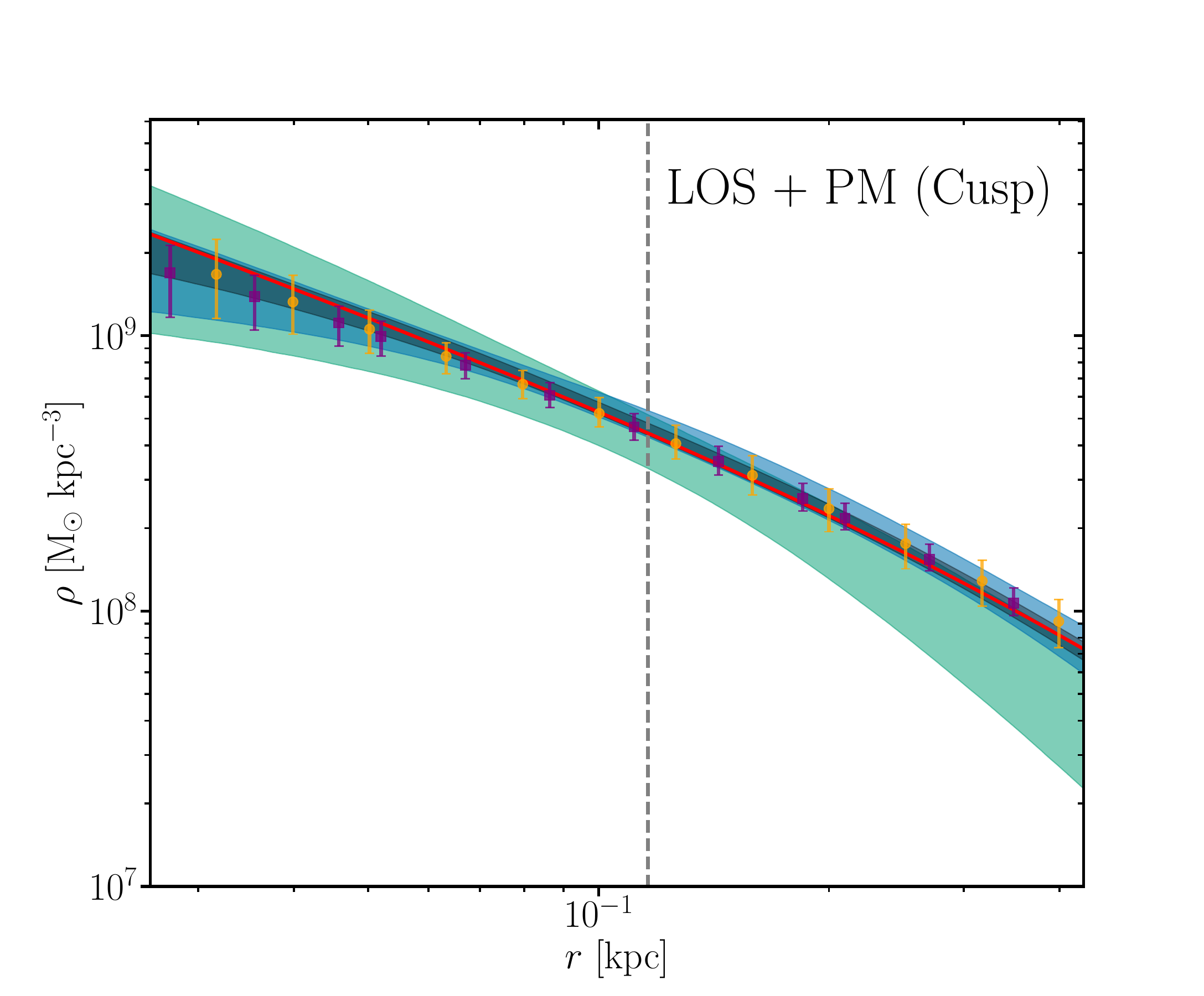}
\includegraphics[width=\ww]{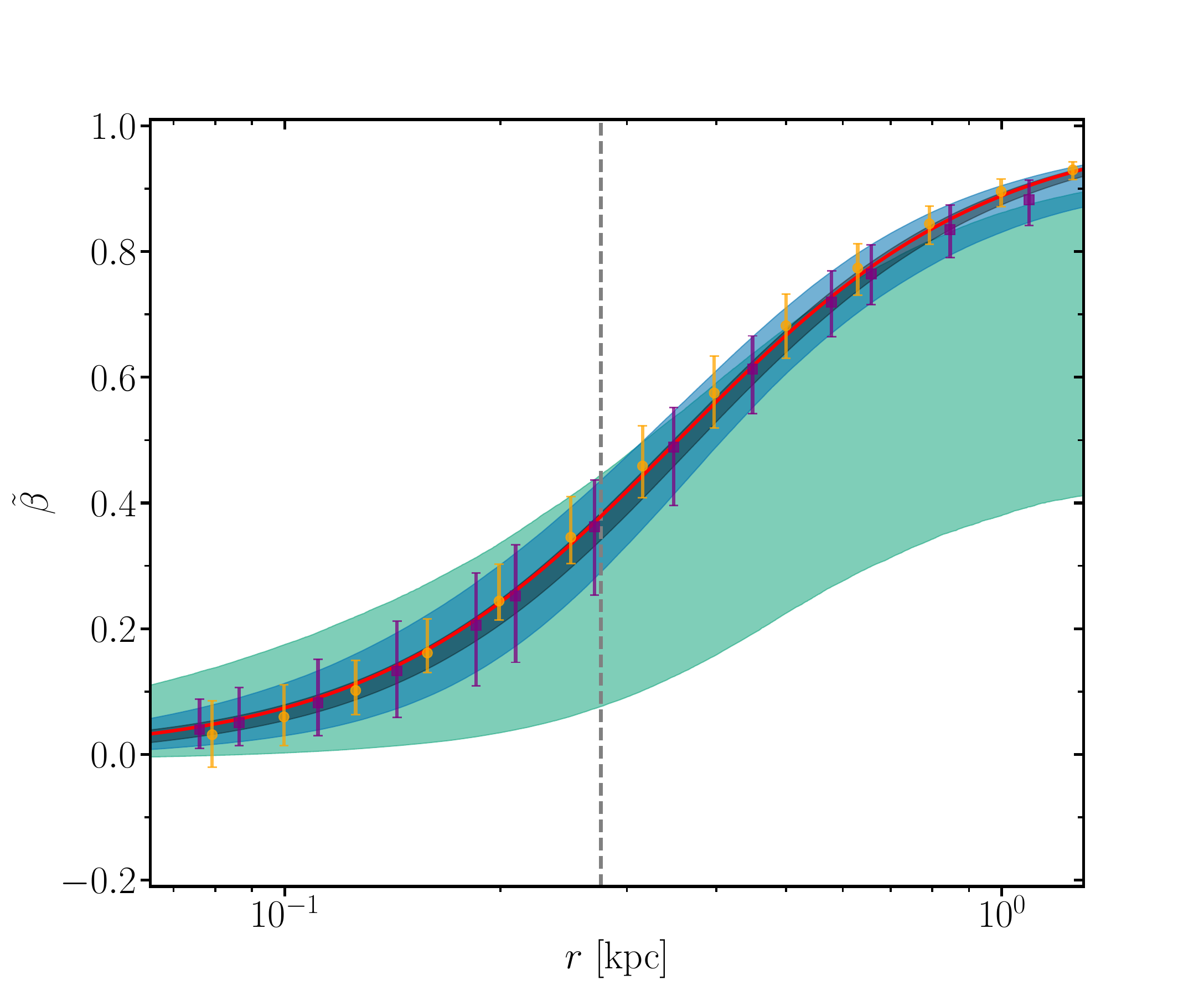} 
\hspace{-0.7cm}
\includegraphics[width=\ww]{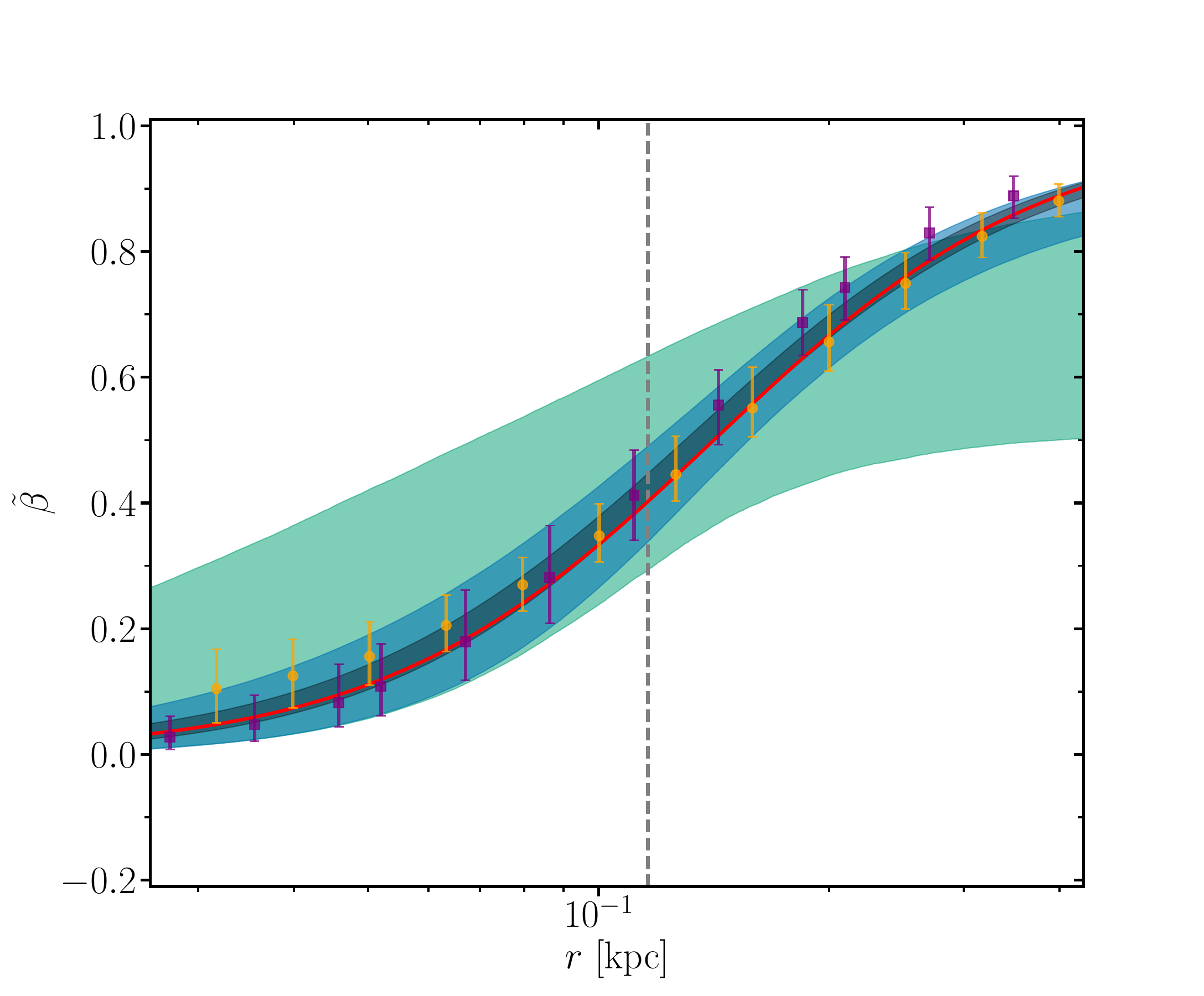}
\end{minipage}\hfill
\begin{minipage}[c]{0.23\textwidth}
\caption{\emph{Top-left:} 3D dark matter density  profile for the PlumCoreOM mock galaxy, bands denote the 95\% CL regions from the median for different numbers of stellar tracers including both LOS and PM velocities. The results from \cite{2021MNRAS.501..978R} for \textsc{Agama} $f(E, L)$ models, and from \cite{2021MNRAS.505.5686C} for \gs1.5 (including LOS VSPs) for 1,000 tracers with their respective median values with 95\% CL errors are shown for comparison. The red line denotes the true analytical solution from which the mock was generated. The gray, dashed line corresponds to the projected half-light radius ($R_{1/2})$.
 \emph{Top-right:} Same as the left hand side but for the PlumCuspOM model.
\emph{Bottom-left:} Symmetrized anisotropy profile for the PlumCoreOM model. The red line again denotes the true solution from which the mock was generated, corresponding to an Osipkov-Merritt profile.
\emph{Lower right:} Same as the left hand side but for the PlumCuspOM model. }
\label{fig:rhoanipm} 
\end{minipage}
\end{figure*}
  
\begin{figure*}
    \centering
\begin{minipage}[c]{0.75\textwidth}
\includegraphics[width=\ww]{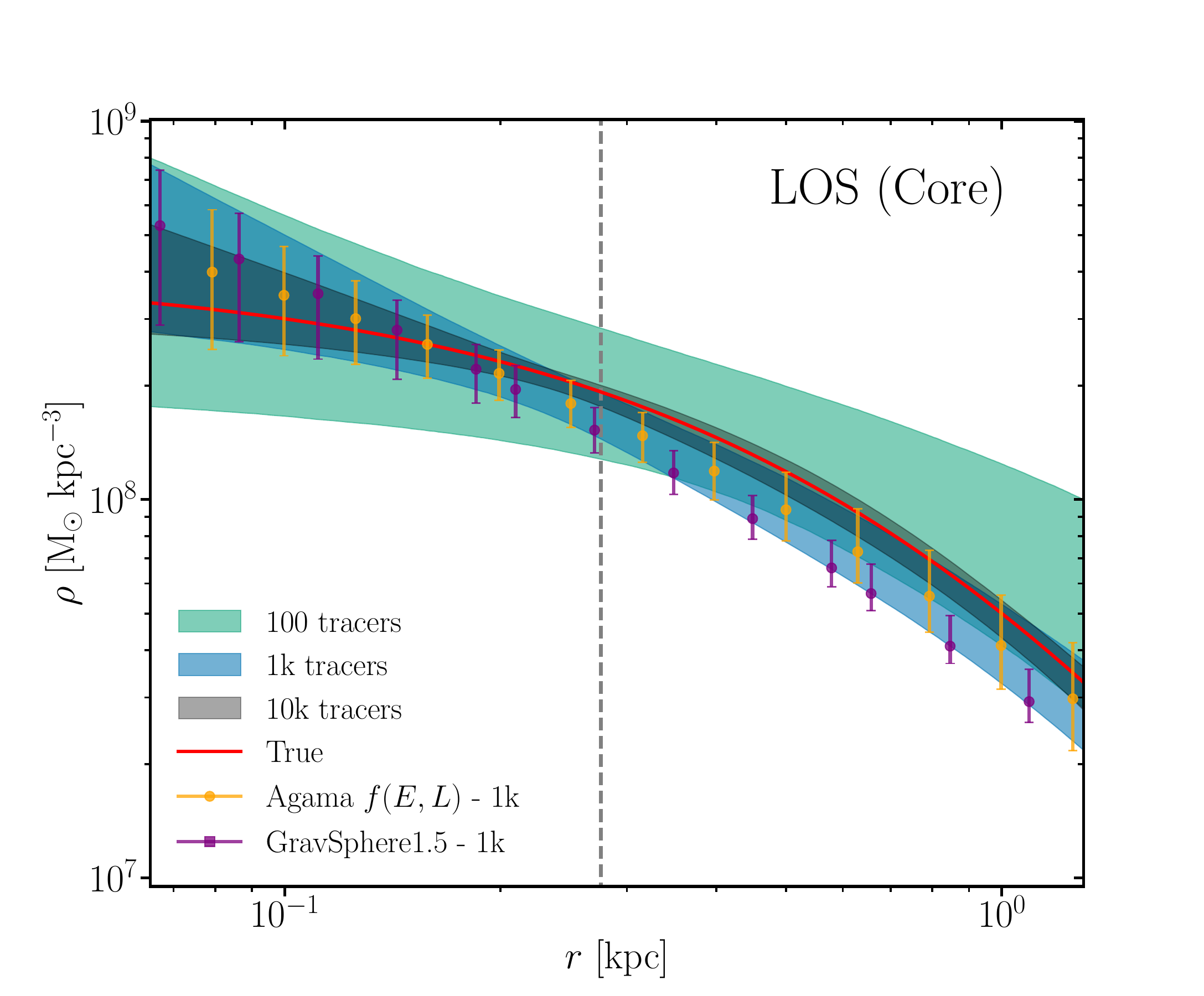}
\hspace{-0.7cm}
\includegraphics[width=\ww]{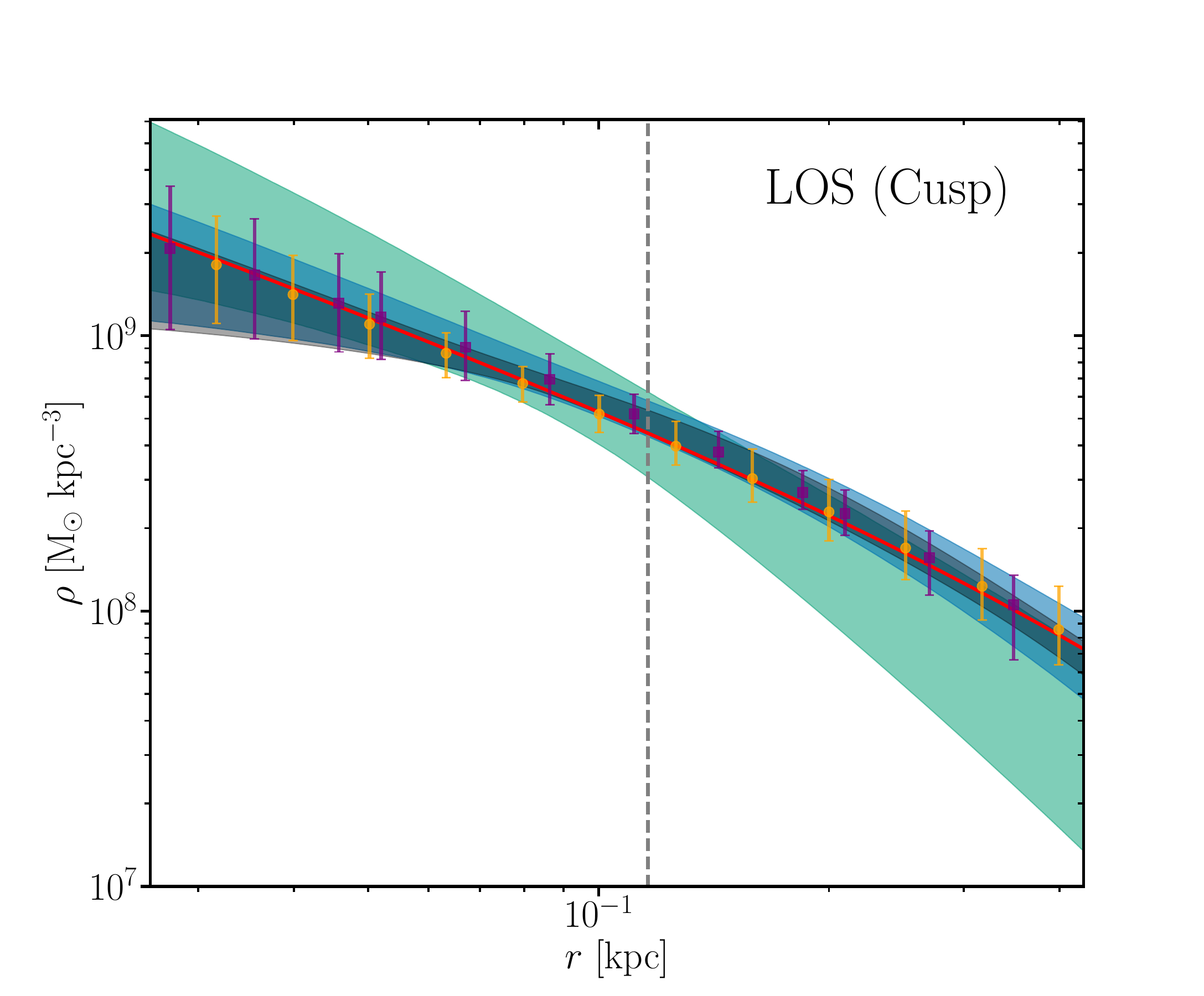}
\includegraphics[width=\ww]{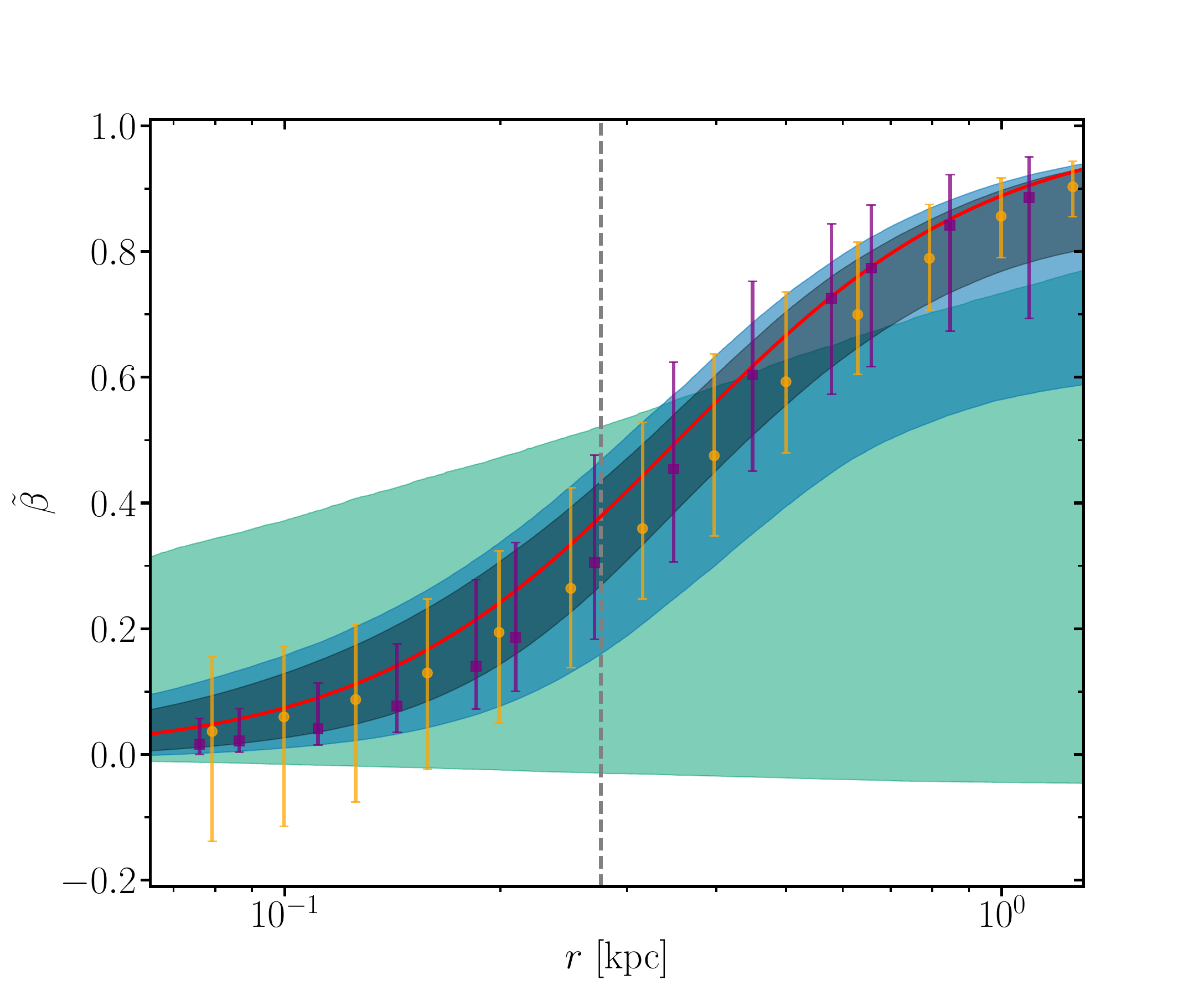} 
\hspace{-0.7cm}
\includegraphics[width=\ww]{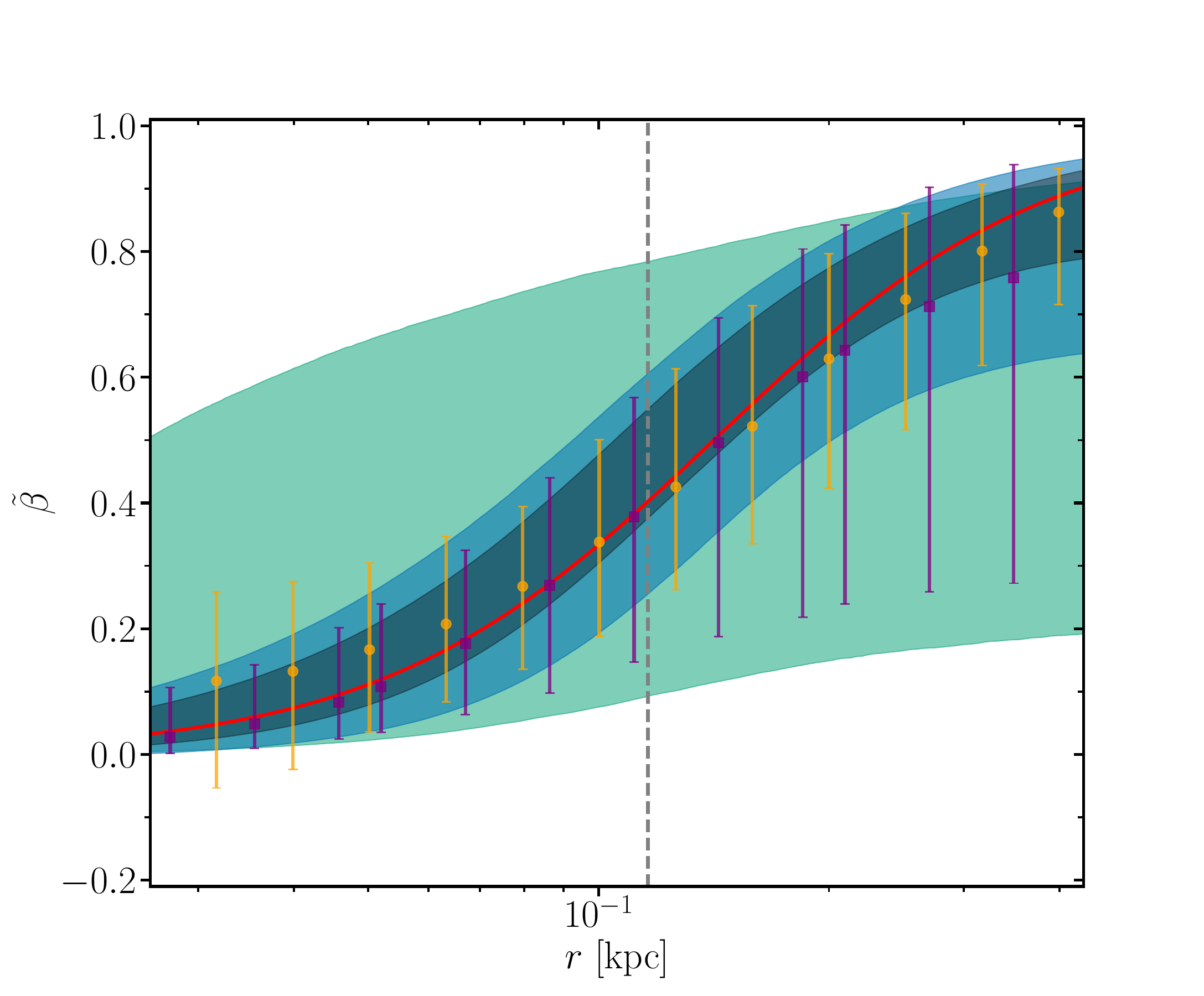}
\end{minipage}\hfill
\begin{minipage}[c]{0.23\textwidth}
\caption{Same as Fig.~\ref{fig:rhoanipm} but only with LOS tracers. \emph{Top-left:} \gs2 recovery for the PlumCoreOM mock 3D mass density. 
 \emph{Top-right:} Same as the left but for the PlumCuspOM model.
\emph{Bottom-left:} Symmetrized anisotropy profile for the PlumCoreOM model.
\emph{Bottom-right:} Same as the left  but for the PlumCuspOM model. 
\label{fig:rhoanilos}} 
\end{minipage}
\end{figure*}

\begin{figure*}
    \centering
\begin{minipage}[c]{0.75\textwidth}
\includegraphics[width=\ww]{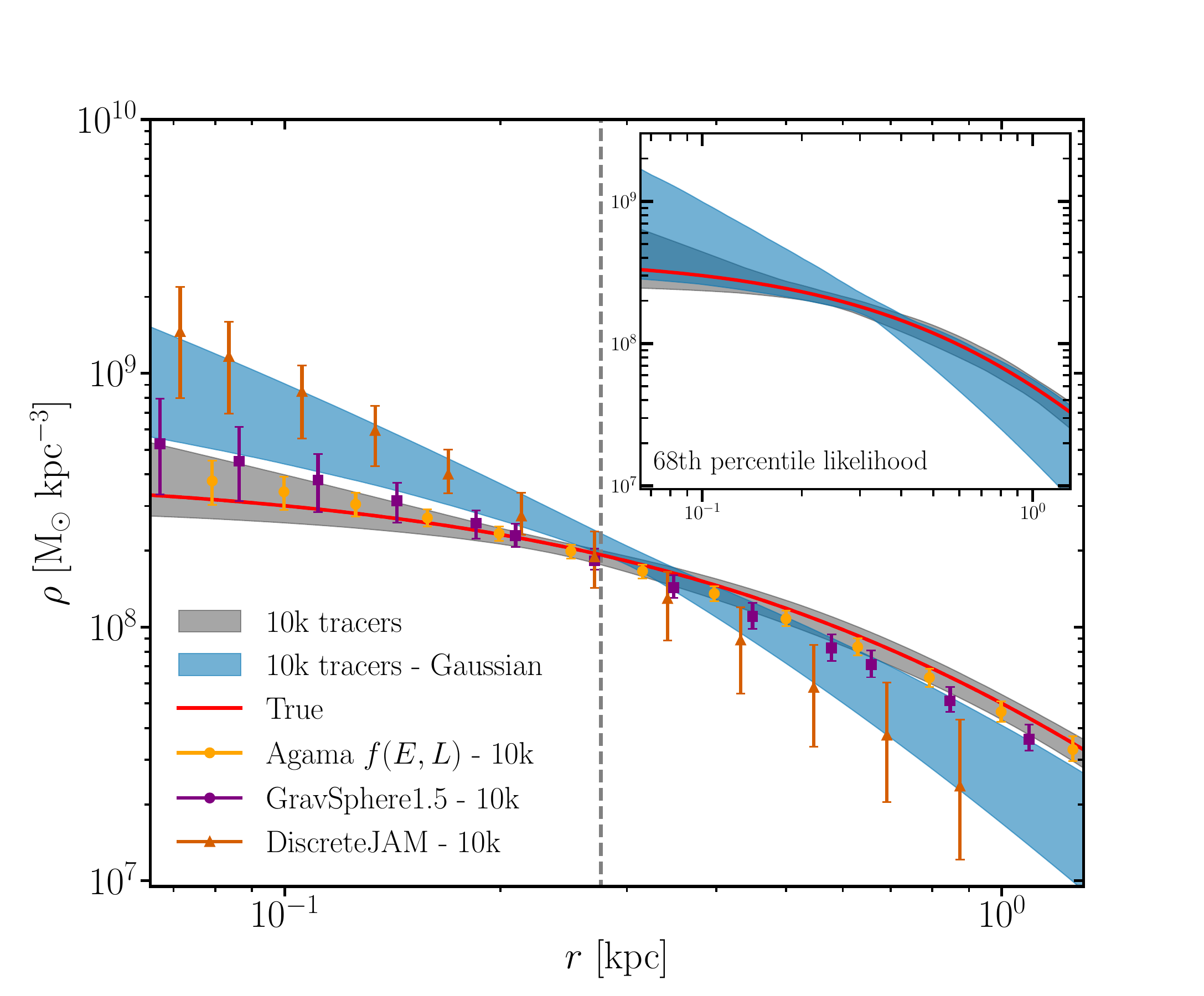}
\hspace{-0.5cm}
\includegraphics[width=\ww]{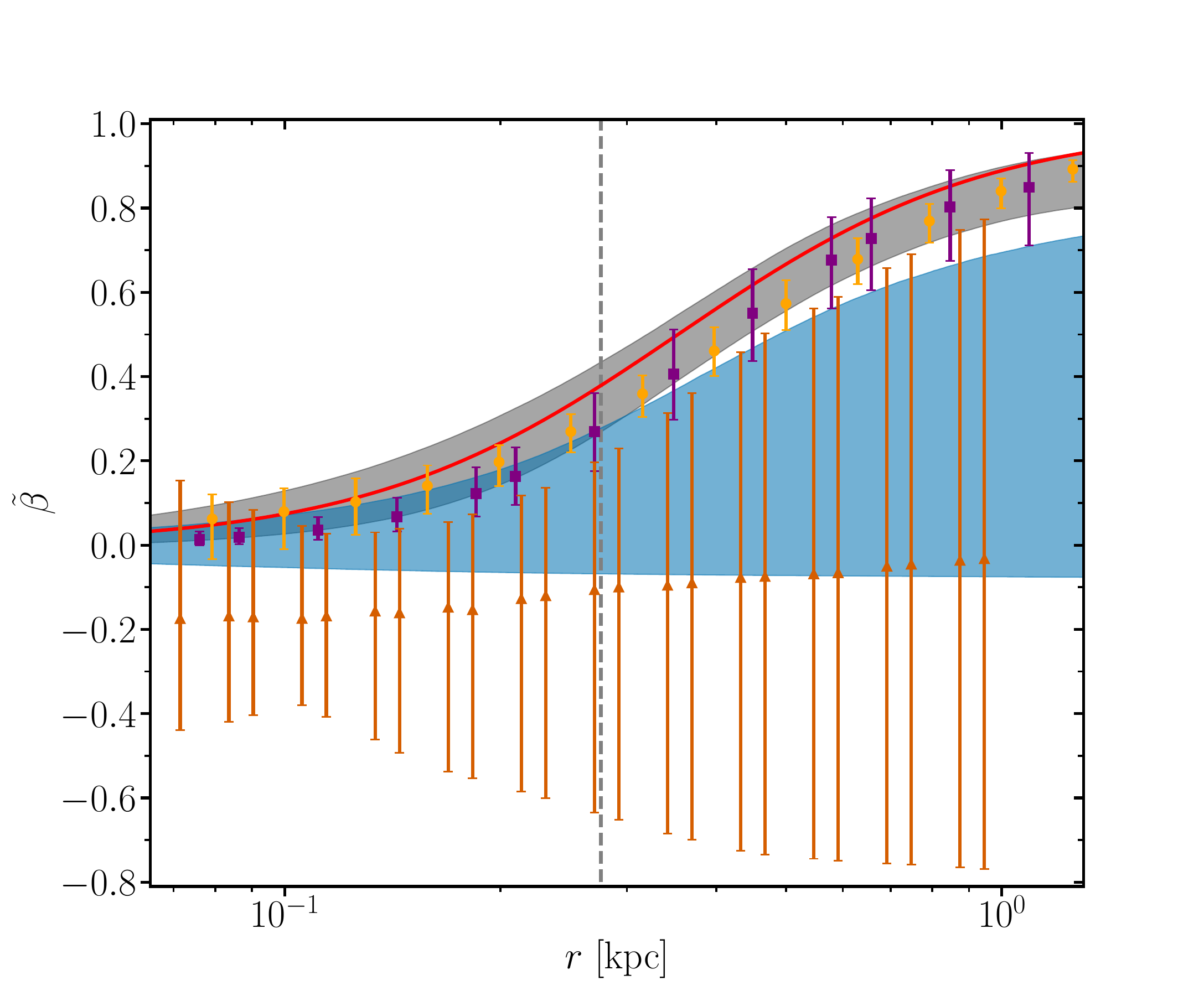}
\end{minipage}\hfill
\begin{minipage}[c]{0.23\textwidth}
\caption{Same as Fig.~\ref{fig:rhoanilos} but for 10,000 tracers in the PlumCoreOM mock, including in the comparison with other models. We have included the \textsc{DiscreteJAM} results from \cite{2021MNRAS.501..978R}.
\emph{Left:} Mass density profiles. Blue band shows our results when Gaussianity is assumed and fourth-order moments are excluded. For comparison, the inset shows instead bands obtained from the posterior distribution of likelihoods when computing extrema of all models falling within the 68th percentile from the maximum likelihood (see text for details). Using this definition, there is a wide range of equally good-fitting models that encompasses the true solution, while the posterior becomes biased due to a relative abundance of cuspier models in the allowable hypervolume. 
 \emph{Right:} Same as left, but for symmetrized anisotropy profile.}
\label{fig:rhoaniglos} 
\end{minipage}
\end{figure*}

Figs.~\ref{fig:rhoanipm} and \ref{fig:rhoanilos} show the results for the mass density and anisotropy recovery of the two models considered, with and without the inclusion of PMs. These are shown as the regions in the posterior spanning 95\% of the distribution centered at the median values of the profiles. For reference, we included the results of \gs1.5 (originally from \citealt{2021MNRAS.505.5686C}),\footnote{Note that \cite{2021MNRAS.505.5686C} only present their results for LOS tracers. We have, however, also included their (until now unpublished) results including also PMs.} the most recent version of \gs \: until now, and the results presented in \cite{2021MNRAS.501..978R} for the \textsc{Agama} $f(E, L)$ distribution function-based models, which was in most cases the best-performing method (or amongst the best-performing ones) of the multiple ones considered in \cite{2021MNRAS.501..978R}. This distribution function method is part of the widely-used \textsc{Agama} library \citep{2019MNRAS.482.1525V}. These cases shown for comparison use 1,000 tracers.

While \gs1.5 shows better performance than \gs1 in most cases for this mock sample (cf. \citealt{2021MNRAS.501..978R}), it still suffers from significant bias, particularly for the cored model, where densities are systematically underestimated beyond $\sim R_{1/2}$, even when PMs are included (Figs.~\ref{fig:rhoanipm} and \ref{fig:rhoanilos}). 
\gs2, on the other hand, is able to recover the true solutions within the 95\% confidence-level (CL) region from the median within  $ 0.1 \lesssim r/R_{1/2} \lesssim 10$ that encloses the vast majority of the tracers. The only exception to this concerns cases with limited data, where we can see that for the core model with 100 LOS tracers (Fig.~\ref{fig:rhoanilos}), the model underestimates the anisotropy beyond $\sim 2 R_{1/2}$, but is nonetheless able to recover the density everywhere at the specified confidence. For the 1,000-tracer case, the true solution also lies at the edge of the $95\%$ CL region beyond $R_{1/2},$ though these mild biases disappear when the number of tracers is increased or PMs are included, which is not the case with \gs1.5. The performance is otherwise excellent and in all cases competitive with sophisticated modeling approaches such as \textsc{Agama}, which shows a similar performance in most cases. We even see some instances where \gs2 shows less bias than \textsc{Agama} (e.g., Fig.~\ref{fig:rhoanipm}, upper left) at larger radii, and reduced errors overall.\footnote{We emphasize, however, that we are referring to the \textsc{Agama} model applied in \cite{2021MNRAS.501..978R}. This means that the biases that we see strictly speaking apply only to the phase-space distribution function parametrization adopted by these authors.}

Another illustrative case is shown in Fig.~\ref{fig:rhoaniglos}, where we compare the core model with 10,000 LOS tracers. Here all the models considered by \cite{2021MNRAS.501..978R} were biased in either the density or anisotropy recovery, with \gs1.5, \gs1, and \textsc{Agama} showing a degree of bias in both. \gs2, on the other hand, is able to obtain a 95\% CL recovery everywhere throughout. Models that  do not capture deviations of Gaussianity in this case suffer particularly strong bias. To illustrate this, we included the results of \textsc{DiscreteJAM} axisymmetric Jeans models, which assume Gaussian velocity PDFs. We then compared this with the result of imposing a Gaussian PDF in our models and not considering higher moments. This yields a similar result in both cases with a clearly biased model showing significantly greater slopes and uncertainty. \gs1.5 and \gs1, with VSPs, and \textsc{Agama}, which also allows more generality in the velocity PDF modeling, are able to achieve better recovery than the other models considered in \cite{2021MNRAS.501..978R}, but still suffer from a degree of bias across much of the radial range shown ($0.25 \lesssim r/R_{1/2} \lesssim 4$). We also note that the anisotropy priors imposed by \cite{2021MNRAS.505.5686C} on \gs1.5, which we also impose, mean that our models are barred from large negative anisotropies and are close to isotropic at the center. Similarly, \textsc{Agama} models are fixed to be purely radially anisotropic at large radii. This partly explains why these models tend to be more tightly constrained toward the edges of our anisotropy plots, but, at least for \gs2, our results are otherwise insensitive to these priors. 

Our results show that the biases we observe can be overcome in \gs2 by including a full treatment of higher moments via the extended Jeans equation (rather than using VSPs, which can be prone to bias themselves when determining them; see the discussion at the end of Sect.~\ref{sec:3d_gs2}). The inclusion of these higher moments also removes negative-moment models at fourth order, while we extract more information from the data than \gs1.5 by fitting the individual velocities with our generalized PDF modeling. For distribution function-based models, the challenge consists in creating sufficiently generalized functional forms so as to eliminate any biases. On the other hand, it is apparent that models which assume Gaussian velocity PDFs will be prone to biased recovery if the velocity distributions of the system,  as in the case of Fig.~\ref{fig:rhoaniglos}, are intrinsically non-Gaussian.

\begin{figure*}
    \centering
\begin{minipage}[c]{0.75\textwidth}
\includegraphics[width=\ww]{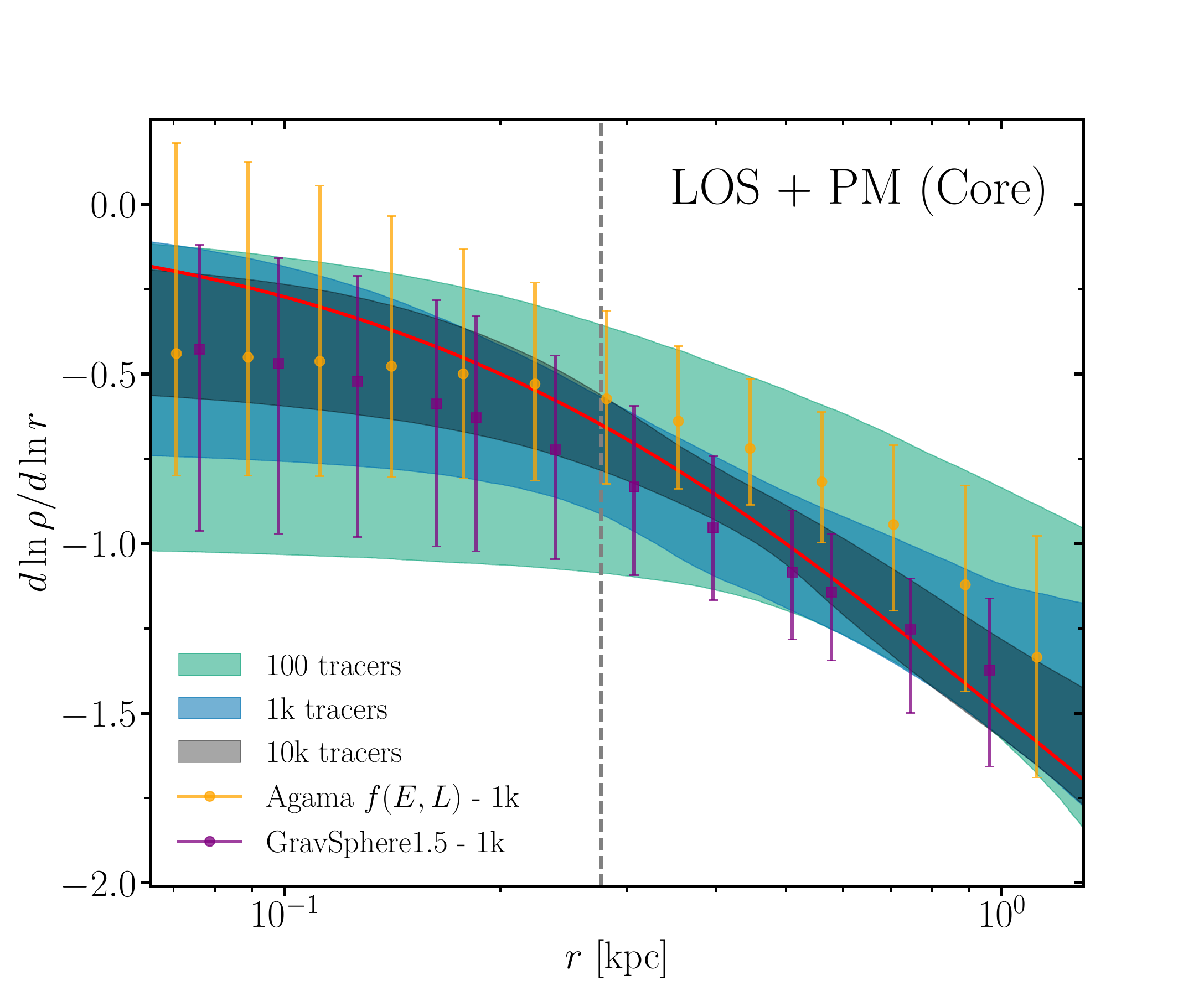}
\hspace{-0.7cm}
\includegraphics[width=\ww]{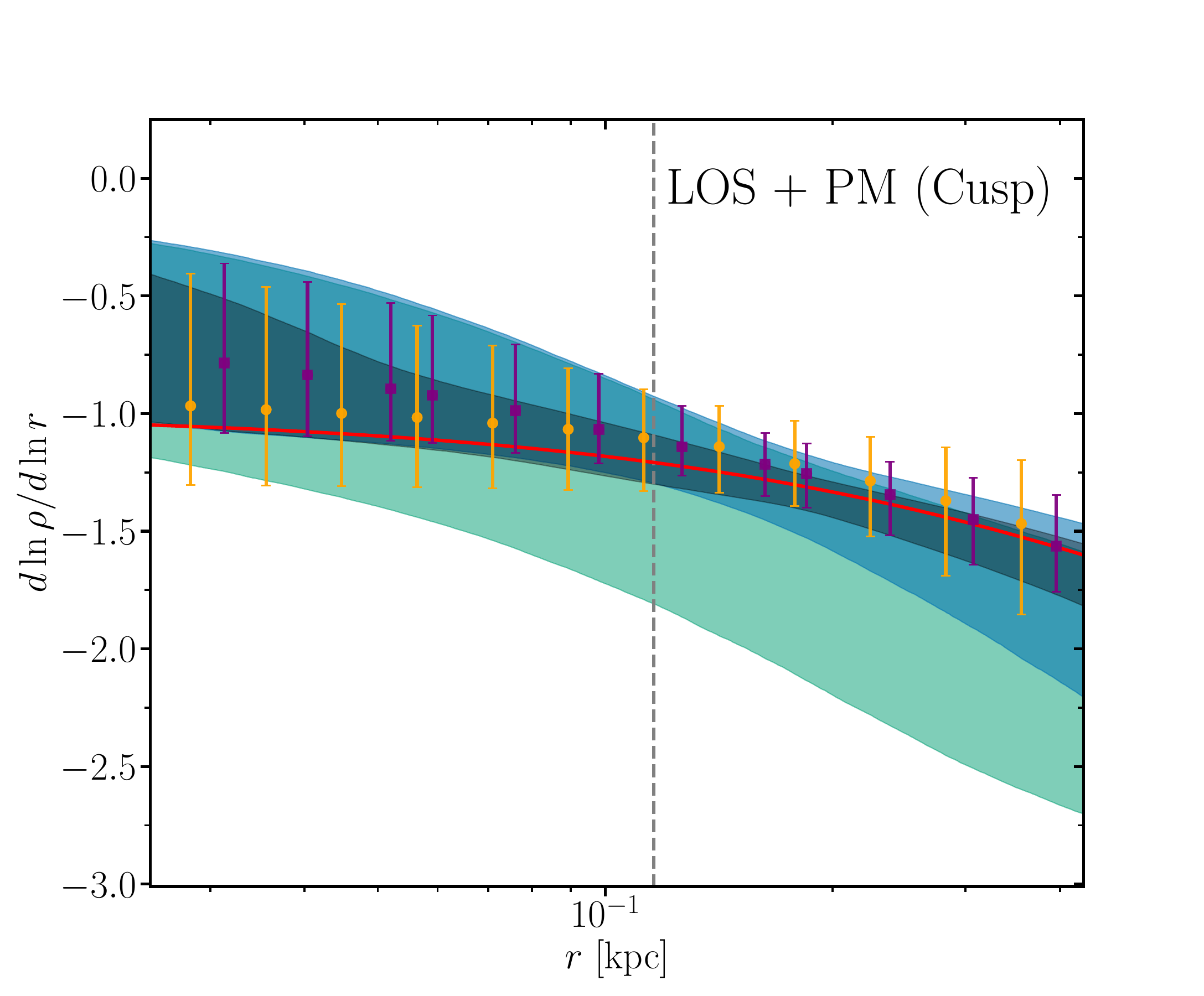}
\includegraphics[width=\ww]{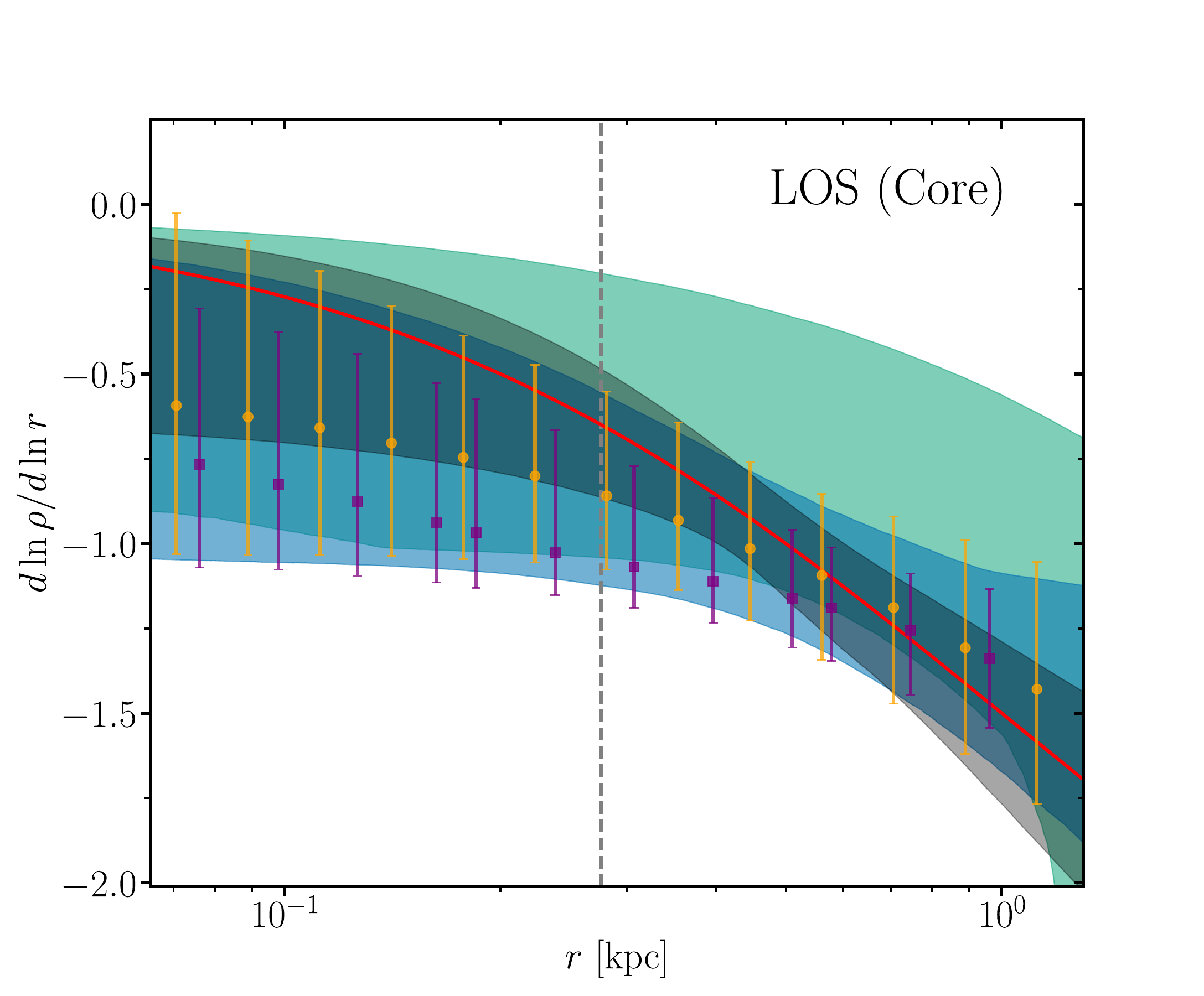}
\hspace{-0.7cm}
\includegraphics[width=\ww]{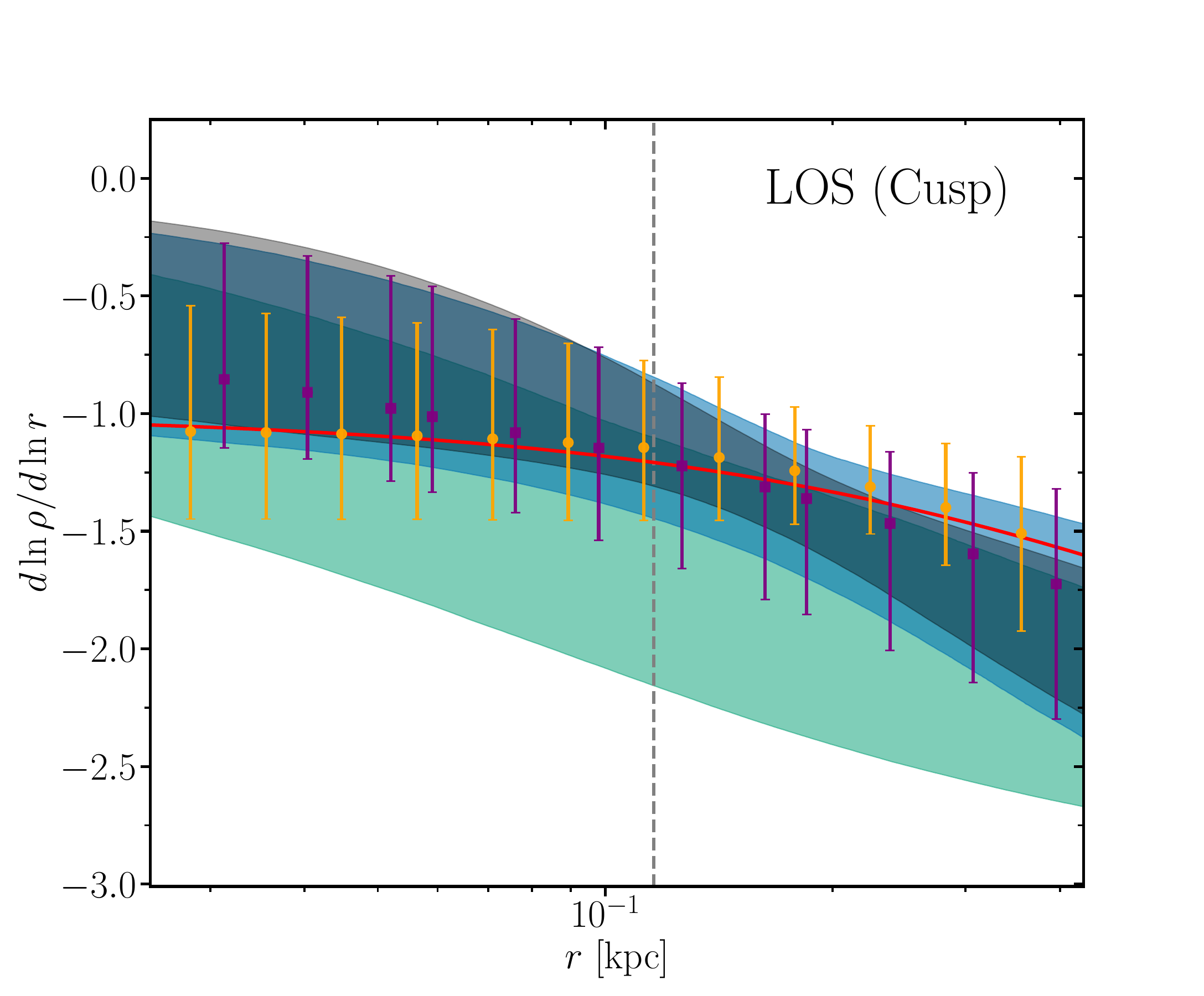}
\end{minipage}\hfill
\begin{minipage}[c]{0.23\textwidth}
 \caption{\gs2 recovery of the logarithmic density slope profile for the \textsc{Gaia Challenge} mock galaxies. Upper panels include LOS and PM tracers, whilst lower panels include only LOS ones. The left panels correspond to the PlumCoreOM case, whilst the right ones to the PlumCuspOM case. The bands, errors and results presented follow the same notation as Fig.~\ref{fig:rhoanipm}. 
 \label{fig:slopes}
 }
\end{minipage}
\end{figure*}

In Fig.~\ref{fig:slopes}, we show the recovery of the logarithmic density slope for the Gaia Challenge mocks. In all cases, \gs2 is able to recover these within its 95\% confidence intervals within $2 R_{1/2},$ and in most cases everywhere in the plot and beyond ($0.1 \lesssim r / R_{1/2} \lesssim 10$). For the 1,000-tracer case, only \gs2 recovers the true solution everywhere within the plots in all mocks. \gs1.5  systematically overestimates the steepness of the slope below $R_{1/2}$ for the LOS core model. \textsc{Agama}, on the other hand, is able to recover the slope profile well for all cases except for the core model with PM data beyond $\gtrsim R_{1/2}$, where moderate bias is observed (and also with higher uncertainties toward the center at $ r \lesssim R_{1/2}$). 
The only case where a mild degree of bias is observed for \gs2 is for the LOS-only cusp model beyond $2 R_{1/2}$, where the true slope profile narrowly falls outside the 95\% CL region. 
 In this case, we verified that this was due to the \textsc{coreNFWtides} model slightly overestimating slopes in these regions. If we adopt instead an $\alpha \beta \gamma$ model (Eq.~\ref{eq:plum}), which was used to generate the mocks (and also adopted for the \textsc{Agama} results), the profile was recovered at all radii without any bias. This test is shown in Appendix~\ref{app:gc_insp}.\footnote{Note: it is not very surprising that biases will be naturally reduced when assuming the same mass model as the one used to generate the mock data. For real data, of course, we cannot know what the true underlying model is, so it is prudent to explore a couple of different options where possible.}

To better understand the reason for the biases observed with other methods, we also show in Fig.~\ref{fig:rhoaniglos} (inset) the results of taking the maxima and minima of models falling within the 68th percentile from the maximum likelihood taken from the posterior distribution. These results were achieved as follows: (i) we compute the posterior of the likelihoods; (ii) determine the limiting value corresponding to the 32nd percentile, so that 68\% of models are equal or above this likelihood (the 100th percentile corresponding to the maximum likelihood); (iii) for each radius, we calculate the minima and maxima of all models with likelihoods greater or equal than our threshold value. This approach differs from the classical Bayesian posterior and we have chosen it as a measure to define the space of models which can be said to demonstrate an "equally good fit." It is more reminiscent of frequentist statistics and we intend to explore it in more detail in upcoming works as a means of testing prior and posterior hypervolume shape dependence in models.

For the Gaussian model, this approach significantly increases the spread and encompasses the true solution, while for the default fourth-order \gs2 model, the effect is more moderate. This indicates that the Gaussian model, rather than unable to fit the correct solution, is not able to discriminate between the true model and those that exhibit substantial deviations from it, while the \gs2 model is. Since the posterior hypervolume of the Gaussian model contains large regions of model configurations with comparable likelihood that are skewed away from the true model, this biases the posterior distribution. This suggests that a similar effect may be taking place in other cases, such as with \gs1.5. 

The latter result leads us to propose a diagnostic that may be important to better understand differences between quoted results for, for instance, the slopes and density profiles of dwarf galaxies, an active topic of debate and discussion in the literature (e.g., \citealt{2018MNRAS.481..860R, 2020ApJ...904...45H, 2024ApJ...970....1V, 2025A&A...699A.347A}). If bounds from the posterior confidence intervals show significantly less spread (i.e., appear more aggressive) than the ones resulting from  the extremal values of models of comparable likelihoods (with our reference test being the 68th percentile from the maximum), this indicates that the results are dominated by the shape of the posterior hypervolume, which is sensitive to priors and model parameterizations. Consequently, one should on these occasions proceed with caution when interpreting the results, which might (without careful consideration) overstate the capacity of the analysis to truly distinguish among competing hypotheses. Conversely, if one does not find such a discrepancy, then it is indicative of the fact that the favored models are indeed better-fitting than the disfavored ones and not merely more abundant in the posterior hypervolume.

As noted in Sect.~\ref{sec:llhood}, however, the priors we adopt in our models are generally very broad compared to the results favored by the data, which we show explicitly in Appendix~\ref{app:priorless}. This shows that the posterior hypervolume dependence that we observe with our diagnostic is not limited to cases where the posterior parameter space spans the entire prior volume, but more generally to degenerate regions allowed by the data within sub-volumes of the prior space, especially when the true solution lies at the edge of the allowable hypervolume. This can be conditioned by a biased likelihood, which is partly at play in our example assuming Gaussianity, and also by the prior itself (noting the hypervolume region is the intersection of the models with allowable likelihoods that lie in the prior space). Also, even in cases where the model correctly recovers the true solution, meaning that the hypervolume is not skewed, a large spread when comparing with our diagnostic can indicate that the confidence of the model is overstated by posterior statistics (i.e.,  that the posterior hypervolume happens to have more models clustering around the true solution). In this way, our diagnostic based purely on the likelihood does not attempt to compute posterior statistics based on the relative abundance of models, as with typical Bayesian approaches, but merely computes the spread on a given quantity allowable within a likelihood threshold, and as a result will not be conditioned by these potential biases.

As noted in Sect.~\ref{sec:llhood}, we  also included marginalized plots of parameters in a public repository for all mocks considered in this study. We note that individual parameters are well constrained in most cases with respect to the prior limits, and even in cases where this is not the case, physical quantities of interest, such as the mass density in regions of interest, are insensitive to these degeneracies (c.f. Appendix~\ref{app:priorless}). 

Lastly, we included binned dispersion and kurtosis profiles for reference in Appendix~\ref{app_bin}. The binned profiles are generally consistent with those derived from our \gs2 mass models and show meaningful deviations from Gaussianity. This, once again, stresses the importance of using models which are able to capture these deviations.

\subsection{Simulated dwarf galaxies}

\begin{figure*}
    \centering
\begin{minipage}[c]{0.75\textwidth}
\includegraphics[width=\ww]{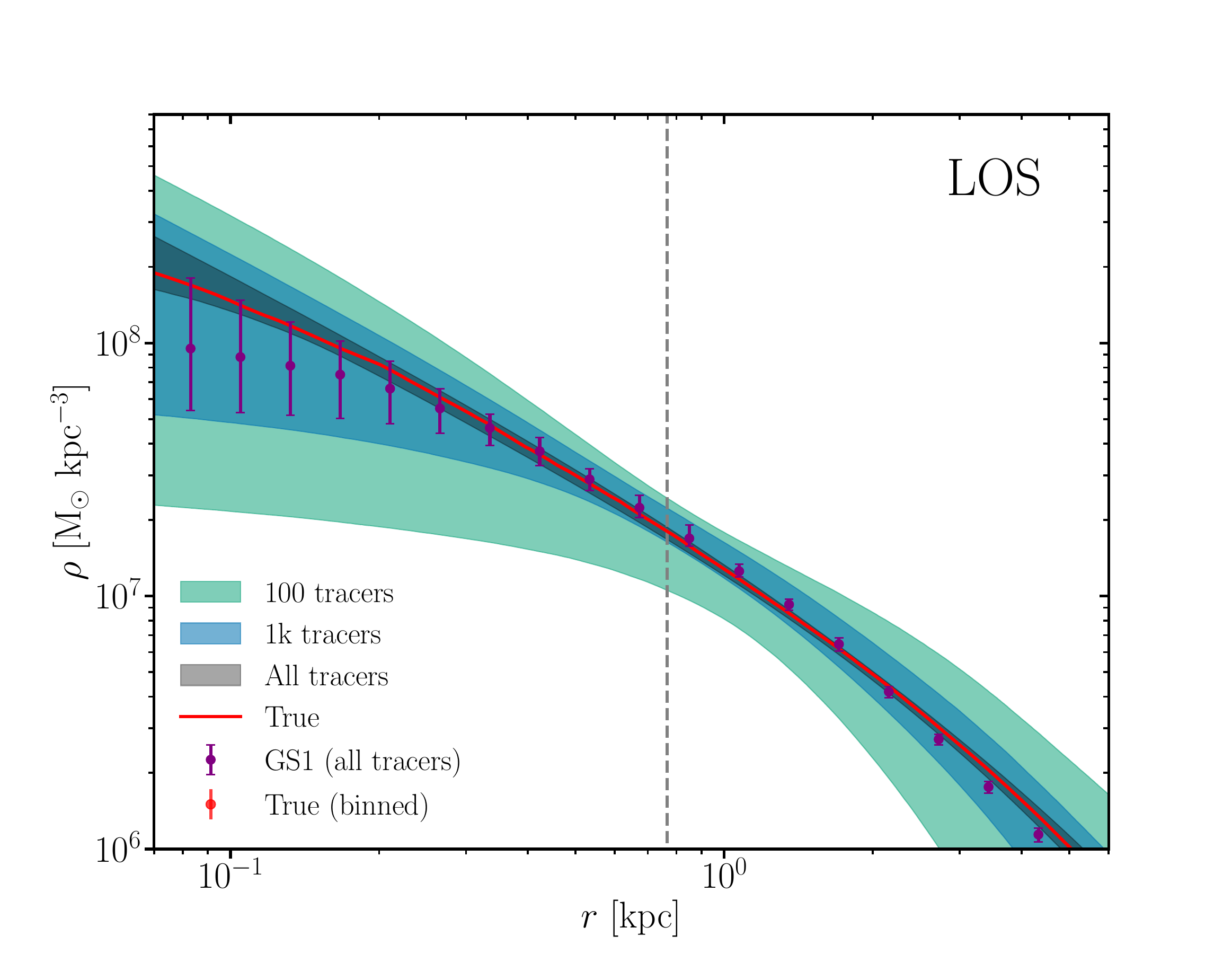}
\hspace{-0.7cm}
\includegraphics[width=\ww]{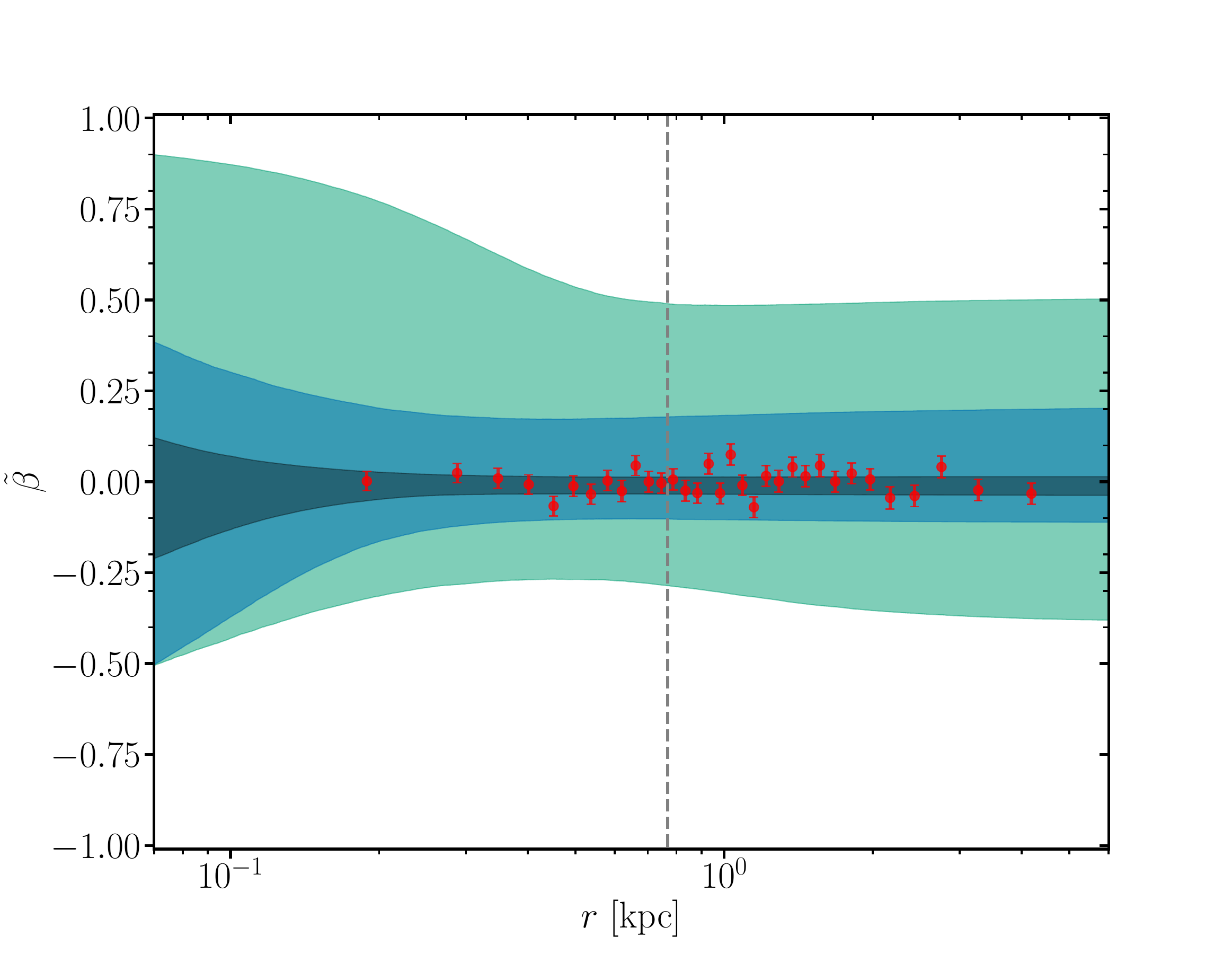}
\includegraphics[width=\ww]{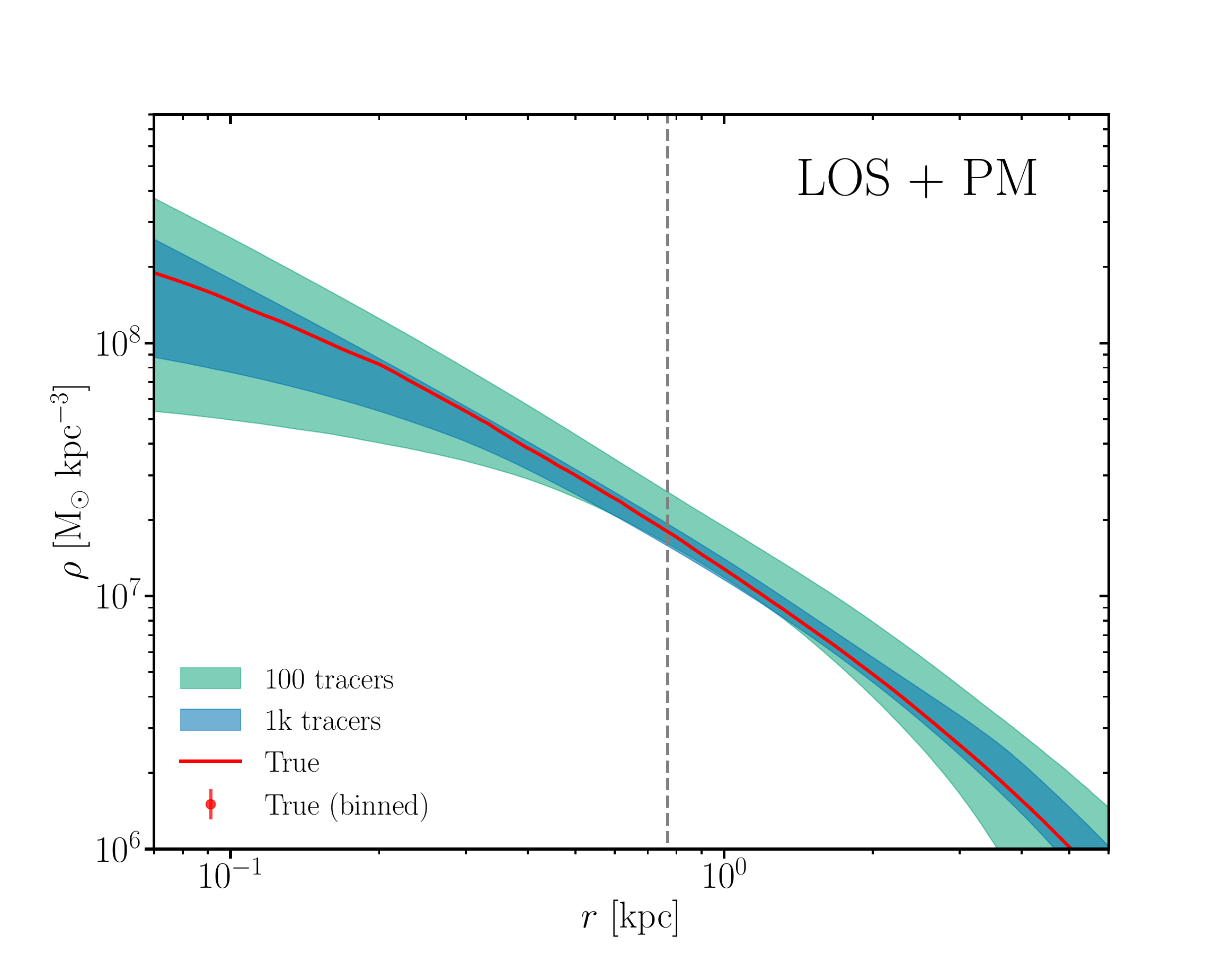} 
\hspace{-0.7cm}
\includegraphics[width=\ww]{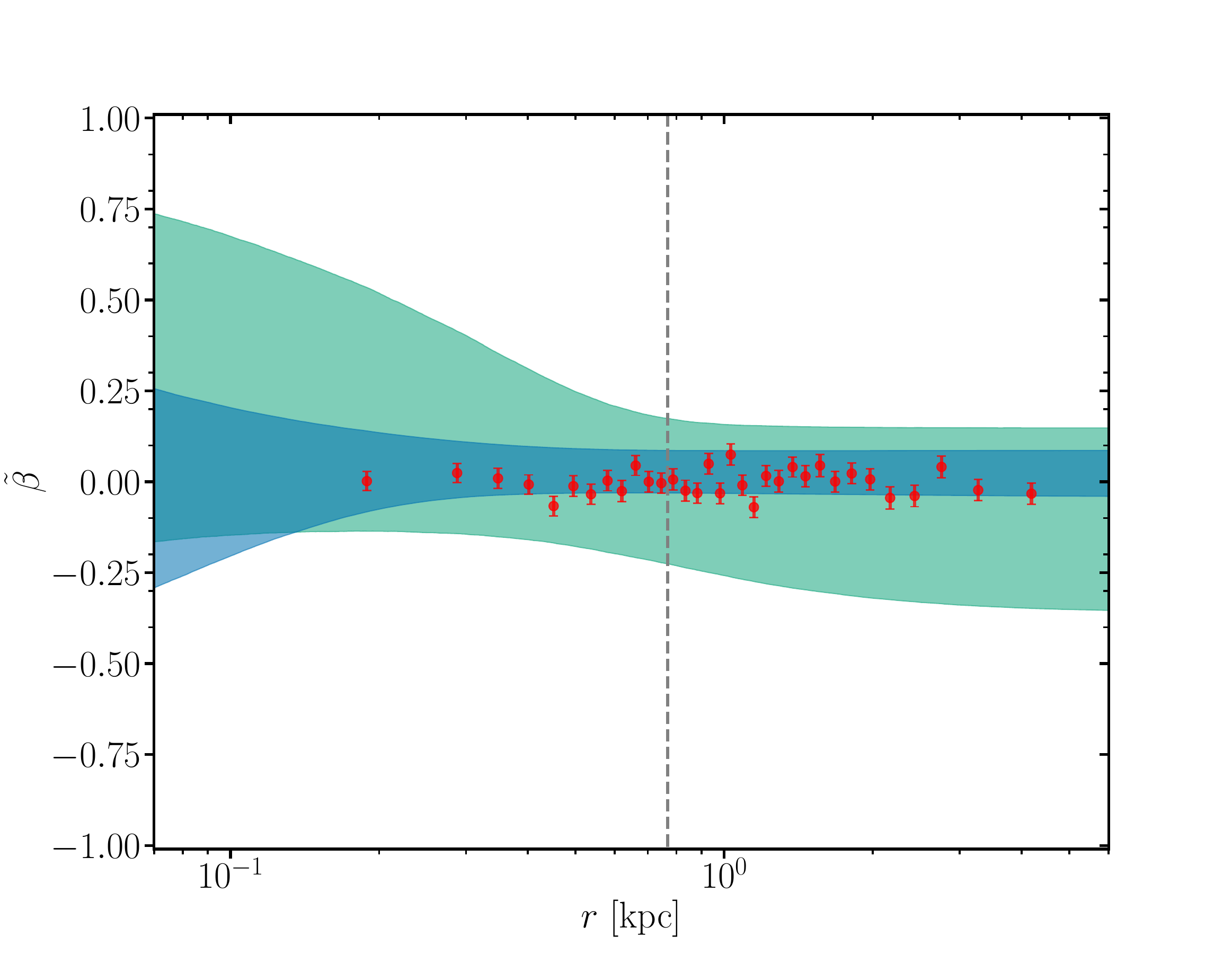}
\end{minipage}\hfill
\begin{minipage}[c]{0.23\textwidth}
 \caption{\emph{Top-left:} \gs2 recovery of the 3D dark matter density  profile for the Fornax-like simulated galaxy, where the bands denote the 95\% CL regions for different numbers of stellar tracers, as marked, including only LOS velocities. The total sample of bound tracers is 34,158 (gray band). The purple points correspond to the median values with 95\% CL errors from \cite{2025arXiv250418617T} using all LOS tracers with \gs1. The red, solid line corresponds to the true density profile from the simulation (smoothened with a Savitzky-Golay filter). The gray, dashed line corresponds to the projected half-light radius ($R_{1/2}$).
 \emph{Top-right:} Symmetrized anisotropy profile results. The red points denote the binned 3D anisotropies obtained from bins of equal numbers of tracers and directly fitting the generalized velocity PDF from Eqs.~\eqref{eq:unik} and \eqref{eq:lapk} along the three spherical-coordinate directions.
\emph{Lower left:} Same as Upper Left but including both LOS and PM velocities.
\emph{Lower right:} Same as Upper Right but with both LOS and PM velocities.
 \label{fig:rhoanipmf} 
 }
\end{minipage}
\end{figure*}

\begin{figure*}
    \centering
\begin{minipage}[c]{0.75\textwidth}
\includegraphics[width=\ww]{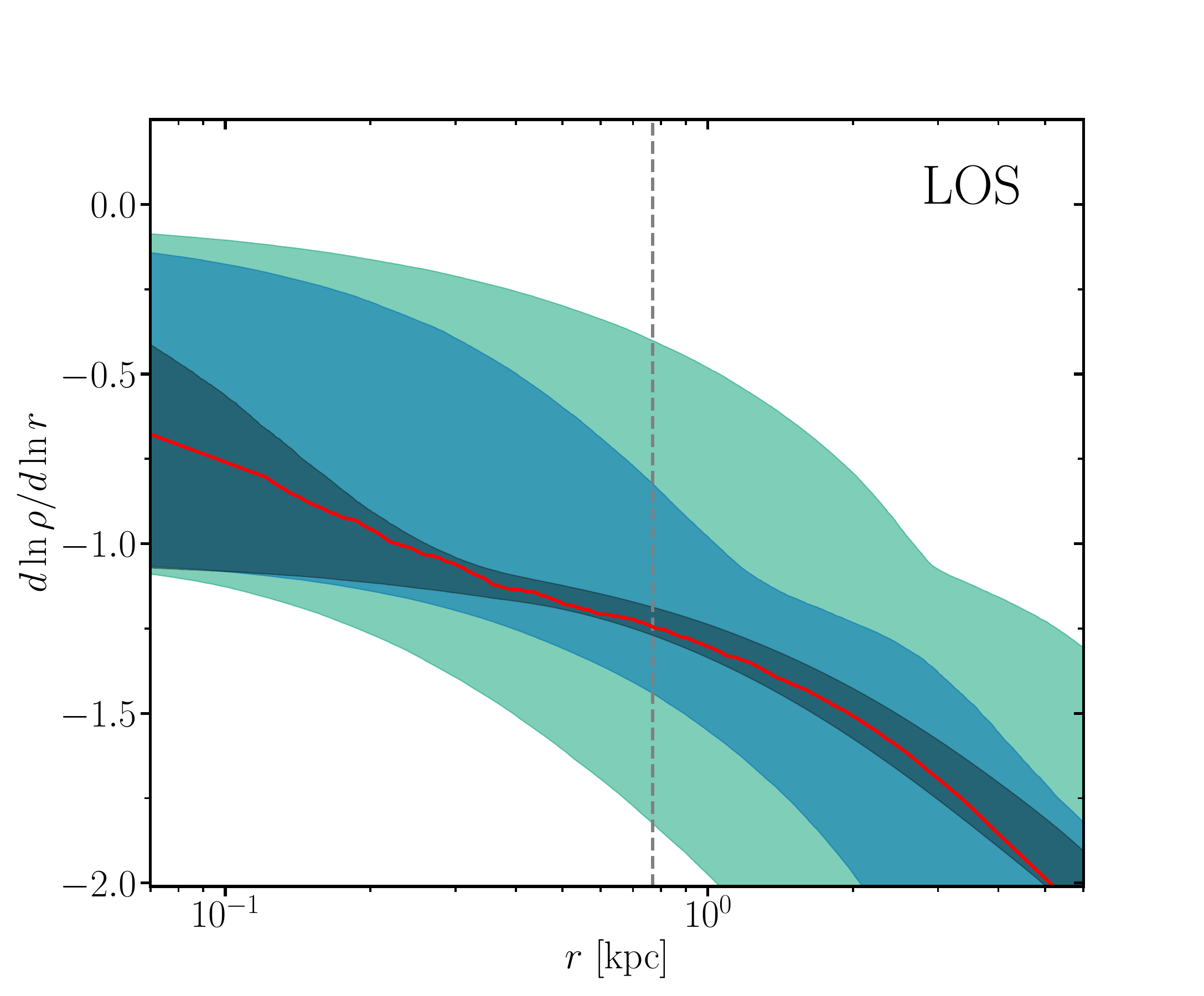}
\hspace{-0.7cm}
\includegraphics[width=\ww]{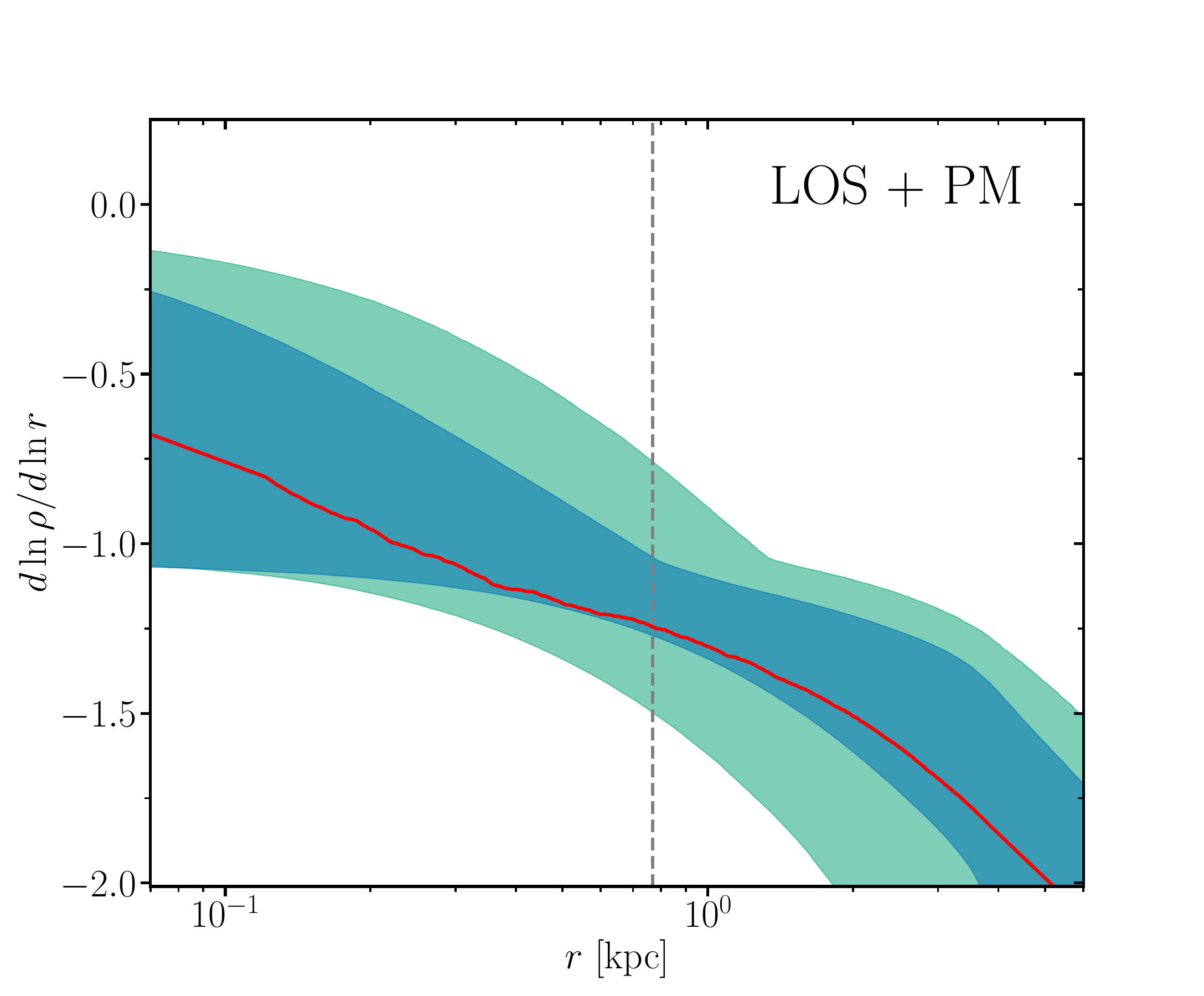}
\end{minipage}\hfill
\begin{minipage}[c]{0.23\textwidth}
 \caption{\emph{Left:} \gs2 recovery of the logarithmic density slope profile for the simulated Fornax-like dwarf. The bands denote the 95\% CL regions for different numbers of stellar tracers (100, 1,000, and 34,518, same as in Fig.~\ref{fig:rhoanipmf}) including only LOS velocities, as marked. The red, solid line corresponds to the slope from the smoothened density profile from the simulation.
 \emph{Right:} Same as the left panel but including both LOS and PM velocities.
 \label{fig:slopef} 
 }
\end{minipage}
\end{figure*}

\begin{figure*}
    \centering
\begin{minipage}[c]{0.75\textwidth}
\includegraphics[width=\ww]{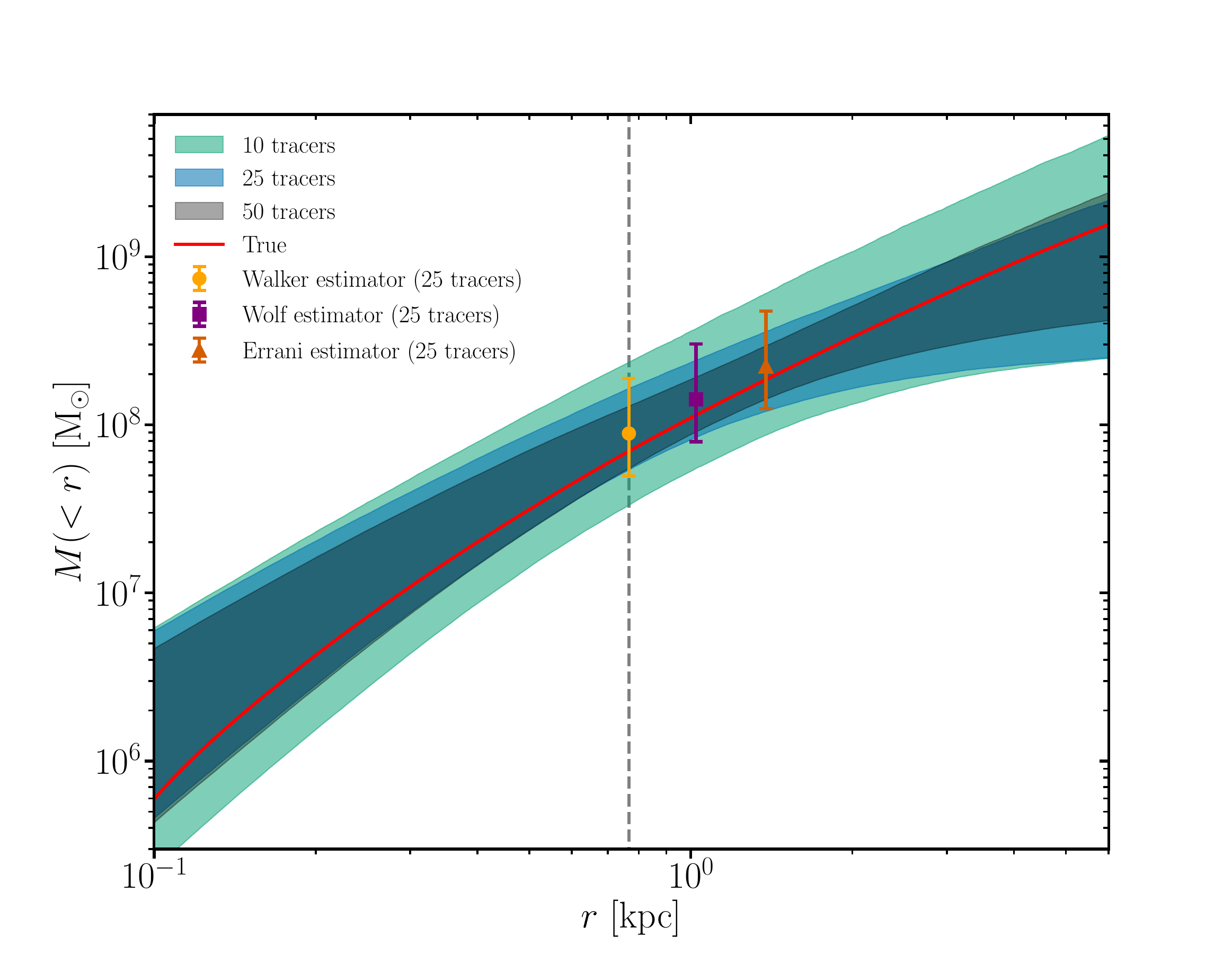}
\hspace{-0.7cm}
\includegraphics[width=\ww]{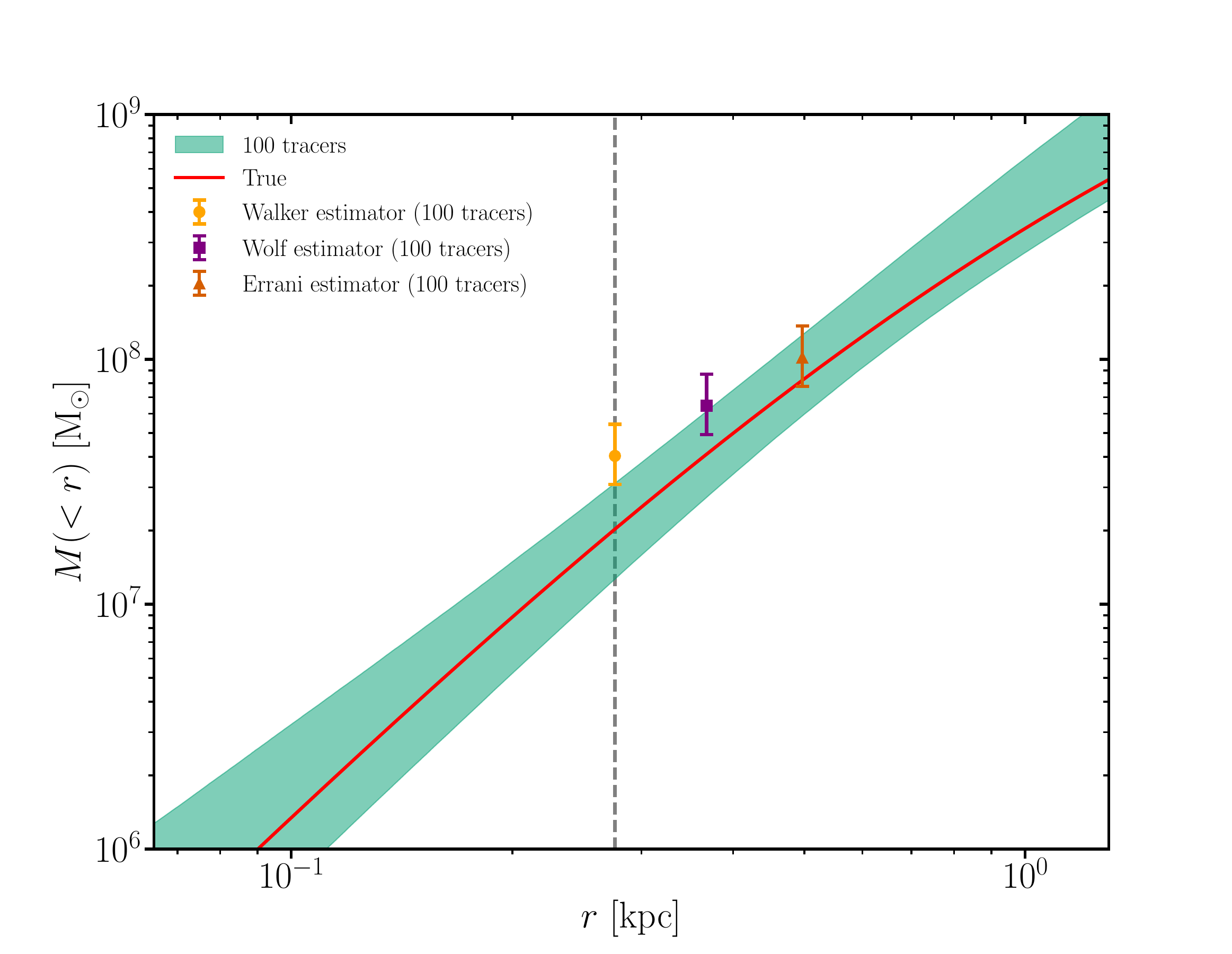}
\end{minipage}\hfill
\begin{minipage}[c]{0.23\textwidth}
 \caption{
 \emph{Left:} \gs2 recovery of the mass profile for the Fornax simulated galaxy with only LOS velocities for low stellar tracer numbers (10, 25, 50). The bands display the 95\% CL values. We show the results obtained using the mass estimators from \cite{2009ApJ...704.1274W, 2010MNRAS.406.1220W, 2018MNRAS.481.5073E}. 
 \emph{Right:} Same, but for the PlumCoreOM mock with 100 tracers and Gaussian PDF binning. 
 \label{fig:slopef2} 
 }
\end{minipage}
\end{figure*}

Fig.~\ref{fig:rhoanipmf} shows the \gs2 recovery of the mass density and anisotropy profiles for the Fornax-like simulated dwarf galaxy with and without PMs in addition to the LOS velocities, whilst Fig.~\ref{fig:slopef} shows the recovery of the logarithmic density slope profiles for these models. 
The recovery of the density, anisotropy and slope using all available bound tracers with LOS velocities is excellent. Using only a small subset of these still yields very good recovery that is further improved by including PMs, leading to a comparable result to the \textsc{Gaia Challenge} equilibrium mocks from Sect.~\ref{sec:gc}, and with true values falling everywhere ($0.1 < r/R_{1/2} < 10$) within the 95\% CL regions. This, on the whole, suggests that these simulated dwarf galaxies do not show significant departures from dynamical equilibrium, spherical symmetry, or the assumption of no rotation. Comparing our results for the all-tracer LOS model with \gs1 from \cite{2025arXiv250418617T} (purple points in Fig.~\ref{fig:rhoanipmf}, upper left), \gs2 is significantly less biased than \gs1 for the density recovery, whilst also showing substantially smaller uncertainties overall (particularly toward the central region). Whilst \cite{2025arXiv250418617T} show that their biases can be substantially reduced by removing VSPs from their fits, this is done at the expense of higher uncertainties. They also obtained less biased results when weakening the monotonicity prior constraint on their nonparametric mass model, though some bias still remained (see their Fig.~D1, also Fig.~8, where their biases are most apparent).

Lastly, we included the projected binned velocity dispersion and kurtosis profiles in Appendix~\ref{app_bin_sim}. These are consistent with the ones inferred from our models using individual velocities and are close to isotropic, not showing large deviations from Gaussianity. The isotropy of the distribution is most evident from the right hand side of Fig.~\ref{fig:rhoanipmf}. This is less evident in \cite{2025arXiv250418617T}, but our binning procedure effectively removes the stochastic noise by averaging over large numbers of tracers, revealing a profile which is close to isotropic.

\subsection{Low tracer numbers and comparisons with mass estimators}

Some of the most interesting objects to study at any given time are either extremely intrinsically faint or distant, meaning that they are (and perhaps always will be) scarce in tracer numbers.\footnote{This should not be interpreted as prognosticating pessimism for future observations, but rather that with increasing observations it appears likely that interest will shift toward ever fainter systems that lie at the limit of the capabilities at any given time.}  Some current examples include the discovery of the faintest known Milky Way satellite \citep{2024ApJ...961...92S} or the dwarf galaxies in Andromeda \citep{2021MNRAS.505.5686C, 2025arXiv250504475P}. In these cases, one is limited to, at most, a few tens of kinematic tracers. It is therefore of considerable interest to study the behavior of \gs2 in this low tracer limit, comparing it with other commonly-applied methods such as the mass estimators from \cite{2009ApJ...704.1274W, 2010MNRAS.406.1220W, 2017MNRAS.469.2335C, 2018MNRAS.481.5073E}. We have done this for the case of the Fornax simulated dwarf, showing our results in Fig.~\ref{fig:slopef2} (left).\footnote{We have omitted showing the mass estimator result from \cite{2017MNRAS.469.2335C} as it produced a very similar result that overlapped with the one from \cite{2018MNRAS.481.5073E}.} For this case, we computed the global velocity dispersion by fitting \gs2's non-Gaussian PDFs from Sect.~\ref{sec:PDF} (i.e., creating one bin for all the tracers), although Gaussian PDF instead did not change our results (in this case). The errors on the estimator were then propagated from the posterior distribution of the global dispersion directly, assuming no other sources of error (systematic or otherwise). In this case, all estimators show good agreement with \gs2, albeit with somewhat larger errors, with all recovering the true solution within the specified 95\% CL regions.

Fig.~\ref{fig:slopef2} (right), on the other hand, shows a different case, where all estimators show a degree of bias that is not present in \gs2. A major source of bias in this case stems from binning  velocities with a Gaussian PDF, as commonly done in the literature.  In Appendix~\ref{app:estimator} we show how using non-Gaussian PDFs instead substantially improves these biases. In this case, all estimators still show a degree of bias, but for the Errani estimator this only becomes apparent in the mock with 10,000 tracers.




Therefore, while the nominal errors of mass estimators are comparable to those of \gs2, the simple estimators suffer from potentially significant systematic biases that are not present in \gs2. We conclude that even for very low numbers of tracers (on the order of 10) it is still worth using \gs2 rather than these estimators, which has the additional advantage of giving constraints on the mass profile at all radii, rather than just one point at $\sim R_{1/2}$. Alternatively, particularly for more robust choices such as the Errani estimator, biases can be substantially reduced when adopting  non-Gaussian PDFs, but these can still creep in as the number of tracers is increased.       

We note that in our main analysis, we  assumed a good knowledge of the photometric tracer profile, using our default value of 10,000 tracers for most tests so far. While it is often the case that the photometric light profile is far better constrained than the kinematic one, one may ask what the effects of restricting ourselves to a low number of photometric tracers might be on the mass estimates. Interestingly, repeating our analysis using the kinematic tracers as the only source of photometric information yielded essentially identical results. This indicates that the advantage of \gs2 over the mass estimators does not stem from additional information provided by a good knowledge of the tracer profile and that the uncertainty in the kinematic information dominates the uncertainty in our mass estimates.

\section{Discussion and conclusion}
\label{sec:disconc}

In this work, we introduced and tested \gs2, an extended Jeans modeling tool that includes full treatment of the higher order Jeans equations of fourth-order moments. We introduced fourth-order PMs as a new observable to break model degeneracies, using a general radially dependent fourth-order anisotropy term as part of our analysis. This approach builds on previous works, which have only considered higher moments for LO velocities and/or made restrictive assumptions about the form of the higher order counterpart of the anisotropy.

In contrast to previous versions of \gs, our methodology does not rely on VSPs which can become biased (see the discussion at the end of Sect.~\ref{sec:3d_gs2}). It is also completely "bin-free," with a generalized velocity PDF that we used to fit discrete data star-by-star. \gs2 includes an improved tracer density model that has the option of fitting tracer positions individually, along with an improved sampling technique. 
\gs2 does not rely on assumptions about the form of the phase-space distribution function, as it implicitly constrains it through its velocity moments via the second- and fourth-order Jeans equations. In this way, our method is able to retain the flexibility and bias-free nature of usual Jeans analyses by not needing to specify any particular form for the distribution function, while gaining constraining power from higher order velocity information,  avoiding negative moments that are unphysical (at fourth order), and offering all the other general advantages of having a distribution function.

We tested these improvements using equilibrium mock data from the \textsc{Gaia Challenge} suite, comparing our results to those from \cite{2021MNRAS.501..978R} and \cite{2021MNRAS.505.5686C}, where multiple methods were tested, including \gs1.5, the (until now) latest version of \gs, \textsc{Agama}, a popular distribution function-based method, and \textsc{DiscreteJAM}, an axisymmetric Jeans-based method. We also tested our method on simulated dwarf galaxy data recently presented in \cite{2025arXiv250418617T}, which were also used to test \gsa\ (referred to in the present paper
as \gs1).

We also tested \gs2 on mocks with a small number of tracers (10 - 100). We compared its performance with simple mass estimators, finding that the \gs2 recovery is not only good, but generally superior. In particular, mass estimators can become significantly biased, while \gs2 not only remains unbiased, but provides constraints on the mass profile at all radii, rather than just one point at $\sim R_{1/2}$ with equal or lower errors. This suggests that even with just ten tracer velocities, it is still preferable to perform full \gs2 modeling (or at least include more flexible, non-Gaussian PDFs such as the ones we use here) over simpler, more commonly used approaches.

While previous studies have suggested that the effectiveness of higher order Jeans analyses is compromised without assumptions about the fourth-order anisotropy counterpart (e.g., \citealt{2013MNRAS.432.3361R, 2014MNRAS.441.1584R}), we find that, not only does a general fourth-order analysis yield comparable (if not improved) constraining power to previous methods for the same data, but, in many cases, a substantially less biased recovery of the anisotropy, mass density, and slope profiles. These are all recovered at 95\% confidence across almost all mocks over a wide range ($0.1 \lesssim r/R_{1/2} \lesssim 10$), regardless of tracer numbers. This stands in contrast to previous methods, including \gs1.5 and \gsa, which are prone to notable biases, particularly at small ($R \lesssim 0.25 R_{1/2}$) and large ($R \gtrsim 2R_{1/2}$) radii.

Using 1,000 tracers without PMs, we recovered the logarithmic density slope at $R_{1/2}$, with $12\%$ (25 \%) statistical errors for 
cuspy (cored) mock data and with comparable absolute errors at lower radii. When including PMs, these can be distinguished with $8\%$ (12 \%) errors. With only 100 tracers and no PMs, we were still able to recover slopes with $\sim 30\%$ (20\%) errors. This indicates that we are able to meaningfully distinguish between cusped and cored models even with limited data, which is particularly timely in light of the present interest for such searches in the literature.

In summary, while PMs lead to substantially improved constraints on our mock data, LOS velocities alone are able to break the mass-anisotropy degeneracy and significantly reduce biases using our framework. Thus, we have established higher order Jeans analysis as a viable alternative to other methods, such as VSPs, which show significantly greater bias across our mock data, irrespective of tracer numbers. Improvements to our method over VSPs likely stem from some combination of biases in the VSPs themselves (see Sect.~\ref{sec:3d_gs2}) and a full star-by-star, fourth-order treatment leading to stronger constraints that disfavor biased models (cf. Fig.~\ref{fig:rhoaniglos}). 

Finally, we contrasted \gs2 with classic second-order Jeans modeling assuming Gaussianity. We demonstrated that due to the mass-anisotropy degeneracy, classic Jeans models are unable to distinguish cusps from cores even with 10,000 tracers. Indeed, as reported previously in the literature (e.g., \citealt{2021MNRAS.501..978R}), classic Jeans can become seemingly highly biased in its recovery of the density profile, appearing to strongly favor a cusp when the mock is actually cored. We show that this bias owes to the presence of many more cuspy models than cored models in the solution hypervolume, rather than being due to cored models giving a poorer fit to the data. As such, apparent biases on this level can be misleading, owing to the shape of the prior volume and unphysical model degeneracies (which \gs2 is able to resolve), rather than the data. We suggest a new diagnostic test to determine when posterior intervals are biased in this way: plotting the spread of models of comparable likelihood, using the 68th percentile from the maximum as a reference threshold. We showed that this diagnostic test removes the apparent bias, giving a more faithful representation of the span of models that well-represent the data. 

Since \gs2 has a distribution function, it can self-consistently model uncertain membership probabilities of individual tracers, robustly including a foreground model, which we will include in an upcoming publication Bañares-Hernández et al. (in prep.). Further additions to \gs2 are in preparation and will be presented in upcoming papers, including a mixture model for binary stars (following \citealt{2025arXiv250914316G}) and a validation of \gs2 on rotating mocks.
Additionally, we have already incorporated in the public version of \gs2 the ability to include self-consistent escape velocities of stars as an additional constraint to the potential (cf. \citealt{2024Natur.631..285H}). 

Given the present challenges concerning the structure of nearby dwarf galaxies and star clusters, \gs2 will be an excellent new tool for studying these systems and making the most of their ever-growing data sets. This will help us to address key outstanding questions in the field, such as whether nearby dSphs have dark matter cusps or cores (e.g., \citealt{2015MNRAS.452.3650O, 2017ARA&A..55..343B, 2019MNRAS.484.1401R, 2020JCAP...06..027K, 2020MNRAS.495...58S, 2024MNRAS.535.1015D}) and whether nearby GCs host central IMBHs (e.g., \citealt{2017IJMPD..2630021M, 2020ARA&A..58..257G, 2020A&ARv..28....4N, 2023MNRAS.522.5740V, 2023arXiv231112118A, 2024Natur.631..285H, 2025A&A...693A.104B}). We will apply \gs2 to data for nearby dSphs and GCs in forthcoming publications to address these questions.

\begin{acknowledgements} 

We thank the anonymous referee for prompt and helpful comments that have helped improving this work. We are grateful to Jason Sanders for helpful comments on the use of velocity distribution functions used for this work and on the manuscript. We are grateful to Anna Genina for facilitating us access to simulated data and \textsc{PyGravSphere} mock data results used in this work.
We would like to thank Eugene Vasiliev, Jorge Martín Camalich, Giuseppina Battaglia, José María Arroyo Polonio, Raffaele Pascale, Eduardo Vitral, Amery Gration, and Dashuang Ye for helpful discussions and their interest in our work. We also thank the organizers of the ``Valencia Workshop on the Small-Scale Structure of the Universe and Self-Interacting Dark Matter,'' where this work was presented.
The authors wish to acknowledge the contribution of the IAC High-Performance Computing support team and hardware facilities to the results of this research. ABH acknowledges funding received from the European Union through the grant ``UNDARK'' of the Widening participation and spreading excellence program (project number 101159929).
ABH acknowledges support from the MICINN through
the grant “DarkMaps” PID2022-142142NB-I00. JIR would like to acknowledge support from STFC grants ST/Y002865/1 and ST/Y002857/1.
This work has made use of the following software packages: \textsc{dynesty} \citep{2020MNRAS.493.3132S}, \textsc{corner.py} \citep{corner}, \textsc{NumPy} \citep{Harris_2020}, \textsc{SciPy} \citep{2020SciPy-NMeth}, \textsc{Matplotlib} \citep{Hunter:2007}, \textsc{Jupyter Notebook} \citep{2016ppap.book...87K}, \textsc{gh\_alternative} \citep{2020MNRAS.499.5806S}. 
\end{acknowledgements} 
\bibliographystyle{aa} 
\bibliography{references}

\begin{appendix}

\section{Gaia Challenge logarithmic density slope profile recovery with an $\alpha \beta \gamma$ mass model}

\label{app:gc_insp}

\begin{figure}[h!]
    \centering
\includegraphics[width=\columnwidth]{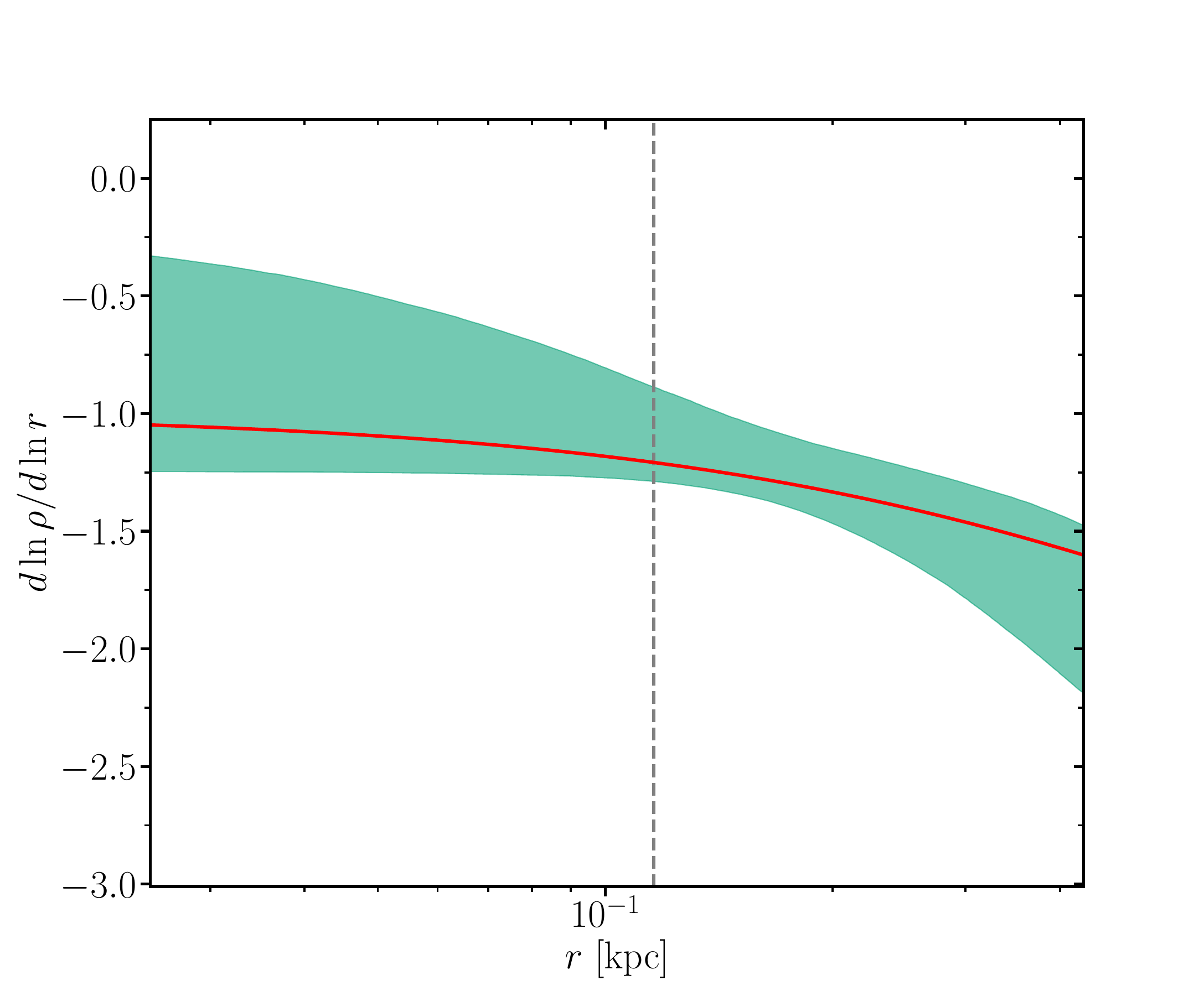}

 \caption{\gs2 recovery of the logarithmic density slope profile for the PlumCuspOM model with 10,000 LOS tracers, using an $\alpha \beta \gamma$ model (Eq.~\eqref{eq:plum}) for the mass density profile instead of the default \textsc{coreNFWtides} model. The true profile (red) is recovered everywhere in the plot within the 95\% CL region (green), yielding less bias than the \textsc{coreNFWtides} model, as expected since the mock was set up using an $\alpha \beta \gamma$ model. 
 \label{fig:innabg} 
 }
\end{figure}

In this appendix, we consider how well \gs2 can recover the PlumCuspOM mock assuming an $\alpha \beta \gamma$ functional form for the enclosed mass profile rather than the default \textsc{coreNFWtides} model. The $\alpha \beta \gamma$ profile was used to generate the mock, and so should yield less bias. The results are shown in Fig.~\ref{fig:innabg}. We note that this does indeed yield less bias as compared to using the \textsc{coreNFWtides} model (cf. with Fig.~\ref{fig:slopes} (lower right), where the recovery is very slightly biased outside its 95\% CL region).

\section{Recovery comparison with priors}

\label{app:priorless}

\begin{figure}[h!]
    \centering
\includegraphics[width=\columnwidth]{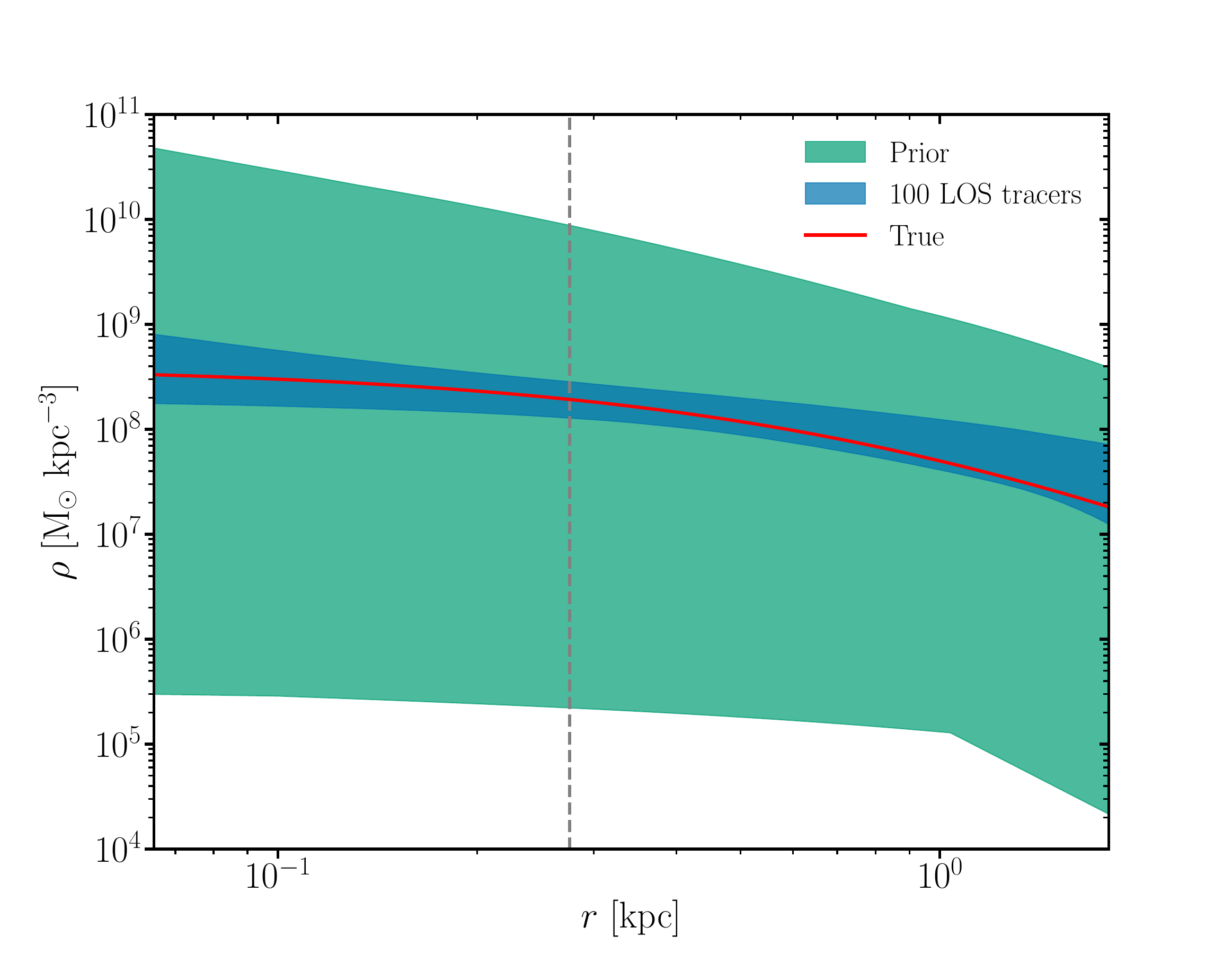}

 \caption{Comparison of \gs2 recovery (95\% CL from the median) with just 100 LOS tracers and the range spanned by the priors on the mass model for the PlumCoreOM mock galaxy from the Gaia Challenge suite (same as in Fig.~\ref{fig:rhoaniglos}).
 \label{fig:priorless} 
 }
\end{figure}

\begin{figure}[h!]
    \centering
\includegraphics[width=\columnwidth]{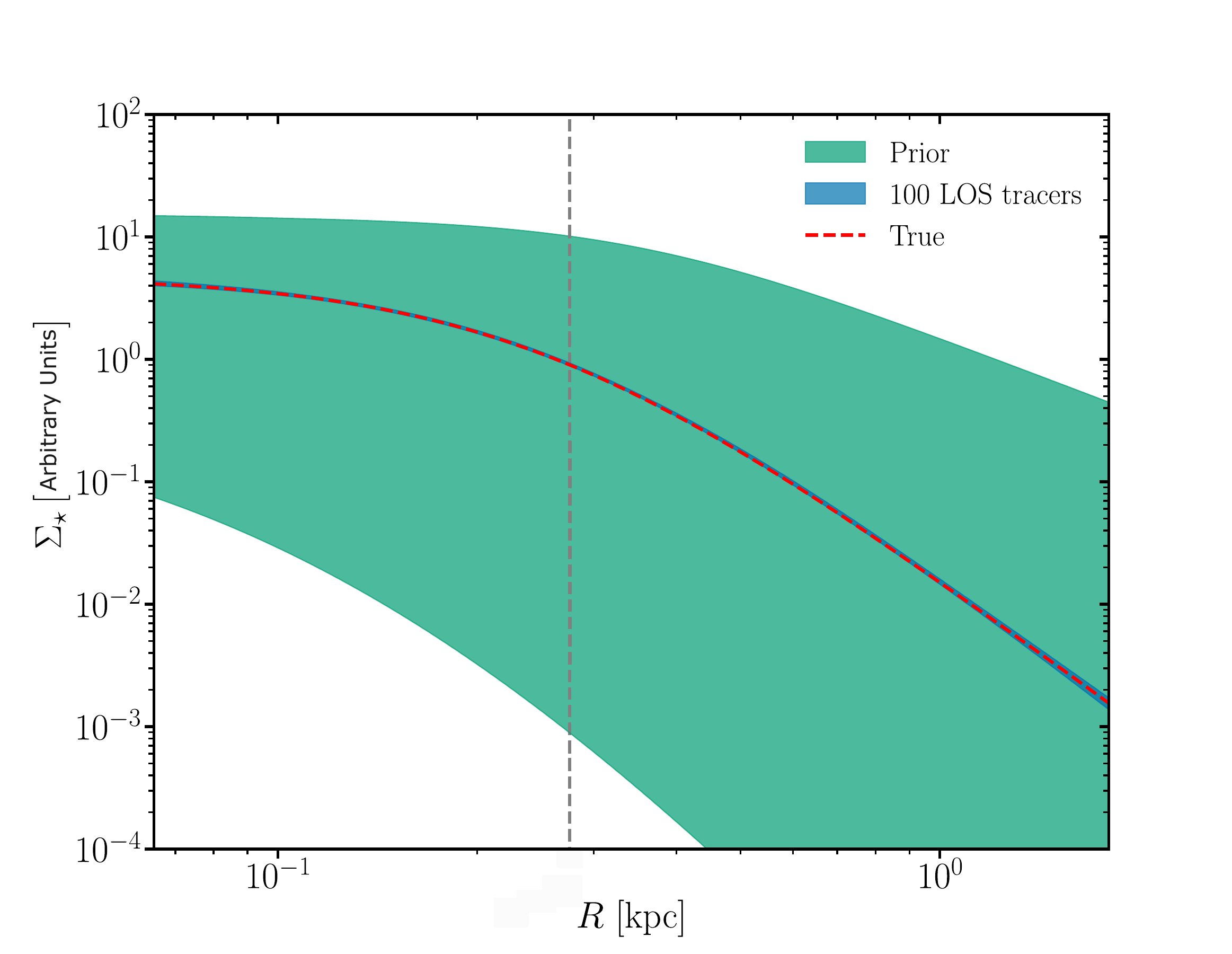}

 \caption{Comparison of \gs2 recovery (95\% CL from the median) with just 100 LOS tracers (10,000 for the binned photometric component) and the range spanned by the priors on the tracer surface density model for the PlumCoreOM mock galaxy from the Gaia Challenge suite (same as in Fig.~\ref{fig:rhoaniglos}).
 \label{fig:priorless_phot} 
 }
\end{figure}

In Fig.~\ref{fig:priorless} we show as an example how, even for our least constraining model of the Gaia Challenge sample (just the LOS velocities of 100 tracers), the range of densities allowed by our priors (which is very similar across all the galaxies considered in our mock samples) spans several orders of magnitude and therefore has little effect on the result when using the data to constrain the mass model. We showed this by running a fit without any likelihood (just sampling the prior space) and plotting the range of densities spanned by the model in this case.
Fig.~\ref{fig:priorless_phot} shows the recovery of the tracer surface density for the same mock galaxy using a binned profile with 10,000 photometric tracers. It is clear that the recovery is excellent and far in excess of the allowable range by the prior. In the cases we considered, results will be essentially unchanged if one instead fixes the photometric profile to its best-fit value without additional marginalization. This can be used to accelerate run-times in \gs2. 

\section{Mass estimators with non-Gaussian PDFs and multiple tracers}

\label{app:estimator}

\begin{figure*}
    \centering
\includegraphics[width=\ww]{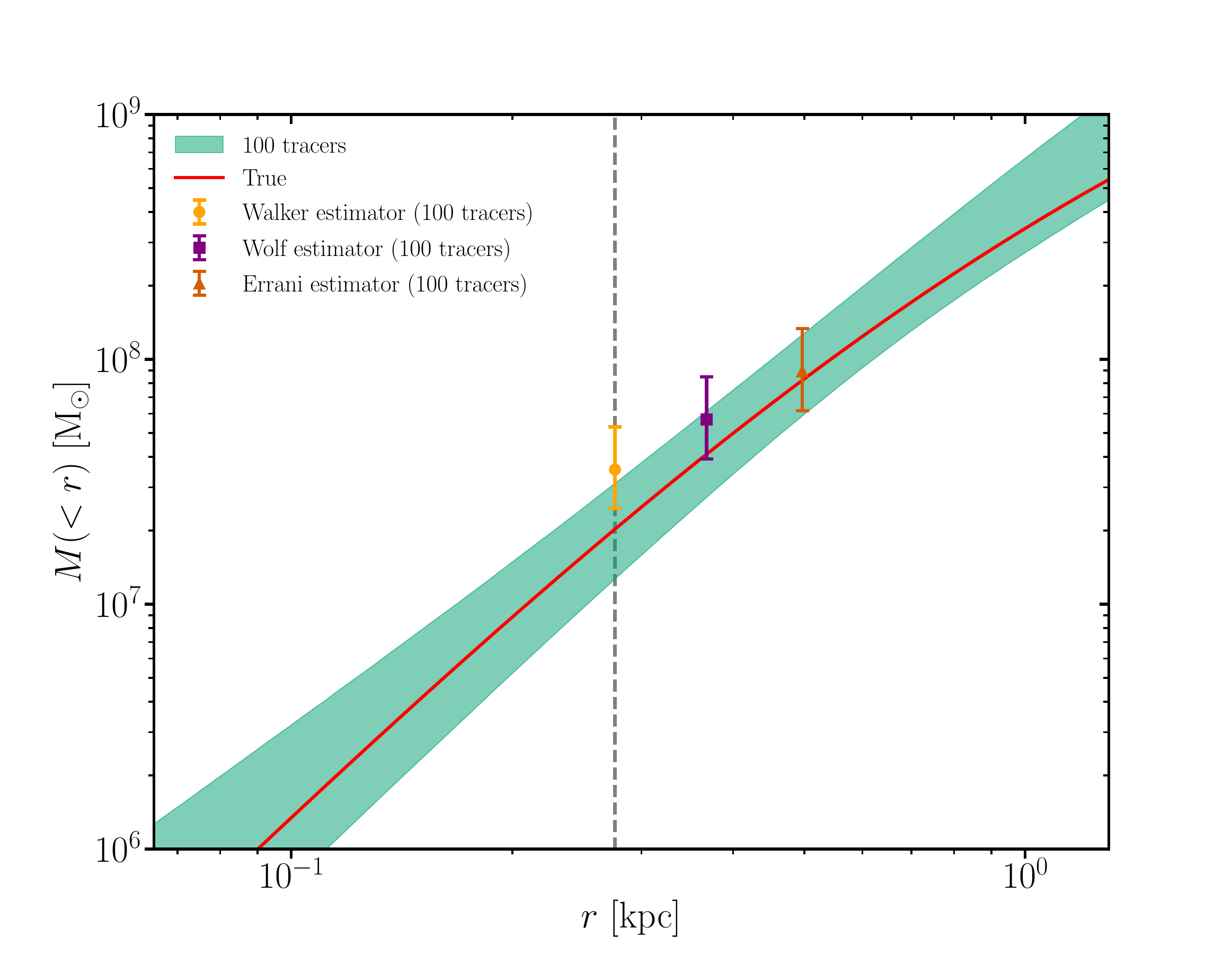}
\includegraphics[width=\ww]{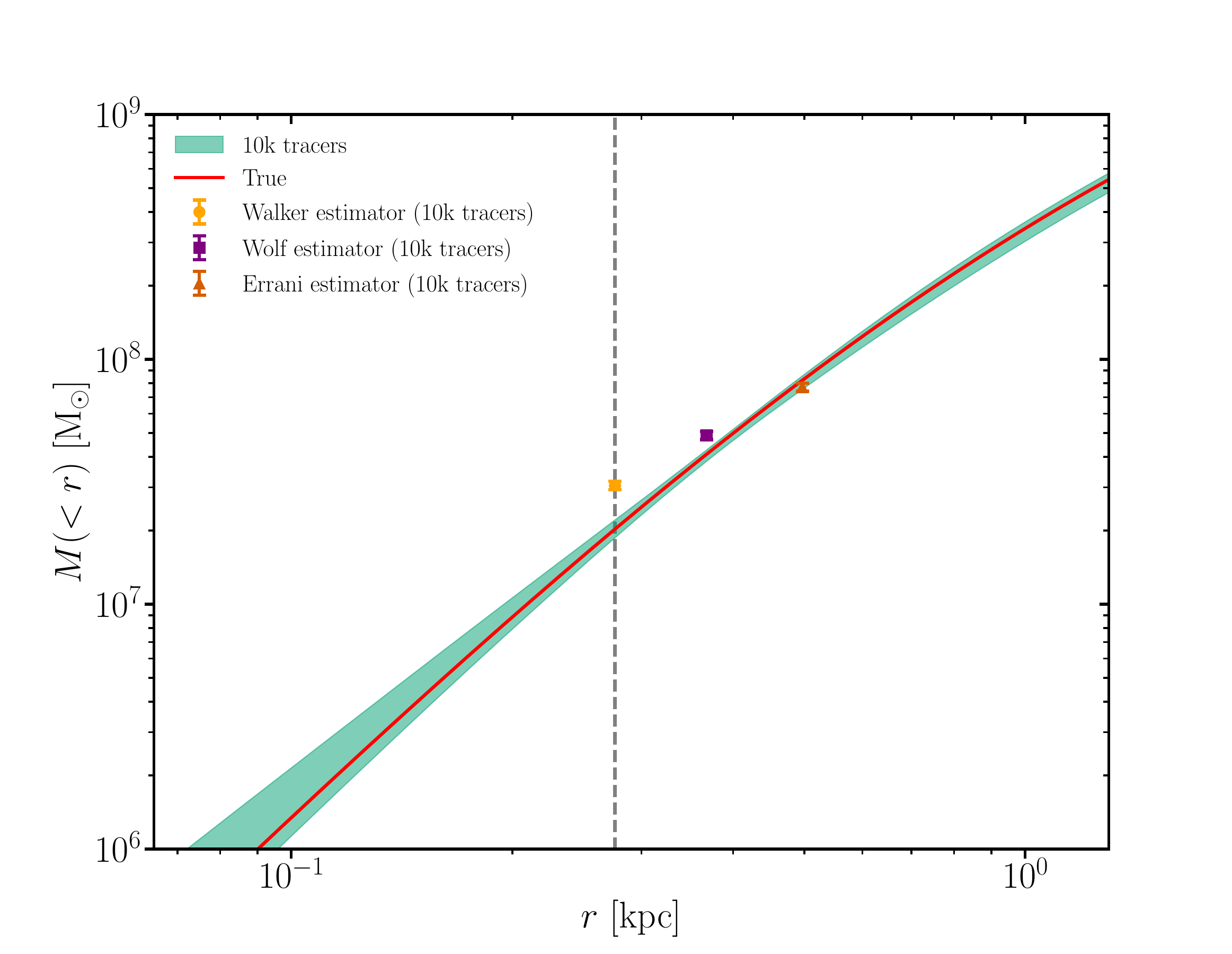}

 \caption{\emph{Left:} \gs2 recovery, using \gs2's non-Gaussian PDFs (Sect.~\ref{sec:PDF}), of the enclosed mass profile for the PlumCoreOM mock galaxy with only LOS velocities for 100 tracers, as compared to the same for simple mass estimators (as marked). The bands show the 95\% CL values. \emph{Right:} Same as left but for 10,000 tracers. While non-Gaussian PDFs improve results, a degree of bias is present in all estimators, though to a lesser extent in the Errani estimator, where the bias becomes negligible (i.e., relative to statistical errors) for 100 tracers.
 \label{fig:lowmass_3} 
 }
\end{figure*}

Here, we explore the behavior of the simple mass estimators using \gs2's non-Gaussian PDFs (Sect.~\ref{sec:PDF}) for both 100 tracers and a much larger number of tracer stars (10,000). The results are shown in Fig.~\ref{fig:lowmass_3}, which compares the \gs2 recovery of the enclosed mass for the PlumCoreOM \textsc{Gaia Challenge} mock with both 100 and 10,000 tracers with the same for simple mass estimators. Notice that while biases improve following these PDFs, they still persist for the Wolf and Walker estimators for 100 tracers and (to a varying extent) in all of them for the 10,000 tracer case, while \gs2 remains unbiased.

\newcommand\x{-0.5cm}
\newcommand\y{6.0cm}
\newcommand\h{-0.0cm}
\newcommand\hb{0.0cm}

\section{Gaia Challenge binned profiles}
\label{app_bin}

\begin{figure*}
    \centering

\includegraphics[width=\y]{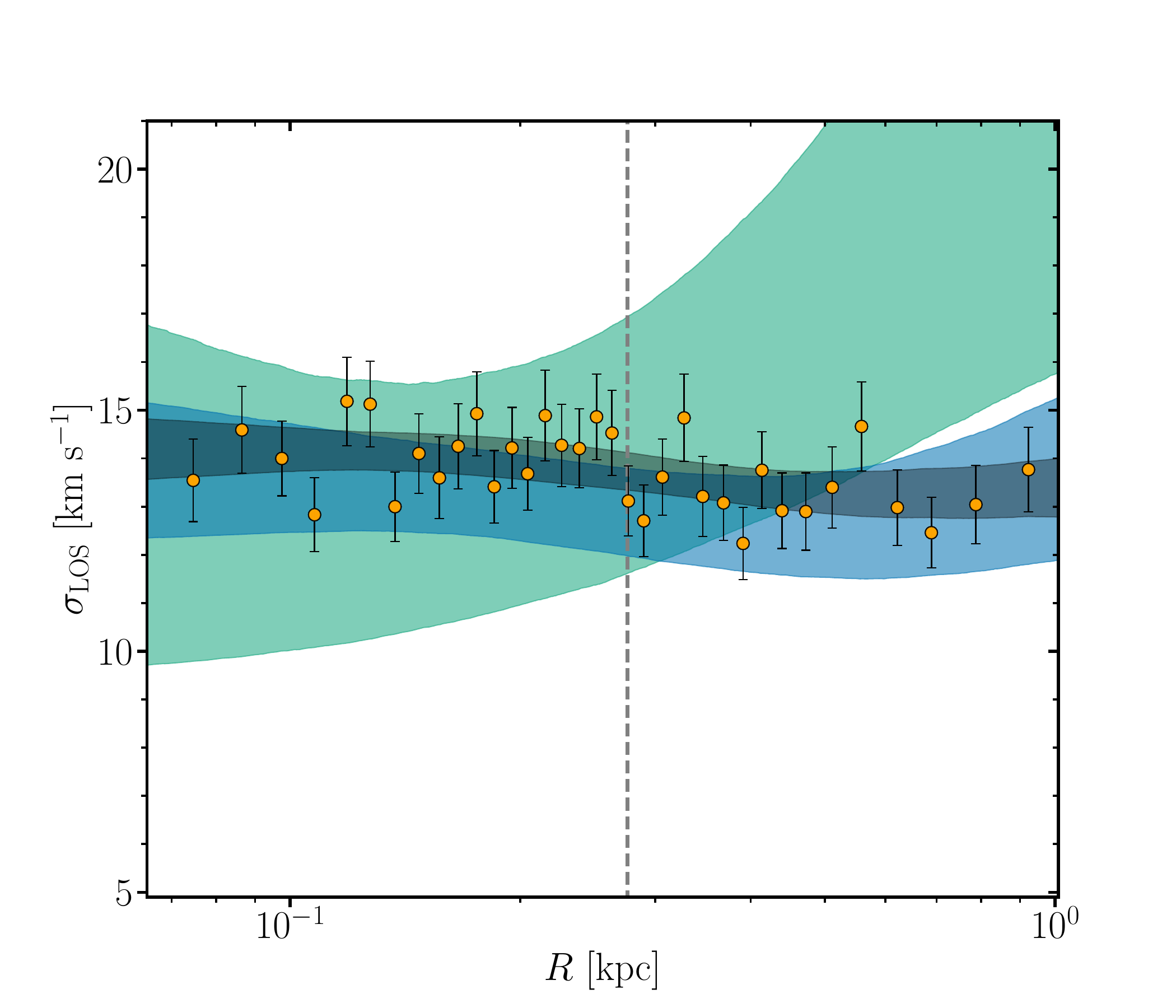}
\hspace{\x}
\includegraphics[width=\y]{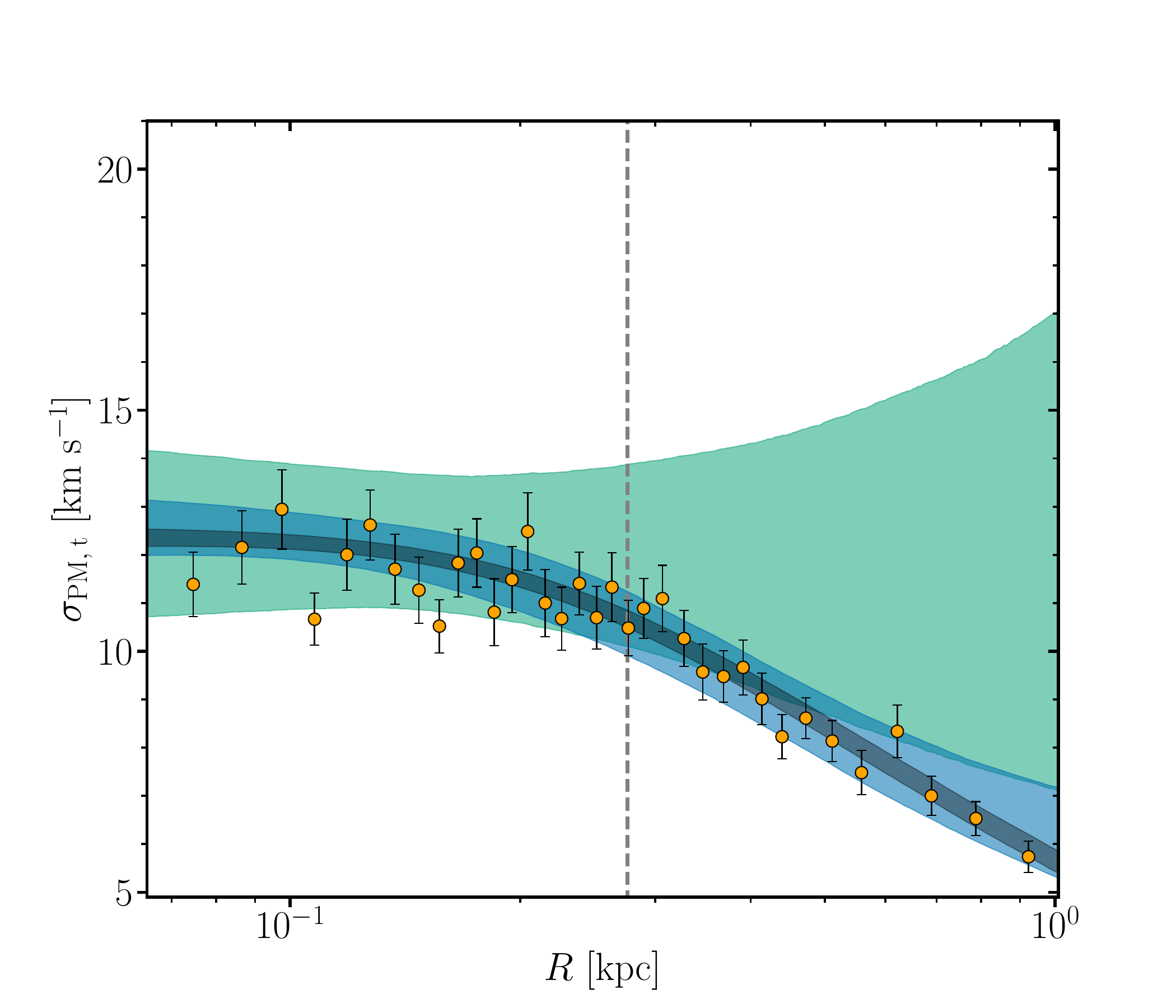}
\hspace{\x}
\includegraphics[width=\y]{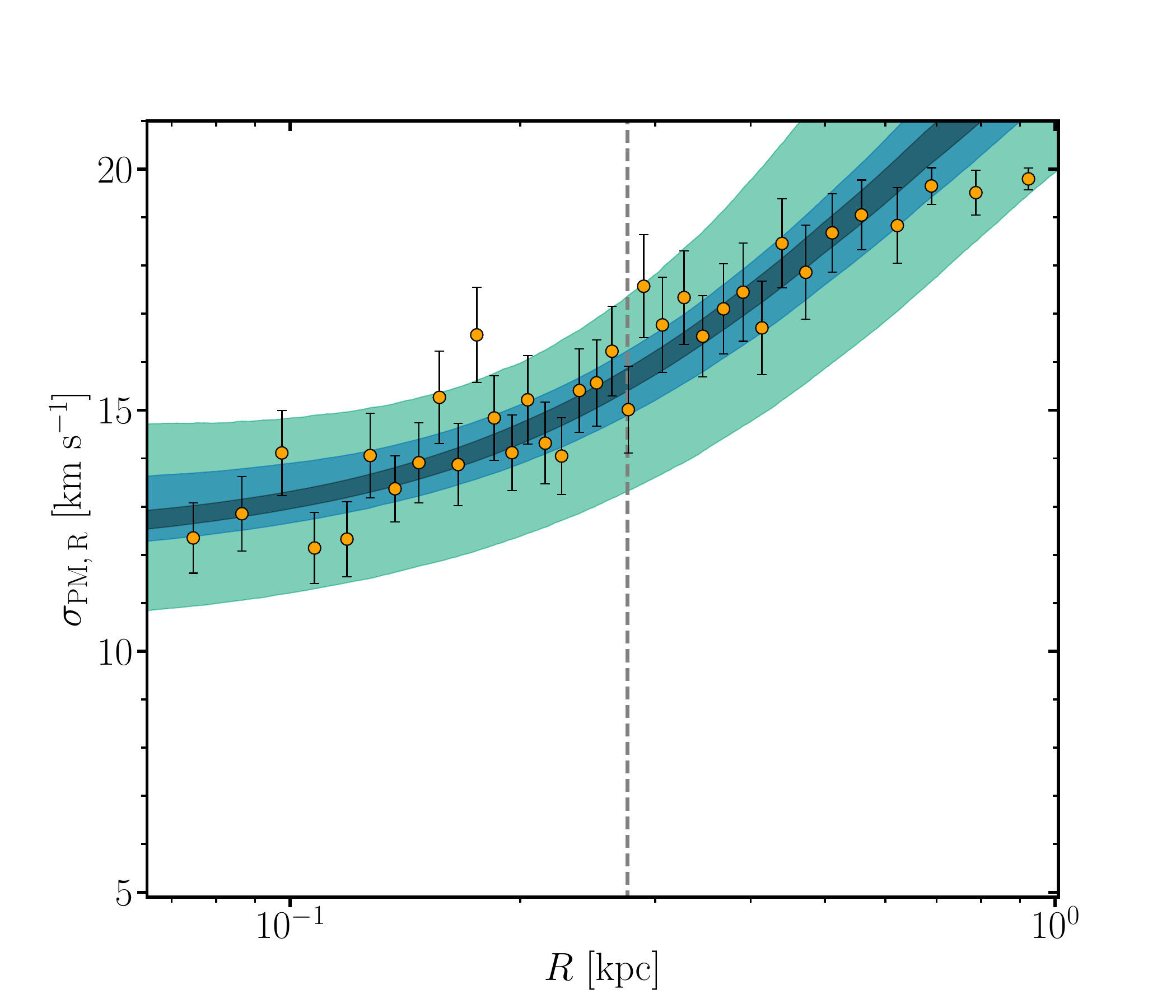}
\vspace{\h}
\includegraphics[width=\y]{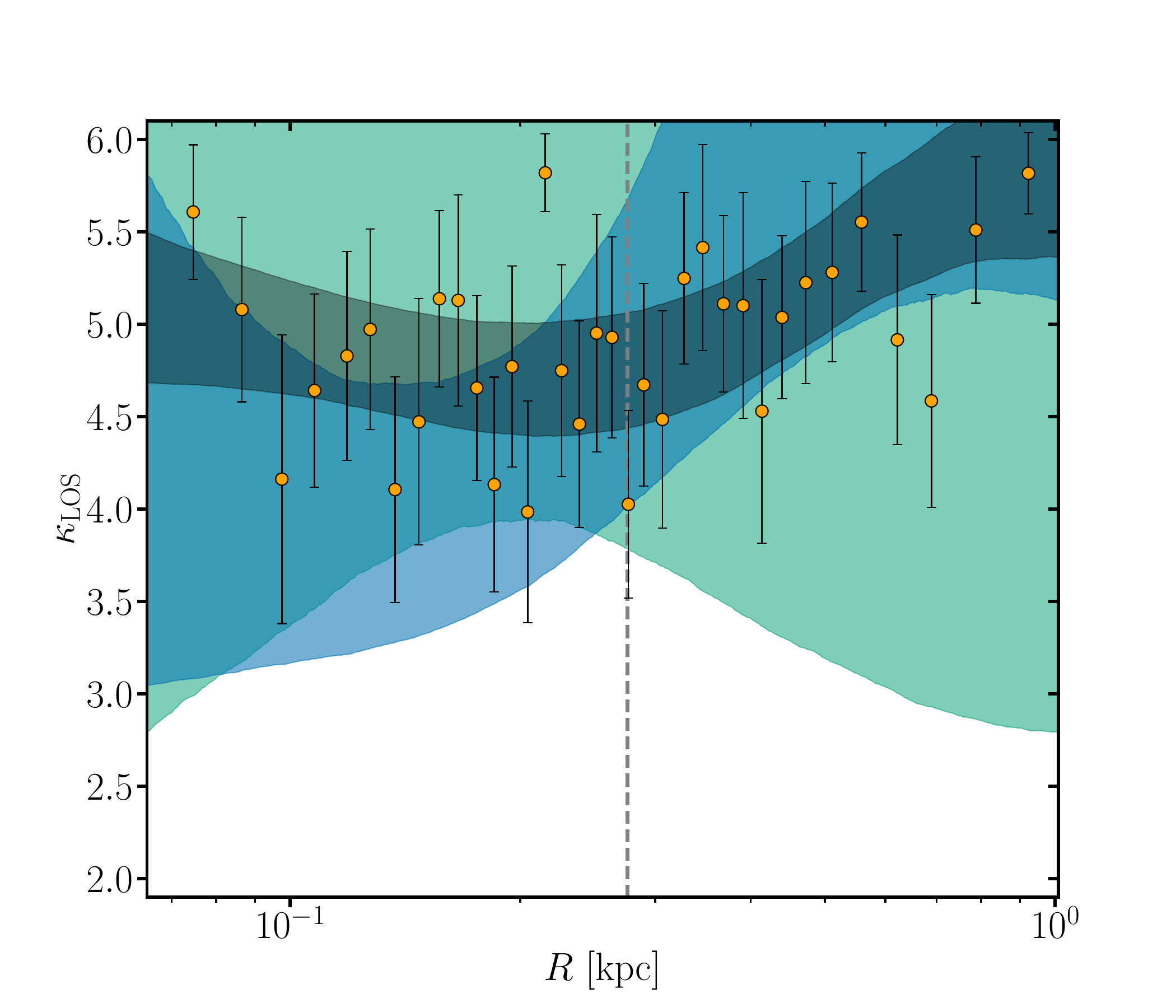}
\hspace{\x}
\includegraphics[width=\y]{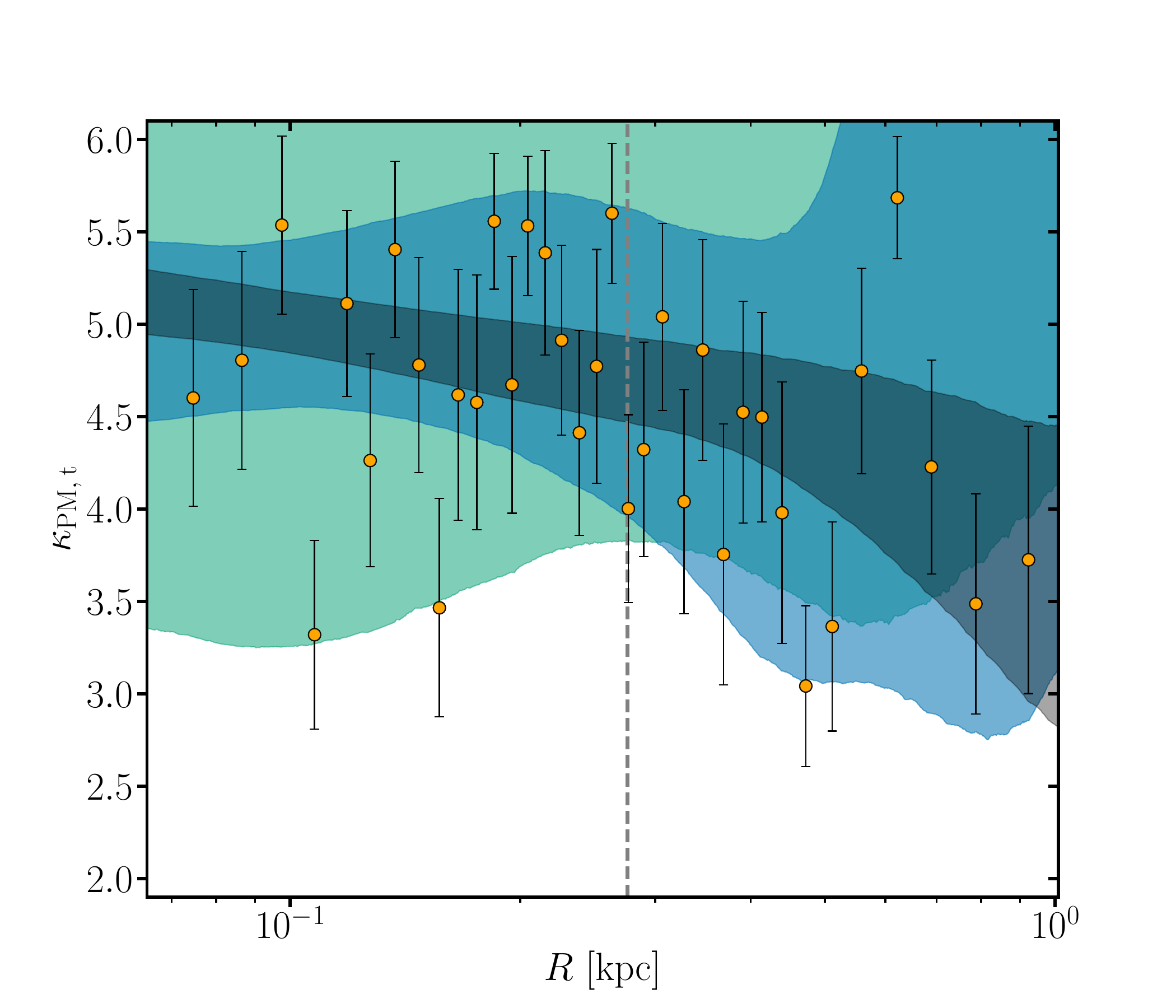}
\hspace{\x}
\includegraphics[width=\y]{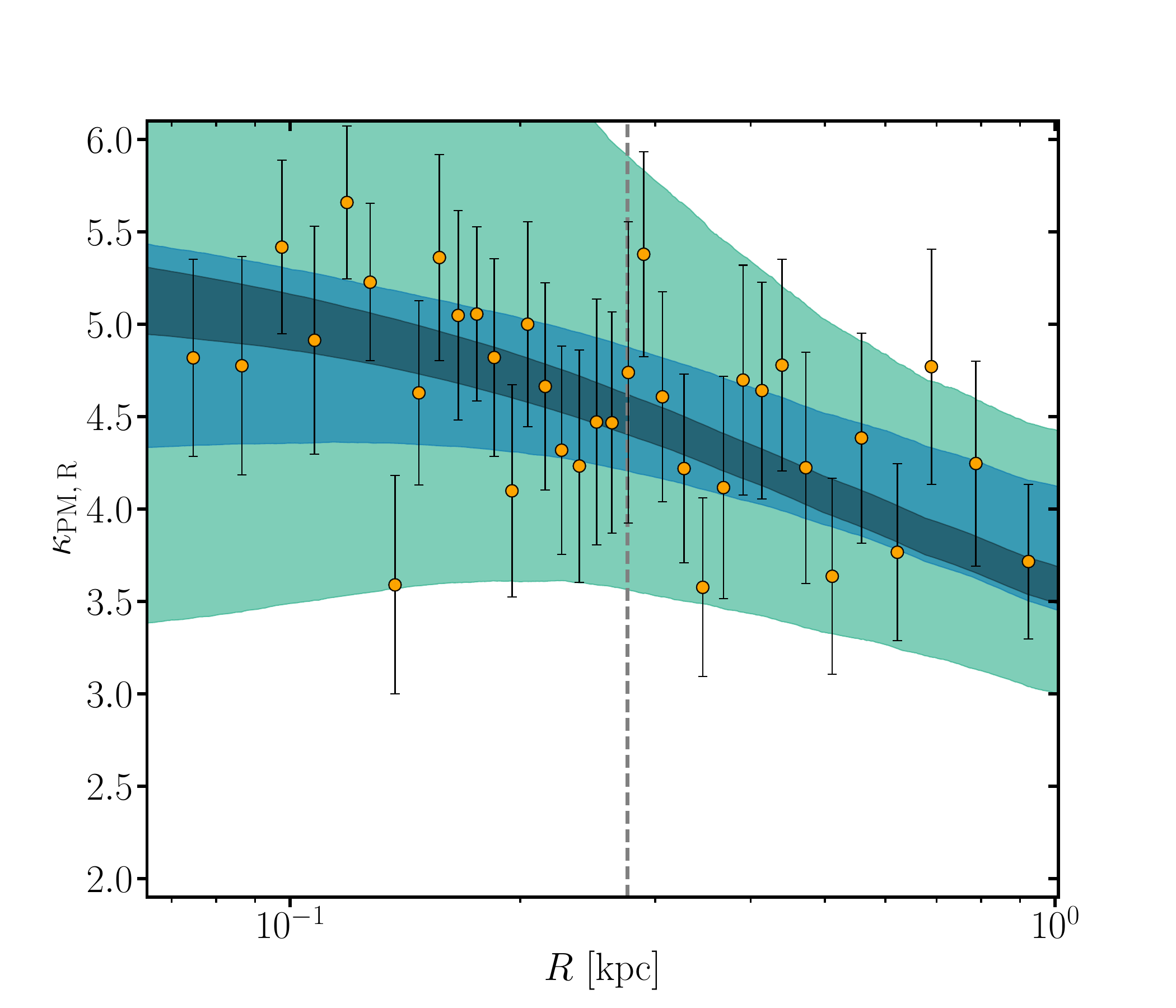}
\vspace{\hb}
 \caption{Projected binned velocity dispersion and kurtosis profiles for the PlumCoreOM \textsc{Gaia Challenge} mock galaxy. LOS and PM velocity dispersion profiles are shown in the upper panels, whilst kurtosis profiles are shown in the bottom ones. The error bars denote 68 \% CL regions from the median, whilst the colored bands denote the 95 \% CL regions for 100 (green), 1,000 (blue), and 10,000 (gray) tracers. The LOS profile results are obtained from the LOS-only model, whilst the PM ones combine both data. Binned profiles are for the 10,000 tracer model in bins with an equal number of stars obtained by fitting the generalized PDFs (Eqs.~\eqref{eq:unik} and  \eqref{eq:lapk}) for a given variance and kurtosis for each bin. The gray, dashed line denotes the projected half-light radius ($R_{1/2}$). \label{fig:kd_core}}
\end{figure*}

\begin{figure*}
    \centering

\includegraphics[width=\y]{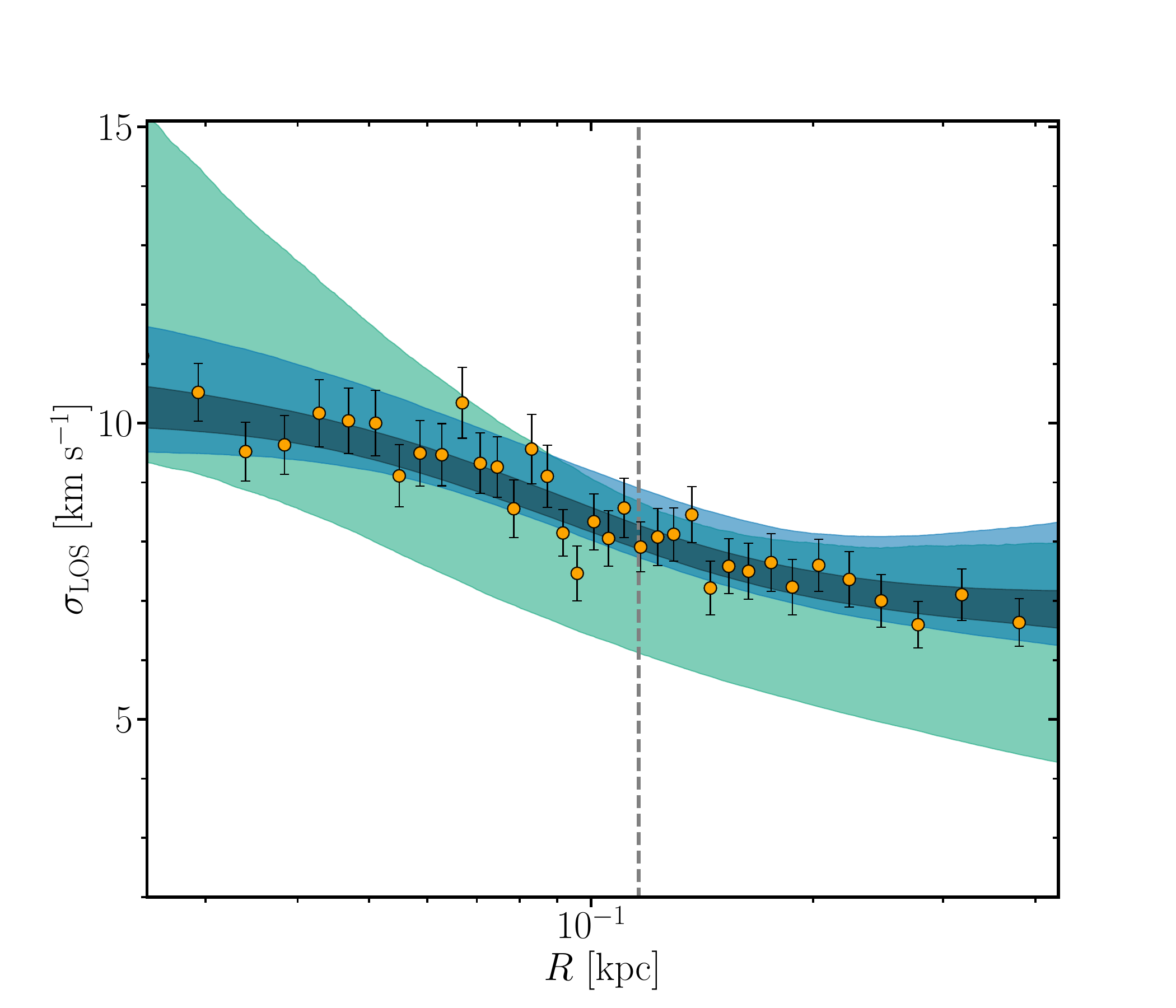}
\hspace{\x}
\includegraphics[width=\y]{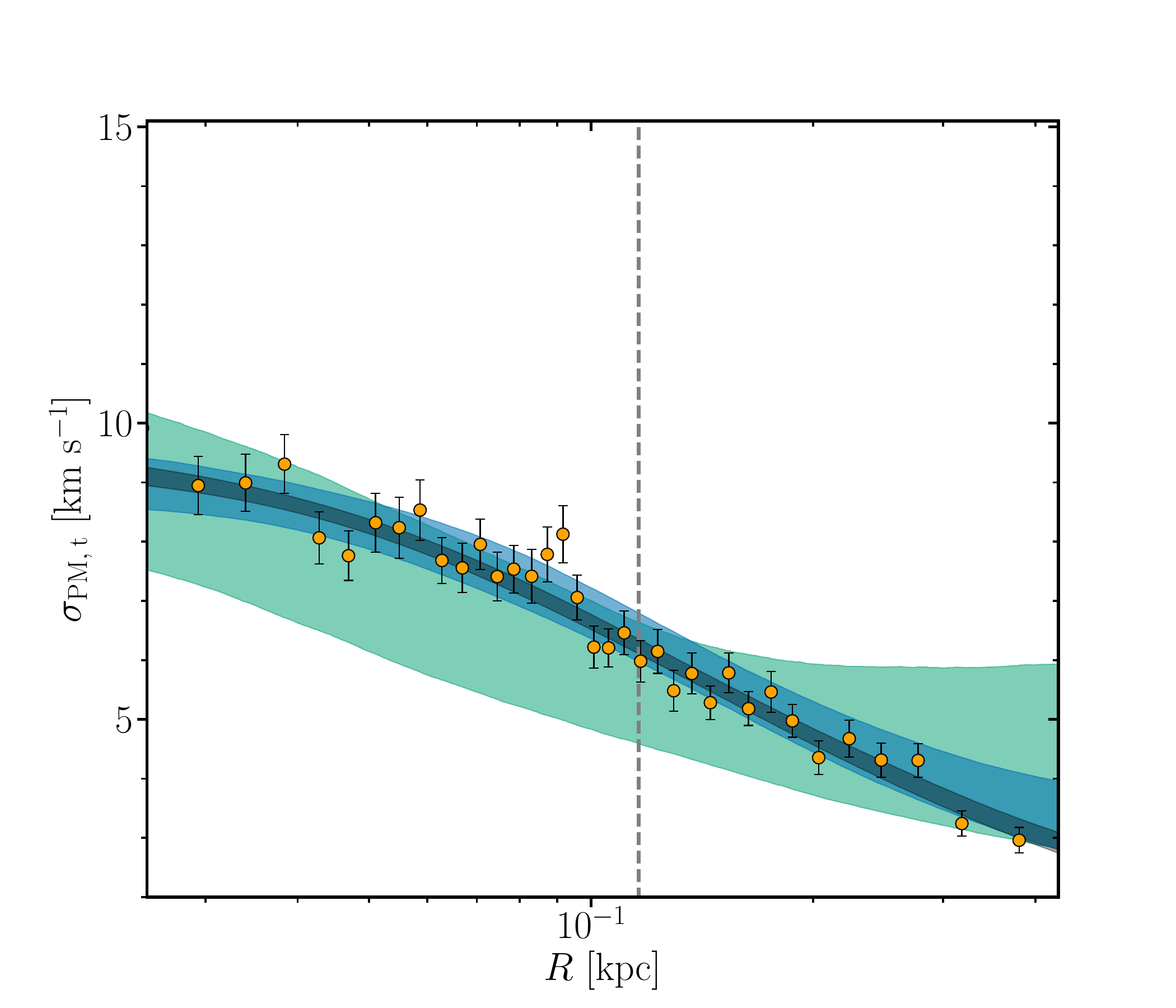}
\hspace{\x}
\includegraphics[width=\y]{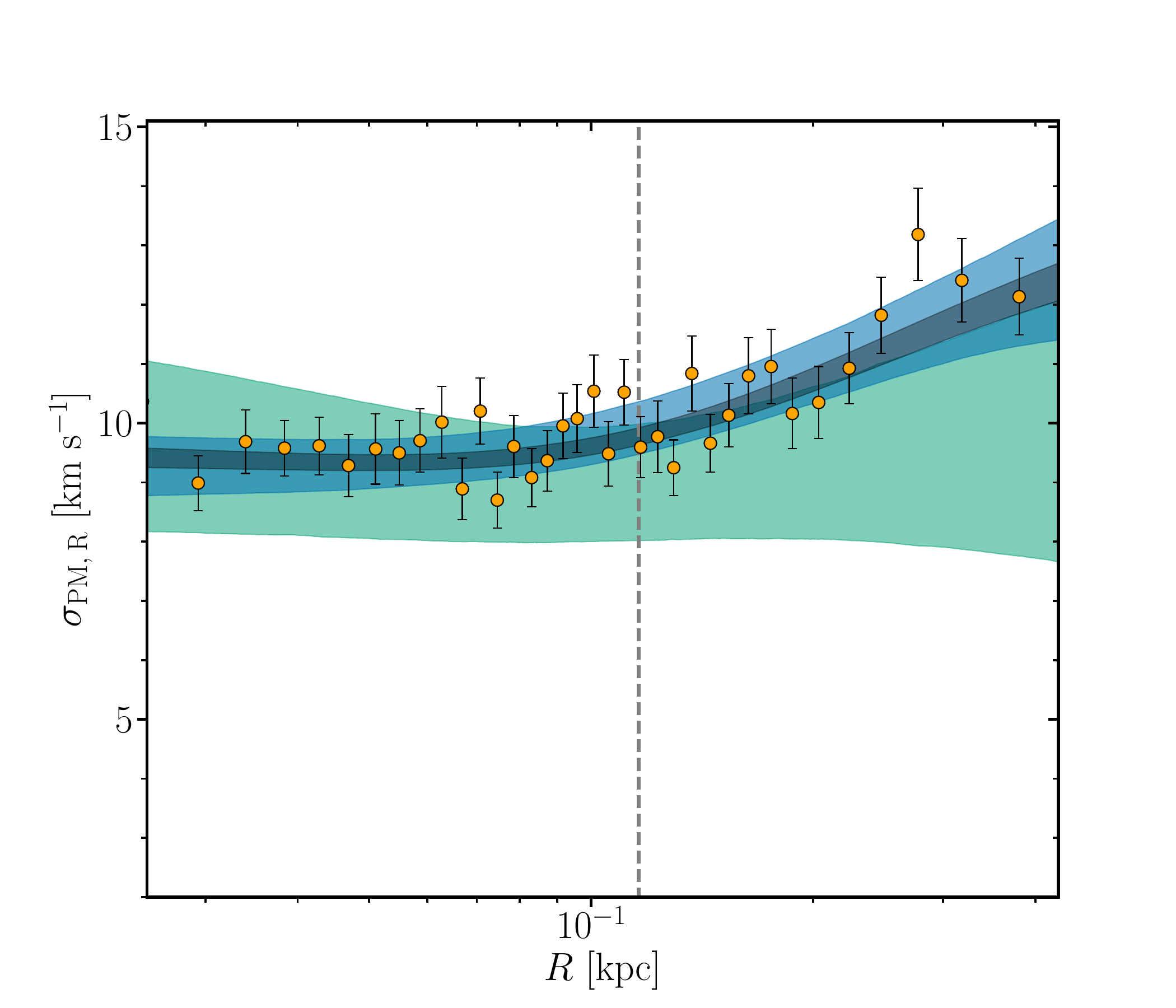}
\vspace{\h}
\includegraphics[width=\y]{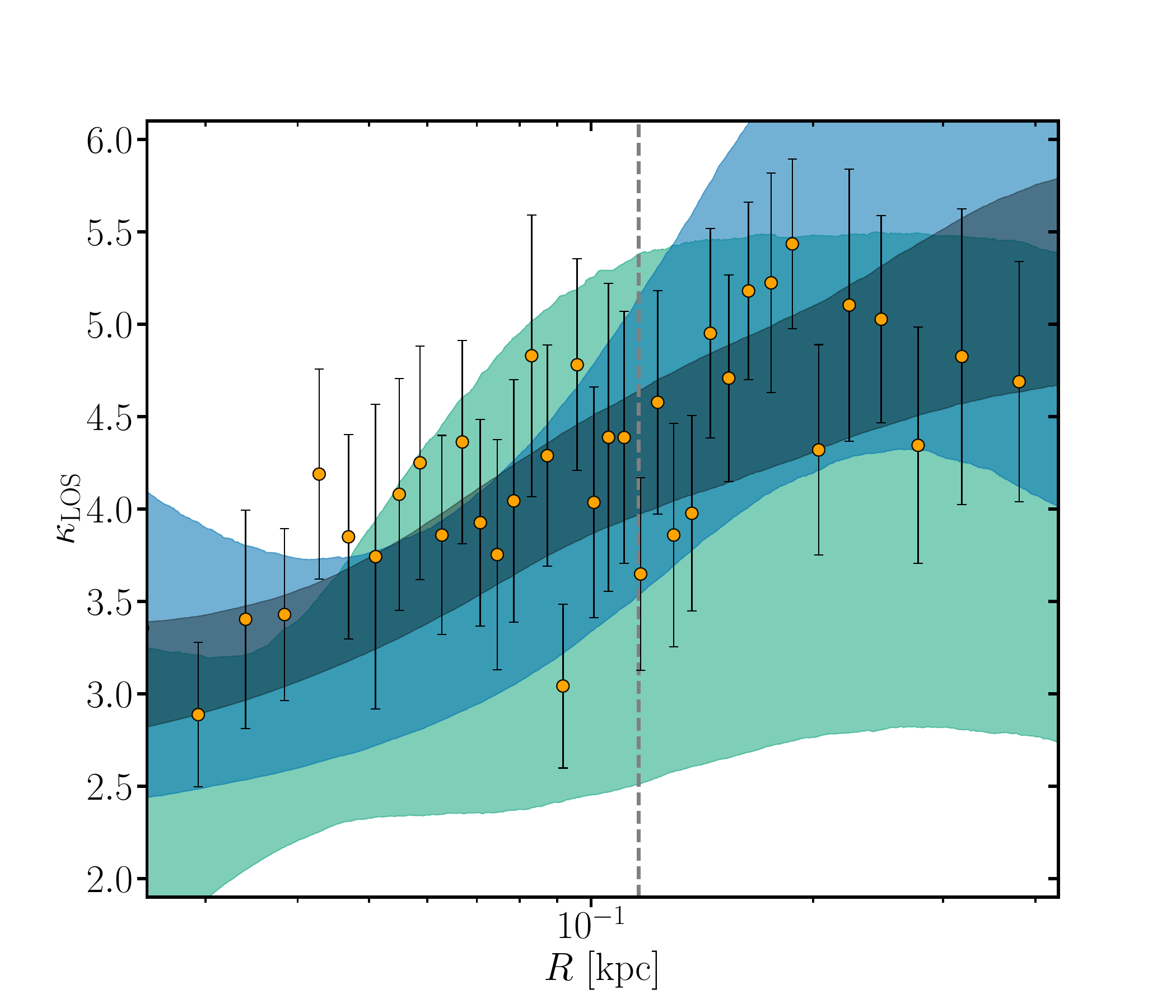}
\hspace{\x}
\includegraphics[width=\y]{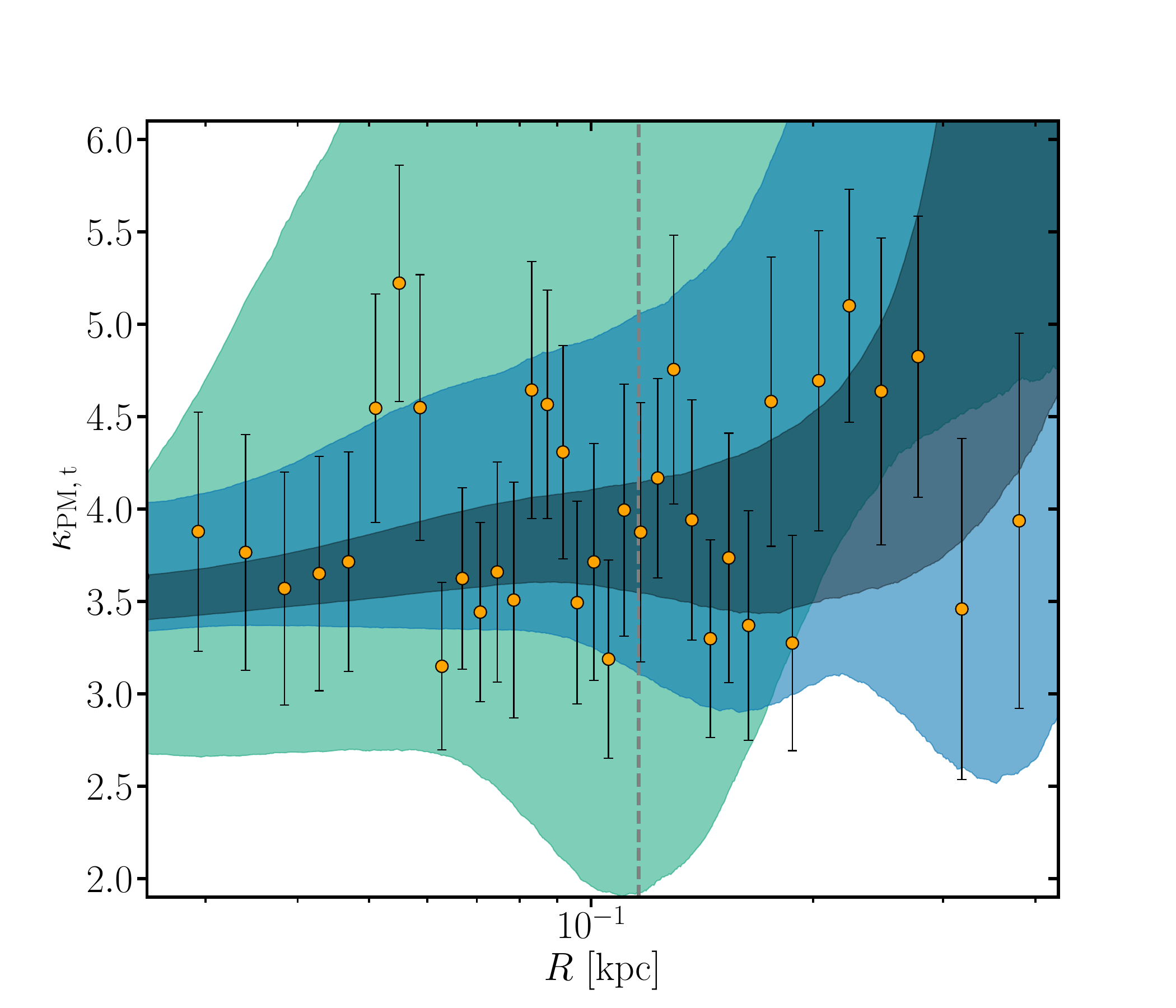}
\hspace{\x}
\includegraphics[width=\y]{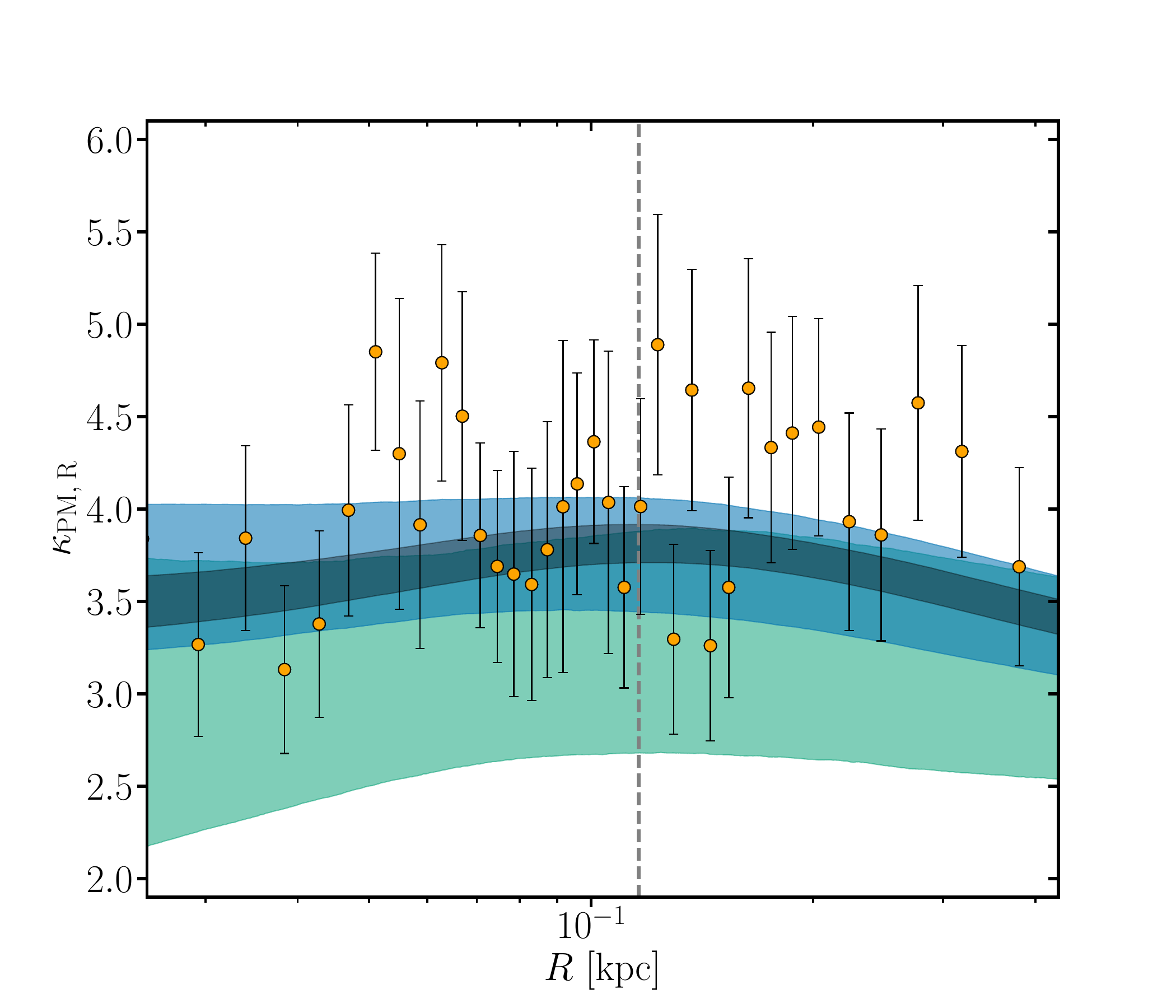}
\vspace{\hb}
 \caption{Same as Fig.~\ref{fig:kd_core} but for the PlumCuspOM \textsc{Gaia Challenge} mock galaxy.
\label{fig:kd_cusp}}
\end{figure*}

Figs.~\ref{fig:kd_core} and \ref{fig:kd_cusp} show binned projected velocity dispersion and kurtosis profiles for the Gaia Challenge mock sample.

\section{Simulated dwarf galaxy binned profiles}
\label{app_bin_sim}
\begin{figure*}
  \centering
\includegraphics[width=\y]{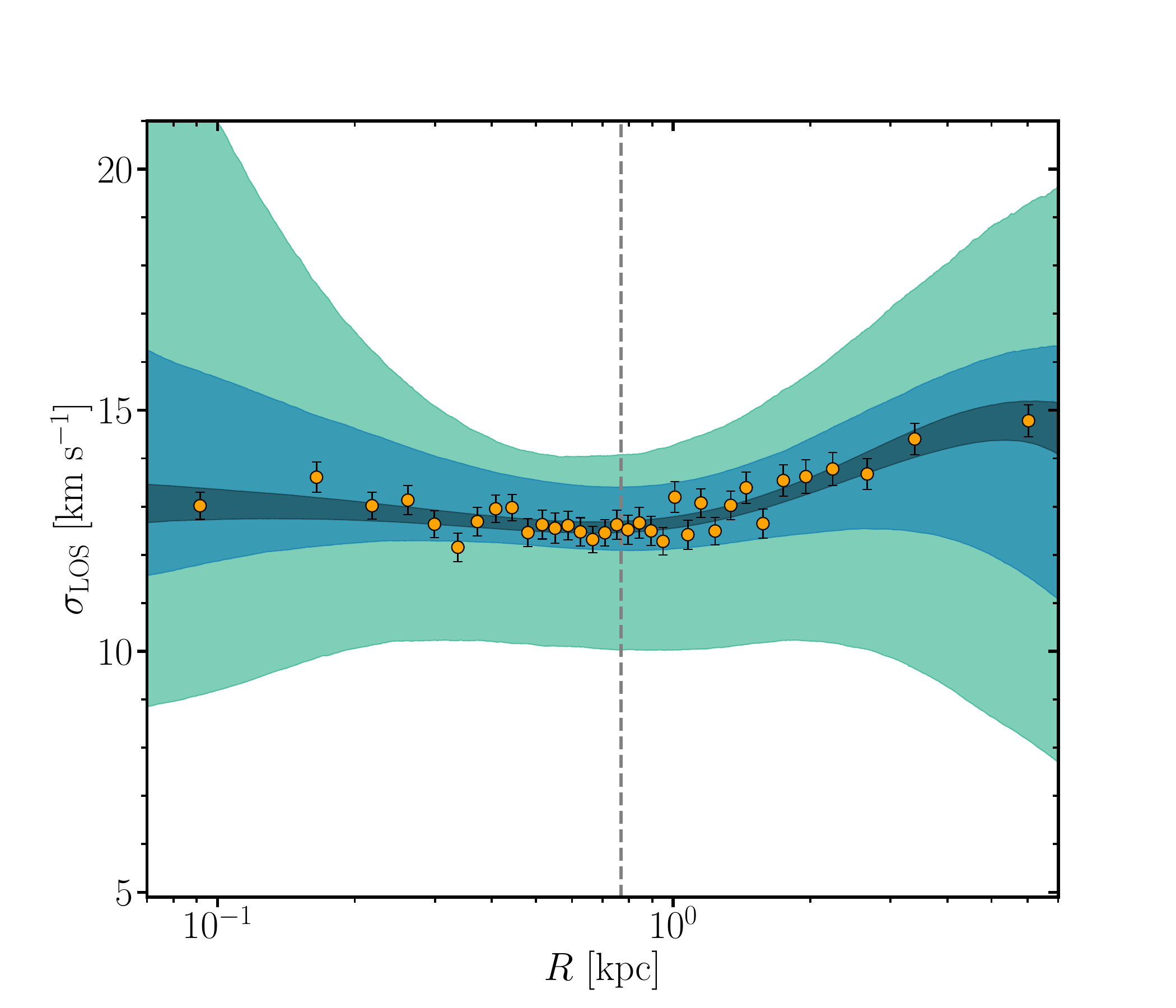}
\hspace{\x}
\includegraphics[width=\y]{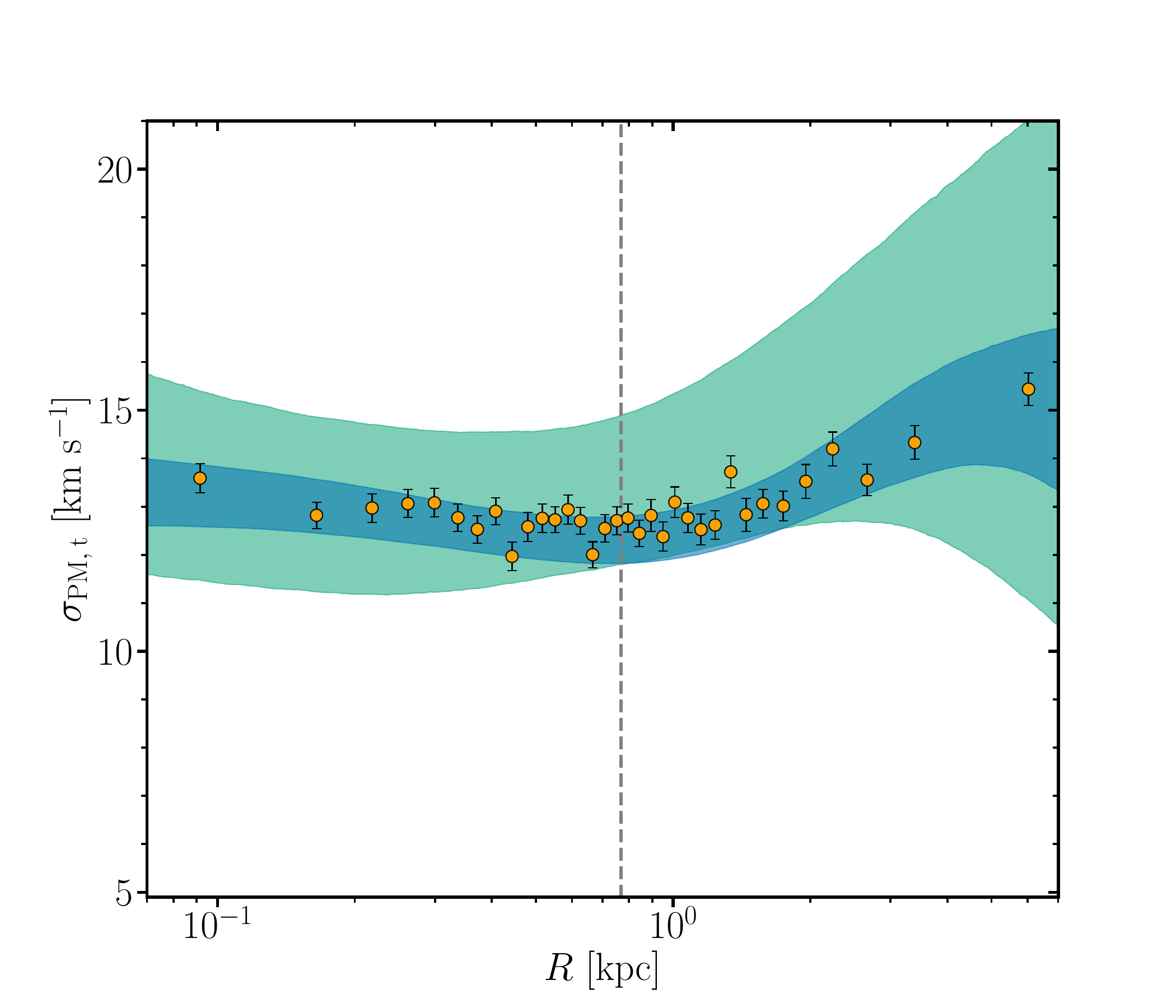}
\hspace{\x}
\includegraphics[width=\y]{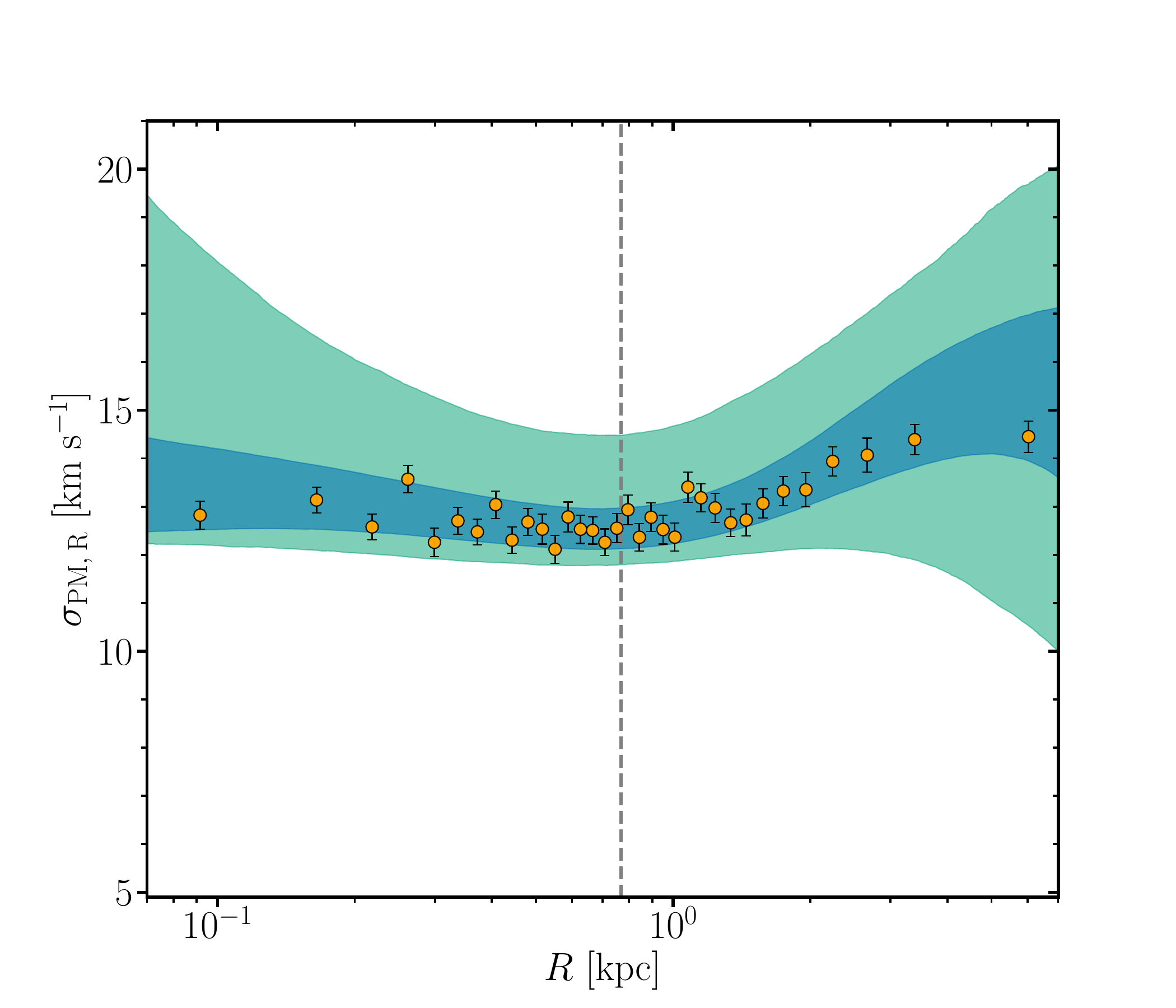}
\vspace{\h}
\includegraphics[width=\y]{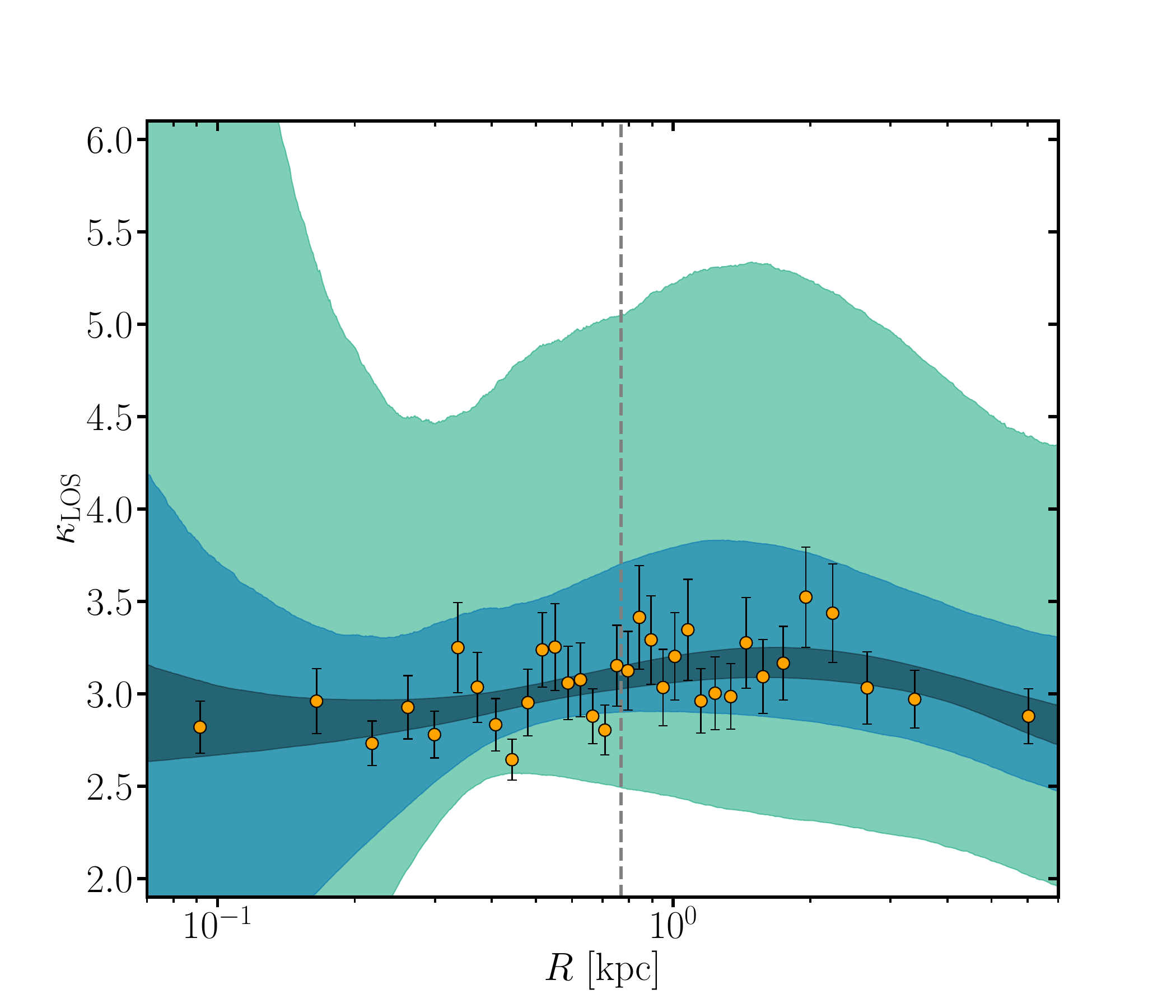}
\hspace{\x}
\includegraphics[width=\y]{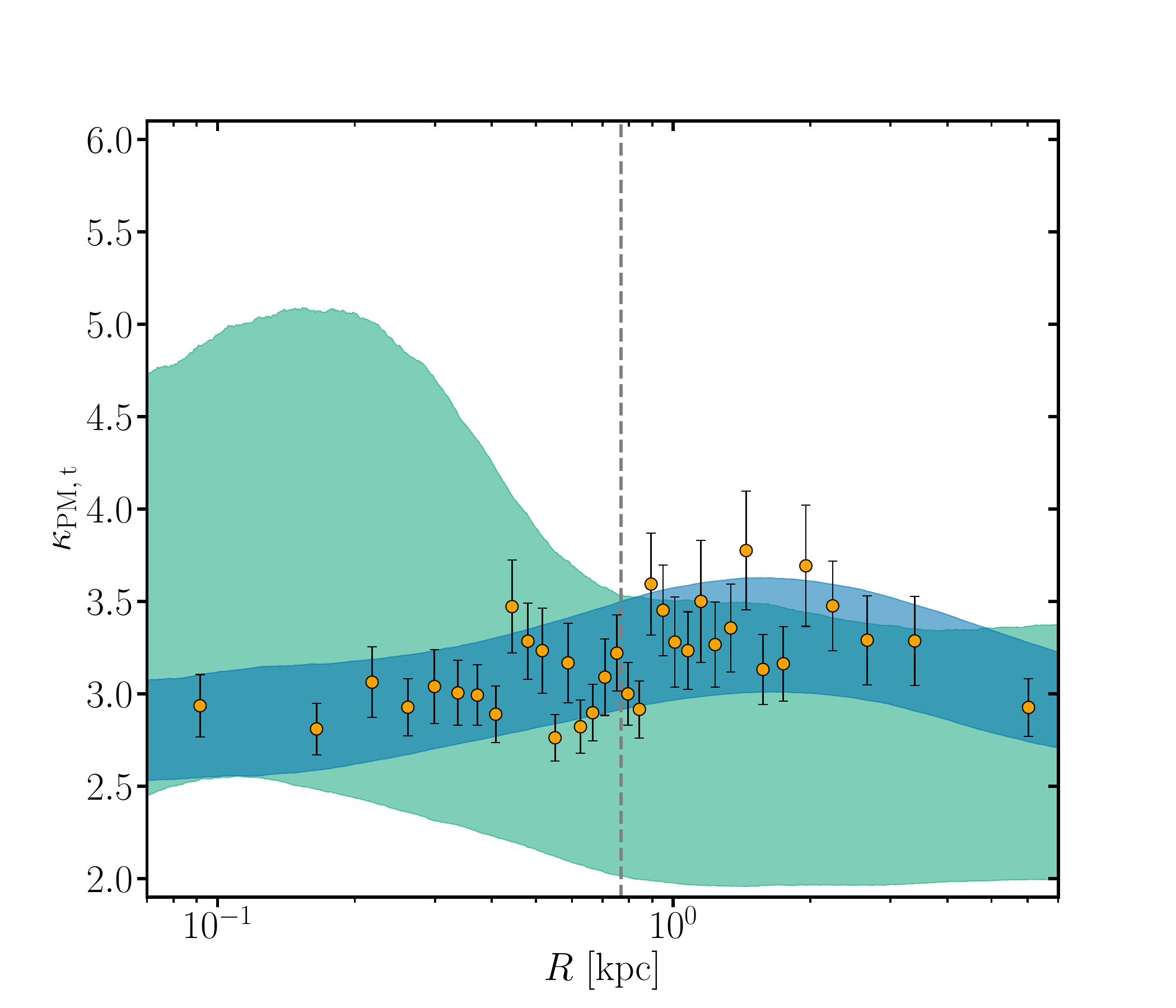}
\hspace{\x}
\includegraphics[width=\y]{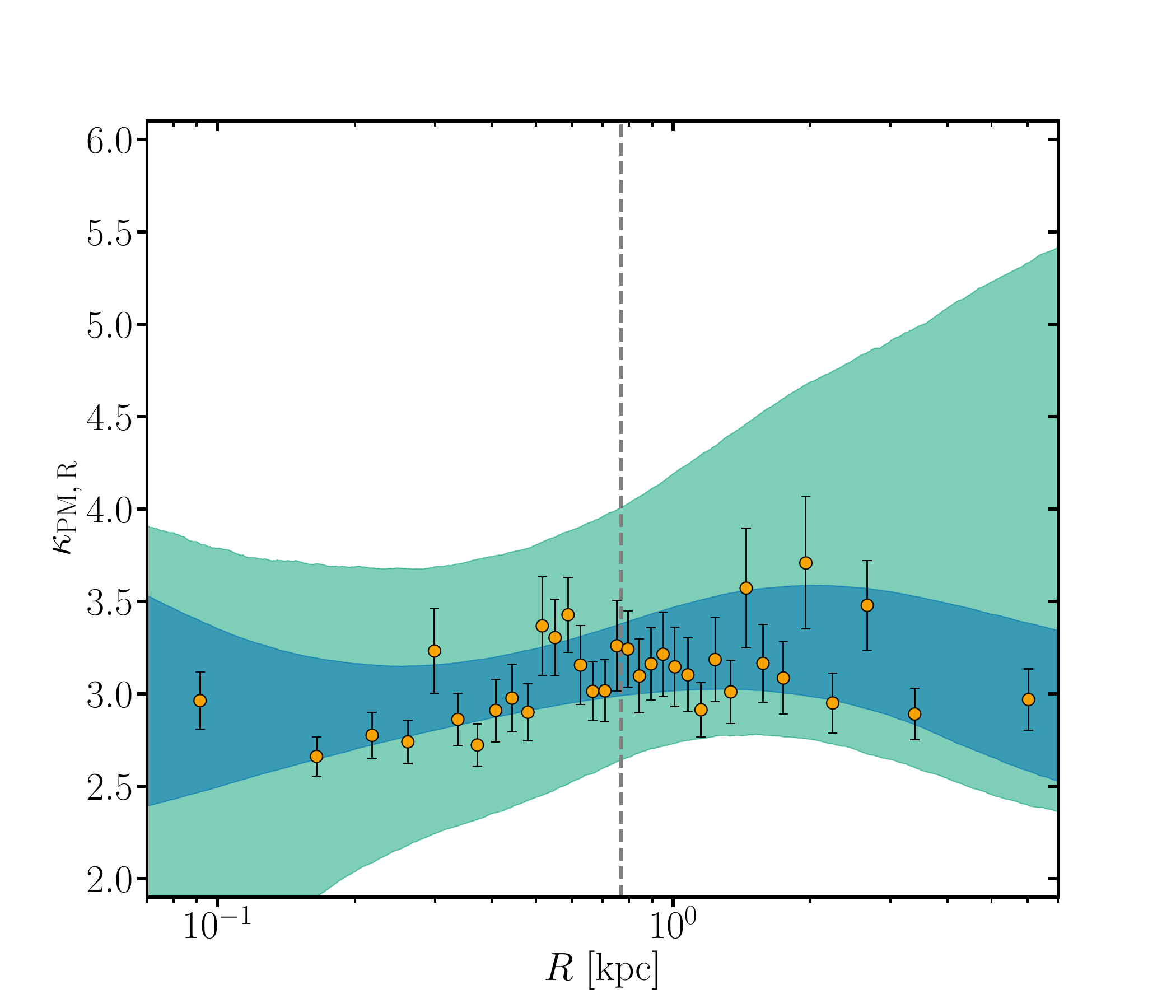}
\vspace{\hb}
 \caption{Projected binned projected velocity dispersion and kurtosis profiles for the Fornax-like simulated dwarf galaxy. The LOS and PM velocity dispersion profiles are shown in the upper panels, whilst kurtosis profiles are shown in the bottom ones. The error bars denote 68 \% CL regions from the median, whilst the colored bands denote the 95 \% CL regions for 100 (green), 1,000 (blue), and all tracers (gray). The LOS profile results are obtained from the LOS-only model, whilst the PM ones combine both data. Binned profiles are derived using all the bound and velocity-clipped tracers from the simulation in bins with of equal number fitting the generalized PDFs (Eqs.~\eqref{eq:unik} and  \eqref{eq:lapk}) for a given variance and kurtosis for each bin. The gray, dashed line denotes the projected half-light radius ($R_{1/2}$). \label{fig:kd_fornax}}
\end{figure*}

Figure~\ref{fig:kd_fornax} shows the projected binned velocity dispersion and kurtosis profiles for the Fornax simulated galaxy. 

\end{appendix}
\end{document}